\documentclass{aa}  

\usepackage{graphicx}
\usepackage{txfonts}
\usepackage{natbib}
\usepackage{hyperref}
\usepackage{booktabs}
\usepackage{mathtools}

\newcommand\gaia{\text{Gaia}\xspace}
\newcommand\gdrone{\gaia DR1\xspace}
\newcommand\gdrtwo{\gaia DR2\xspace}
\newcommand\gdrthree{\gaia DR3\xspace}
\newcommand\gedrthree{\gaia E-DR3\xspace}
\newcommand\gdrfour{\gaia DR4\xspace}
\newcommand\gdrfive{\gaia DR5\xspace}
\newcommand\gfpr{\gaia FPR\xspace}

\newcommand\secrefalt[1]{Section~\ref{#1}}

\def\arcsec{\ensuremath{^{\prime\prime}}\xspace}

\makeatletter
\renewcommand*\aa@pageof{, page \thepage{} of \pageref*{LastPage}}
\makeatother

\newcommand{\orcit}[1]{\protect\href{https://orcid.org/#1}{\protect\includegraphics[width=8pt]{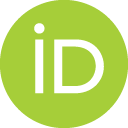}}}

\makeatletter
\renewcommand*\maketitle{%
  \thispagestyle{firstpage}
\begingroup
    \if@wideboxfn
    \setlength\bibindent{1.4\parindent}
    \else
    \setlength\bibindent{\parindent}
    \fi
    \renewcommand*\thefootnote{\@fnsymbol\c@footnote}%
    \renewcommand\@makefntext[1]{%
    \ifaa@longfn\hsize\textwidth\fi
    \noindent
    \hb@xt@\bibindent{\hss\@makefnmark\enspace}##1}
  \ifaa@twocolumn
  \begingroup
    \begin{aa@strip}
          \aa@maketitle
    \end{aa@strip}
    \@thanks            
  \endgroup
  \else
    \begingroup
      \let\thanks\footnote
      \aa@maketitle
    \endgroup
  \fi
\endgroup
  \setcounter{footnote}{0}%
}
\makeatother

\begin{document} 

   \title{Gaia Focused Product Release: Asteroid  orbital solution }

   \subtitle{Properties and assessment}

\author{
{\it Gaia} Collaboration
\and P.        ~David                         \inst{\ref{inst:0001}}
\and F.        ~Mignard                       \inst{\ref{inst:0002}}
\and D.        ~Hestroffer                    \orcit{0000-0003-0472-9459}\inst{\ref{inst:0001}}
\and P.        ~Tanga                         \orcit{0000-0002-2718-997X}\inst{\ref{inst:0002}}
\and F.        ~Spoto                         \orcit{0000-0001-7319-5847}\inst{\ref{inst:0005}}
\and J.        ~Berthier                      \orcit{0000-0003-1846-6485}\inst{\ref{inst:0001}}
\and T.        ~Pauwels                       \inst{\ref{inst:0007}}
\and W.        ~Roux                          \orcit{0000-0002-7816-1950}\inst{\ref{inst:0008}}
\and A.        ~Barbier                       \orcit{0009-0004-0983-931X}\inst{\ref{inst:0008}}
\and A.        ~Cellino                       \orcit{0000-0002-6645-334X}\inst{\ref{inst:0010}}
\and B.        ~Carry                         \orcit{0000-0001-5242-3089}\inst{\ref{inst:0002}}
\and M.        ~Delbo                         \orcit{0000-0002-8963-2404}\inst{\ref{inst:0002}}
\and A.        ~Dell'Oro                      \orcit{0000-0003-1561-9685}\inst{\ref{inst:0013}}
\and C.        ~Fouron                        \inst{\ref{inst:0014}}
\and L.        ~Galluccio                     \orcit{0000-0002-8541-0476}\inst{\ref{inst:0002}}
\and S.A.      ~Klioner                       \orcit{0000-0003-4682-7831}\inst{\ref{inst:0016}}
\and N.        ~Mary                          \inst{\ref{inst:0017}}
\and K.        ~Muinonen                      \orcit{0000-0001-8058-2642}\inst{\ref{inst:0018},\ref{inst:0019}}
\and C.        ~Ordenovic                     \inst{\ref{inst:0002}}
\and I.        ~Oreshina-Slezak               \inst{\ref{inst:0002}}
\and C.        ~Panem                         \inst{\ref{inst:0008}}
\and J.-M.     ~Petit                         \orcit{0000-0003-0407-2266}\inst{\ref{inst:0023}}
\and J.        ~Portell                       \orcit{0000-0002-8886-8925}\inst{\ref{inst:0024},\ref{inst:0025},\ref{inst:0026}}
\and A.G.A.    ~Brown                         \orcit{0000-0002-7419-9679}\inst{\ref{inst:0027}}
\and W.        ~Thuillot                      \orcit{0000-0002-5203-6932}\inst{\ref{inst:0001}}
\and A.        ~Vallenari                     \orcit{0000-0003-0014-519X}\inst{\ref{inst:0029}}
\and T.        ~Prusti                        \orcit{0000-0003-3120-7867}\inst{\ref{inst:0030}}
\and J.H.J.    ~de Bruijne                    \orcit{0000-0001-6459-8599}\inst{\ref{inst:0030}}
\and F.        ~Arenou                        \orcit{0000-0003-2837-3899}\inst{\ref{inst:0032}}
\and C.        ~Babusiaux                     \orcit{0000-0002-7631-348X}\inst{\ref{inst:0033}}
\and M.        ~Biermann                      \orcit{0000-0002-5791-9056}\inst{\ref{inst:0034}}
\and O.L.      ~Creevey                       \orcit{0000-0003-1853-6631}\inst{\ref{inst:0002}}
\and C.        ~Ducourant                     \orcit{0000-0003-4843-8979}\inst{\ref{inst:0036}}
\and D.W.      ~Evans                         \orcit{0000-0002-6685-5998}\inst{\ref{inst:0037}}
\and L.        ~Eyer                          \orcit{0000-0002-0182-8040}\inst{\ref{inst:0038}}
\and R.        ~Guerra                        \orcit{0000-0002-9850-8982}\inst{\ref{inst:0039}}
\and A.        ~Hutton                        \inst{\ref{inst:0040}}
\and C.        ~Jordi                         \orcit{0000-0001-5495-9602}\inst{\ref{inst:0024},\ref{inst:0025},\ref{inst:0026}}
\and U.        ~Lammers                       \orcit{0000-0001-8309-3801}\inst{\ref{inst:0039}}
\and L.        ~Lindegren                     \orcit{0000-0002-5443-3026}\inst{\ref{inst:0045}}
\and X.        ~Luri                          \orcit{0000-0001-5428-9397}\inst{\ref{inst:0024},\ref{inst:0025},\ref{inst:0026}}
\and S.        ~Randich                       \orcit{0000-0003-2438-0899}\inst{\ref{inst:0013}}
\and P.        ~Sartoretti                    \orcit{0000-0002-6574-7565}\inst{\ref{inst:0032}}
\and R.        ~Smiljanic                     \orcit{0000-0003-0942-7855}\inst{\ref{inst:0051}}
\and N.A.      ~Walton                        \orcit{0000-0003-3983-8778}\inst{\ref{inst:0037}}
\and C.A.L.    ~Bailer-Jones                  \inst{\ref{inst:0053}}
\and U.        ~Bastian                       \orcit{0000-0002-8667-1715}\inst{\ref{inst:0034}}
\and M.        ~Cropper                       \orcit{0000-0003-4571-9468}\inst{\ref{inst:0055}}
\and R.        ~Drimmel                       \orcit{0000-0002-1777-5502}\inst{\ref{inst:0010}}
\and D.        ~Katz                          \orcit{0000-0001-7986-3164}\inst{\ref{inst:0032}}
\and C.        ~Soubiran                      \orcit{0000-0003-3304-8134}\inst{\ref{inst:0036}}
\and F.        ~van Leeuwen                   \orcit{0000-0003-1781-4441}\inst{\ref{inst:0037}}
\and M.        ~Audard                        \orcit{0000-0003-4721-034X}\inst{\ref{inst:0038},\ref{inst:0061}}
\and J.        ~Bakker                        \inst{\ref{inst:0039}}
\and R.        ~Blomme                        \orcit{0000-0002-2526-346X}\inst{\ref{inst:0007}}
\and J.        ~Casta\~{n}eda                 \orcit{0000-0001-7820-946X}\inst{\ref{inst:0064},\ref{inst:0024},\ref{inst:0026}}
\and F.        ~De Angeli                     \orcit{0000-0003-1879-0488}\inst{\ref{inst:0037}}
\and C.        ~Fabricius                     \orcit{0000-0003-2639-1372}\inst{\ref{inst:0026},\ref{inst:0024},\ref{inst:0025}}
\and M.        ~Fouesneau                     \orcit{0000-0001-9256-5516}\inst{\ref{inst:0053}}
\and Y.        ~Fr\'{e}mat                    \orcit{0000-0002-4645-6017}\inst{\ref{inst:0007}}
\and A.        ~Guerrier                      \inst{\ref{inst:0008}}
\and E.        ~Masana                        \orcit{0000-0002-4819-329X}\inst{\ref{inst:0026},\ref{inst:0024},\ref{inst:0025}}
\and R.        ~Messineo                      \inst{\ref{inst:0077}}
\and C.        ~Nicolas                       \inst{\ref{inst:0008}}
\and K.        ~Nienartowicz                  \orcit{0000-0001-5415-0547}\inst{\ref{inst:0079},\ref{inst:0061}}
\and F.        ~Pailler                       \orcit{0000-0002-4834-481X}\inst{\ref{inst:0008}}
\and P.        ~Panuzzo                       \orcit{0000-0002-0016-8271}\inst{\ref{inst:0032}}
\and F.        ~Riclet                        \inst{\ref{inst:0008}}
\and G.M.      ~Seabroke                      \orcit{0000-0003-4072-9536}\inst{\ref{inst:0055}}
\and R.        ~Sordo                         \orcit{0000-0003-4979-0659}\inst{\ref{inst:0029}}
\and F.        ~Th\'{e}venin                  \orcit{0000-0002-5032-2476}\inst{\ref{inst:0002}}
\and G.        ~Gracia-Abril                  \inst{\ref{inst:0087},\ref{inst:0034}}
\and D.        ~Teyssier                      \orcit{0000-0002-6261-5292}\inst{\ref{inst:0089}}
\and M.        ~Altmann                       \orcit{0000-0002-0530-0913}\inst{\ref{inst:0034},\ref{inst:0091}}
\and K.        ~Benson                        \inst{\ref{inst:0055}}
\and P.W.      ~Burgess                       \orcit{0009-0002-6668-4559}\inst{\ref{inst:0037}}
\and D.        ~Busonero                      \orcit{0000-0002-3903-7076}\inst{\ref{inst:0010}}
\and G.        ~Busso                         \orcit{0000-0003-0937-9849}\inst{\ref{inst:0037}}
\and H.        ~C\'{a}novas                   \orcit{0000-0001-7668-8022}\inst{\ref{inst:0089}}
\and N.        ~Cheek                         \inst{\ref{inst:0097}}
\and G.        ~Clementini                    \orcit{0000-0001-9206-9723}\inst{\ref{inst:0098}}
\and Y.        ~Damerdji                      \orcit{0000-0002-3107-4024}\inst{\ref{inst:0099},\ref{inst:0100}}
\and M.        ~Davidson                      \orcit{0000-0001-9271-4411}\inst{\ref{inst:0101}}
\and P.        ~de Teodoro                    \inst{\ref{inst:0039}}
\and L.        ~Delchambre                    \orcit{0000-0003-2559-408X}\inst{\ref{inst:0099}}
\and E.        ~Fraile Garcia                 \orcit{0000-0001-7742-9663}\inst{\ref{inst:0104}}
\and D.        ~Garabato                      \orcit{0000-0002-7133-6623}\inst{\ref{inst:0105}}
\and P.        ~Garc\'{i}a-Lario              \orcit{0000-0003-4039-8212}\inst{\ref{inst:0039}}
\and N.        ~Garralda Torres               \inst{\ref{inst:0107}}
\and P.        ~Gavras                        \orcit{0000-0002-4383-4836}\inst{\ref{inst:0104}}
\and R.        ~Haigron                       \inst{\ref{inst:0032}}
\and N.C.      ~Hambly                        \orcit{0000-0002-9901-9064}\inst{\ref{inst:0101}}
\and D.L.      ~Harrison                      \orcit{0000-0001-8687-6588}\inst{\ref{inst:0037},\ref{inst:0112}}
\and D.        ~Hatzidimitriou                \orcit{0000-0002-5415-0464}\inst{\ref{inst:0113}}
\and J.        ~Hern\'{a}ndez                 \orcit{0000-0002-0361-4994}\inst{\ref{inst:0039}}
\and S.T.      ~Hodgkin                       \orcit{0000-0002-5470-3962}\inst{\ref{inst:0037}}
\and B.        ~Holl                          \orcit{0000-0001-6220-3266}\inst{\ref{inst:0038},\ref{inst:0061}}
\and S.        ~Jamal                         \orcit{0000-0002-3929-6668}\inst{\ref{inst:0053}}
\and S.        ~Jordan                        \orcit{0000-0001-6316-6831}\inst{\ref{inst:0034}}
\and A.        ~Krone-Martins                 \orcit{0000-0002-2308-6623}\inst{\ref{inst:0120},\ref{inst:0121}}
\and A.C.      ~Lanzafame                     \orcit{0000-0002-2697-3607}\inst{\ref{inst:0122},\ref{inst:0123}}
\and W.        ~L\"{ o}ffler                  \inst{\ref{inst:0034}}
\and A.        ~Lorca                         \orcit{0000-0002-7985-250X}\inst{\ref{inst:0040}}
\and O.        ~Marchal                       \orcit{ 0000-0001-7461-892}\inst{\ref{inst:0126}}
\and P.M.      ~Marrese                       \orcit{0000-0002-8162-3810}\inst{\ref{inst:0127},\ref{inst:0128}}
\and A.        ~Moitinho                      \orcit{0000-0003-0822-5995}\inst{\ref{inst:0121}}
\and M.        ~Nu\~{n}ez Campos              \inst{\ref{inst:0040}}
\and P.        ~Osborne                       \orcit{0000-0003-4482-3538}\inst{\ref{inst:0037}}
\and E.        ~Pancino                       \orcit{0000-0003-0788-5879}\inst{\ref{inst:0013},\ref{inst:0128}}
\and A.        ~Recio-Blanco                  \orcit{0000-0002-6550-7377}\inst{\ref{inst:0002}}
\and M.        ~Riello                        \orcit{0000-0002-3134-0935}\inst{\ref{inst:0037}}
\and L.        ~Rimoldini                     \orcit{0000-0002-0306-585X}\inst{\ref{inst:0061}}
\and A.C.      ~Robin                         \orcit{0000-0001-8654-9499}\inst{\ref{inst:0023}}
\and T.        ~Roegiers                      \orcit{0000-0002-1231-4440}\inst{\ref{inst:0138}}
\and L.M.      ~Sarro                         \orcit{0000-0002-5622-5191}\inst{\ref{inst:0139}}
\and M.        ~Schultheis                    \orcit{0000-0002-6590-1657}\inst{\ref{inst:0002}}
\and C.        ~Siopis                        \orcit{0000-0002-6267-2924}\inst{\ref{inst:0141}}
\and M.        ~Smith                         \inst{\ref{inst:0055}}
\and A.        ~Sozzetti                      \orcit{0000-0002-7504-365X}\inst{\ref{inst:0010}}
\and E.        ~Utrilla                       \inst{\ref{inst:0040}}
\and M.        ~van Leeuwen                   \orcit{0000-0001-9698-2392}\inst{\ref{inst:0037}}
\and K.        ~Weingrill                     \orcit{0000-0002-8163-2493}\inst{\ref{inst:0146}}
\and U.        ~Abbas                         \orcit{0000-0002-5076-766X}\inst{\ref{inst:0010}}
\and P.        ~\'{A}brah\'{a}m               \orcit{0000-0001-6015-646X}\inst{\ref{inst:0148},\ref{inst:0149}}
\and A.        ~Abreu Aramburu                \orcit{0000-0003-3959-0856}\inst{\ref{inst:0107}}
\and C.        ~Aerts                         \orcit{0000-0003-1822-7126}\inst{\ref{inst:0151},\ref{inst:0152},\ref{inst:0053}}
\and G.        ~Altavilla                     \orcit{0000-0002-9934-1352}\inst{\ref{inst:0127},\ref{inst:0128}}
\and M.A.      ~\'{A}lvarez                   \orcit{0000-0002-6786-2620}\inst{\ref{inst:0105}}
\and J.        ~Alves                         \orcit{0000-0002-4355-0921}\inst{\ref{inst:0157}}
\and R.I.      ~Anderson                      \orcit{0000-0001-8089-4419}\inst{\ref{inst:0158}}
\and T.        ~Antoja                        \orcit{0000-0003-2595-5148}\inst{\ref{inst:0024},\ref{inst:0025},\ref{inst:0026}}
\and D.        ~Baines                        \orcit{0000-0002-6923-3756}\inst{\ref{inst:0162}}
\and S.G.      ~Baker                         \orcit{0000-0002-6436-1257}\inst{\ref{inst:0055}}
\and Z.        ~Balog                         \orcit{0000-0003-1748-2926}\inst{\ref{inst:0034},\ref{inst:0053}}
\and C.        ~Barache                       \inst{\ref{inst:0091}}
\and D.        ~Barbato                       \inst{\ref{inst:0038},\ref{inst:0010}}
\and M.        ~Barros                        \orcit{0000-0002-9728-9618}\inst{\ref{inst:0169}}
\and M.A.      ~Barstow                       \orcit{0000-0002-7116-3259}\inst{\ref{inst:0170}}
\and S.        ~Bartolom\'{e}                 \orcit{0000-0002-6290-6030}\inst{\ref{inst:0026},\ref{inst:0024},\ref{inst:0025}}
\and D.        ~Bashi                         \orcit{0000-0002-9035-2645}\inst{\ref{inst:0174},\ref{inst:0175}}
\and N.        ~Bauchet                       \orcit{0000-0002-2307-8973}\inst{\ref{inst:0032}}
\and N.        ~Baudeau                       \inst{\ref{inst:0014}}
\and U.        ~Becciani                      \orcit{0000-0002-4389-8688}\inst{\ref{inst:0122}}
\and L.R.      ~Bedin                         \inst{\ref{inst:0029}}
\and I.        ~Bellas-Velidis                \inst{\ref{inst:0180}}
\and M.        ~Bellazzini                    \orcit{0000-0001-8200-810X}\inst{\ref{inst:0098}}
\and W.        ~Beordo                        \orcit{0000-0002-5094-1306}\inst{\ref{inst:0010},\ref{inst:0183}}
\and A.        ~Berihuete                     \orcit{0000-0002-8589-4423}\inst{\ref{inst:0184}}
\and M.        ~Bernet                        \orcit{0000-0001-7503-1010}\inst{\ref{inst:0024},\ref{inst:0025},\ref{inst:0026}}
\and C.        ~Bertolotto                    \inst{\ref{inst:0077}}
\and S.        ~Bertone                       \orcit{0000-0001-9885-8440}\inst{\ref{inst:0010}}
\and L.        ~Bianchi                       \orcit{0000-0002-7999-4372}\inst{\ref{inst:0190}}
\and A.        ~Binnenfeld                    \orcit{0000-0002-9319-3838}\inst{\ref{inst:0191}}
\and A.        ~Blazere                       \inst{\ref{inst:0192}}
\and T.        ~Boch                          \orcit{0000-0001-5818-2781}\inst{\ref{inst:0126}}
\and A.        ~Bombrun                       \inst{\ref{inst:0194}}
\and S.        ~Bouquillon                    \inst{\ref{inst:0091},\ref{inst:0196}}
\and A.        ~Bragaglia                     \orcit{0000-0002-0338-7883}\inst{\ref{inst:0098}}
\and J.        ~Braine                        \orcit{0000-0003-1740-1284}\inst{\ref{inst:0036}}
\and L.        ~Bramante                      \inst{\ref{inst:0077}}
\and E.        ~Breedt                        \orcit{0000-0001-6180-3438}\inst{\ref{inst:0037}}
\and A.        ~Bressan                       \orcit{0000-0002-7922-8440}\inst{\ref{inst:0201}}
\and N.        ~Brouillet                     \orcit{0000-0002-3274-7024}\inst{\ref{inst:0036}}
\and E.        ~Brugaletta                    \orcit{0000-0003-2598-6737}\inst{\ref{inst:0122}}
\and B.        ~Bucciarelli                   \orcit{0000-0002-5303-0268}\inst{\ref{inst:0010},\ref{inst:0183}}
\and A.G.      ~Butkevich                     \orcit{0000-0002-4098-3588}\inst{\ref{inst:0010}}
\and R.        ~Buzzi                         \orcit{0000-0001-9389-5701}\inst{\ref{inst:0010}}
\and E.        ~Caffau                        \orcit{0000-0001-6011-6134}\inst{\ref{inst:0032}}
\and R.        ~Cancelliere                   \orcit{0000-0002-9120-3799}\inst{\ref{inst:0209}}
\and S.        ~Cannizzo                      \inst{\ref{inst:0017}}
\and R.        ~Carballo                      \orcit{0000-0001-7412-2498}\inst{\ref{inst:0212}}
\and T.        ~Carlucci                      \inst{\ref{inst:0091}}
\and M.I.      ~Carnerero                     \orcit{0000-0001-5843-5515}\inst{\ref{inst:0010}}
\and J.M.      ~Carrasco                      \orcit{0000-0002-3029-5853}\inst{\ref{inst:0026},\ref{inst:0024},\ref{inst:0025}}
\and J.        ~Carretero                     \orcit{0000-0002-3130-0204}\inst{\ref{inst:0218},\ref{inst:0219}}
\and S.        ~Carton                        \inst{\ref{inst:0017}}
\and L.        ~Casamiquela                   \orcit{0000-0001-5238-8674}\inst{\ref{inst:0036},\ref{inst:0032}}
\and M.        ~Castellani                    \orcit{0000-0002-7650-7428}\inst{\ref{inst:0127}}
\and A.        ~Castro-Ginard                 \orcit{0000-0002-9419-3725}\inst{\ref{inst:0027}}
\and V.        ~Cesare                        \orcit{0000-0003-1119-4237}\inst{\ref{inst:0122}}
\and P.        ~Charlot                       \orcit{0000-0002-9142-716X}\inst{\ref{inst:0036}}
\and L.        ~Chemin                        \orcit{0000-0002-3834-7937}\inst{\ref{inst:0227}}
\and V.        ~Chiaramida                    \inst{\ref{inst:0077}}
\and A.        ~Chiavassa                     \orcit{0000-0003-3891-7554}\inst{\ref{inst:0002}}
\and N.        ~Chornay                       \orcit{0000-0002-8767-3907}\inst{\ref{inst:0037},\ref{inst:0061}}
\and R.        ~Collins                       \orcit{0000-0001-8437-1703}\inst{\ref{inst:0101}}
\and G.        ~Contursi                      \orcit{0000-0001-5370-1511}\inst{\ref{inst:0002}}
\and W.J.      ~Cooper                        \orcit{0000-0003-3501-8967}\inst{\ref{inst:0234},\ref{inst:0010}}
\and T.        ~Cornez                        \inst{\ref{inst:0017}}
\and M.        ~Crosta                        \orcit{0000-0003-4369-3786}\inst{\ref{inst:0010},\ref{inst:0238}}
\and C.        ~Crowley                       \orcit{0000-0002-9391-9360}\inst{\ref{inst:0194}}
\and C.        ~Dafonte                       \orcit{0000-0003-4693-7555}\inst{\ref{inst:0105}}
\and P.        ~de Laverny                    \orcit{0000-0002-2817-4104}\inst{\ref{inst:0002}}
\and F.        ~De Luise                      \orcit{0000-0002-6570-8208}\inst{\ref{inst:0242}}
\and R.        ~De March                      \orcit{0000-0003-0567-842X}\inst{\ref{inst:0077}}
\and R.        ~de Souza                      \orcit{0009-0007-7669-0254}\inst{\ref{inst:0244}}
\and A.        ~de Torres                     \inst{\ref{inst:0194}}
\and E.F.      ~del Peloso                    \inst{\ref{inst:0034}}
\and A.        ~Delgado                       \inst{\ref{inst:0104}}
\and T.E.      ~Dharmawardena                 \orcit{0000-0002-9583-5216}\inst{\ref{inst:0053}}
\and S.        ~Diakite                       \inst{\ref{inst:0249}}
\and C.        ~Diener                        \inst{\ref{inst:0037}}
\and E.        ~Distefano                     \orcit{0000-0002-2448-2513}\inst{\ref{inst:0122}}
\and C.        ~Dolding                       \inst{\ref{inst:0055}}
\and K.        ~Dsilva                        \orcit{0000-0002-1476-9772}\inst{\ref{inst:0141}}
\and J.        ~Dur\'{a}n                     \inst{\ref{inst:0104}}
\and H.        ~Enke                          \orcit{0000-0002-2366-8316}\inst{\ref{inst:0146}}
\and P.        ~Esquej                        \orcit{0000-0001-8195-628X}\inst{\ref{inst:0104}}
\and C.        ~Fabre                         \inst{\ref{inst:0192}}
\and M.        ~Fabrizio                      \orcit{0000-0001-5829-111X}\inst{\ref{inst:0127},\ref{inst:0128}}
\and S.        ~Faigler                       \orcit{0000-0002-8368-5724}\inst{\ref{inst:0174}}
\and M.        ~Fatovi\'{c}                   \orcit{0000-0003-1911-4326}\inst{\ref{inst:0261}}
\and G.        ~Fedorets                      \orcit{0000-0002-8418-4809}\inst{\ref{inst:0018},\ref{inst:0263}}
\and J.        ~Fern\'{a}ndez-Hern\'{a}ndez   \inst{\ref{inst:0104}}
\and P.        ~Fernique                      \orcit{0000-0002-3304-2923}\inst{\ref{inst:0126}}
\and F.        ~Figueras                      \orcit{0000-0002-3393-0007}\inst{\ref{inst:0024},\ref{inst:0025},\ref{inst:0026}}
\and Y.        ~Fournier                      \orcit{0000-0002-6633-9088}\inst{\ref{inst:0146}}
\and M.        ~Gai                           \orcit{0000-0001-9008-134X}\inst{\ref{inst:0010}}
\and M.        ~Galinier                      \orcit{0000-0001-7920-0133}\inst{\ref{inst:0002}}
\and A.        ~Garcia-Gutierrez              \inst{\ref{inst:0026},\ref{inst:0024},\ref{inst:0025}}
\and M.        ~Garc\'{i}a-Torres             \orcit{0000-0002-6867-7080}\inst{\ref{inst:0275}}
\and A.        ~Garofalo                      \orcit{0000-0002-5907-0375}\inst{\ref{inst:0098}}
\and E.        ~Gerlach                       \orcit{0000-0002-9533-2168}\inst{\ref{inst:0016}}
\and R.        ~Geyer                         \orcit{0000-0001-6967-8707}\inst{\ref{inst:0016}}
\and P.        ~Giacobbe                      \orcit{0000-0001-7034-7024}\inst{\ref{inst:0010}}
\and G.        ~Gilmore                       \orcit{0000-0003-4632-0213}\inst{\ref{inst:0037},\ref{inst:0281}}
\and S.        ~Girona                        \orcit{0000-0002-1975-1918}\inst{\ref{inst:0282}}
\and G.        ~Giuffrida                     \orcit{0000-0002-8979-4614}\inst{\ref{inst:0127}}
\and R.        ~Gomel                         \inst{\ref{inst:0174}}
\and A.        ~Gomez                         \orcit{0000-0002-3796-3690}\inst{\ref{inst:0105}}
\and J.        ~Gonz\'{a}lez-N\'{u}\~{n}ez    \orcit{0000-0001-5311-5555}\inst{\ref{inst:0286}}
\and I.        ~Gonz\'{a}lez-Santamar\'{i}a   \orcit{0000-0002-8537-9384}\inst{\ref{inst:0105}}
\and E.        ~Gosset                        \inst{\ref{inst:0099},\ref{inst:0289}}
\and M.        ~Granvik                       \orcit{0000-0002-5624-1888}\inst{\ref{inst:0018},\ref{inst:0291}}
\and V.        ~Gregori Barrera               \inst{\ref{inst:0026},\ref{inst:0024},\ref{inst:0025}}
\and R.        ~Guti\'{e}rrez-S\'{a}nchez     \orcit{0009-0003-1500-4733}\inst{\ref{inst:0089}}
\and M.        ~Haywood                       \orcit{0000-0003-0434-0400}\inst{\ref{inst:0032}}
\and A.        ~Helmer                        \inst{\ref{inst:0017}}
\and A.        ~Helmi                         \orcit{0000-0003-3937-7641}\inst{\ref{inst:0298}}
\and K.        ~Henares                       \inst{\ref{inst:0162}}
\and S.L.      ~Hidalgo                       \orcit{0000-0002-0002-9298}\inst{\ref{inst:0300},\ref{inst:0301}}
\and T.        ~Hilger                        \orcit{0000-0003-1646-0063}\inst{\ref{inst:0016}}
\and D.        ~Hobbs                         \orcit{0000-0002-2696-1366}\inst{\ref{inst:0045}}
\and C.        ~Hottier                       \orcit{0000-0002-3498-3944}\inst{\ref{inst:0032}}
\and H.E.      ~Huckle                        \inst{\ref{inst:0055}}
\and M.        ~Jab\l{}o\'{n}ska              \orcit{0000-0001-6962-4979}\inst{\ref{inst:0306},\ref{inst:0307}}
\and F.        ~Jansen                        \inst{\ref{inst:0308}}
\and \'{O}.    ~Jim\'{e}nez-Arranz            \orcit{0000-0001-7434-5165}\inst{\ref{inst:0024},\ref{inst:0025},\ref{inst:0026}}
\and J.        ~Juaristi Campillo             \inst{\ref{inst:0034}}
\and S.        ~Khanna                        \orcit{0000-0002-2604-4277}\inst{\ref{inst:0010},\ref{inst:0298}}
\and G.        ~Kordopatis                    \orcit{0000-0002-9035-3920}\inst{\ref{inst:0002}}
\and \'{A}     ~K\'{o}sp\'{a}l                \orcit{0000-0001-7157-6275}\inst{\ref{inst:0148},\ref{inst:0053},\ref{inst:0149}}
\and Z.        ~Kostrzewa-Rutkowska           \inst{\ref{inst:0027}}
\and M.        ~Kun                           \orcit{0000-0002-7538-5166}\inst{\ref{inst:0148}}
\and S.        ~Lambert                       \orcit{0000-0001-6759-5502}\inst{\ref{inst:0091}}
\and A.F.      ~Lanza                         \orcit{0000-0001-5928-7251}\inst{\ref{inst:0122}}
\and J.-F.     ~Le Campion                    \inst{\ref{inst:0036}}
\and Y.        ~Lebreton                      \orcit{0000-0002-4834-2144}\inst{\ref{inst:0324},\ref{inst:0325}}
\and T.        ~Lebzelter                     \orcit{0000-0002-0702-7551}\inst{\ref{inst:0157}}
\and S.        ~Leccia                        \orcit{0000-0001-5685-6930}\inst{\ref{inst:0327}}
\and I.        ~Lecoeur-Taibi                 \orcit{0000-0003-0029-8575}\inst{\ref{inst:0061}}
\and G.        ~Lecoutre                      \inst{\ref{inst:0023}}
\and S.        ~Liao                          \orcit{0000-0002-9346-0211}\inst{\ref{inst:0330},\ref{inst:0010},\ref{inst:0332}}
\and L.        ~Liberato                      \orcit{0000-0003-3433-6269}\inst{\ref{inst:0002},\ref{inst:0334}}
\and E.        ~Licata                        \orcit{0000-0002-5203-0135}\inst{\ref{inst:0010}}
\and H.E.P.    ~Lindstr{\o}m                  \orcit{0009-0004-8864-5459}\inst{\ref{inst:0010},\ref{inst:0337},\ref{inst:0338}}
\and T.A.      ~Lister                        \orcit{0000-0002-3818-7769}\inst{\ref{inst:0339}}
\and E.        ~Livanou                       \orcit{0000-0003-0628-2347}\inst{\ref{inst:0113}}
\and A.        ~Lobel                         \orcit{0000-0001-5030-019X}\inst{\ref{inst:0007}}
\and C.        ~Loup                          \inst{\ref{inst:0126}}
\and L.        ~Mahy                          \orcit{0000-0003-0688-7987}\inst{\ref{inst:0007}}
\and R.G.      ~Mann                          \orcit{0000-0002-0194-325X}\inst{\ref{inst:0101}}
\and M.        ~Manteiga                      \orcit{0000-0002-7711-5581}\inst{\ref{inst:0345}}
\and J.M.      ~Marchant                      \orcit{0000-0002-3678-3145}\inst{\ref{inst:0346}}
\and M.        ~Marconi                       \orcit{0000-0002-1330-2927}\inst{\ref{inst:0327}}
\and D.        ~Mar\'{i}n Pina                \orcit{0000-0001-6482-1842}\inst{\ref{inst:0024},\ref{inst:0025},\ref{inst:0026}}
\and S.        ~Marinoni                      \orcit{0000-0001-7990-6849}\inst{\ref{inst:0127},\ref{inst:0128}}
\and D.J.      ~Marshall                      \orcit{0000-0003-3956-3524}\inst{\ref{inst:0353}}
\and J.        ~Mart\'{i}n Lozano             \orcit{0009-0001-2435-6680}\inst{\ref{inst:0097}}
\and J.M.      ~Mart\'{i}n-Fleitas            \orcit{0000-0002-8594-569X}\inst{\ref{inst:0040}}
\and G.        ~Marton                        \orcit{0000-0002-1326-1686}\inst{\ref{inst:0148}}
\and A.        ~Masip                         \orcit{0000-0003-1419-0020}\inst{\ref{inst:0026},\ref{inst:0024},\ref{inst:0025}}
\and D.        ~Massari                       \orcit{0000-0001-8892-4301}\inst{\ref{inst:0098}}
\and A.        ~Mastrobuono-Battisti          \orcit{0000-0002-2386-9142}\inst{\ref{inst:0032}}
\and T.        ~Mazeh                         \orcit{0000-0002-3569-3391}\inst{\ref{inst:0174}}
\and P.J.      ~McMillan                      \orcit{0000-0002-8861-2620}\inst{\ref{inst:0045}}
\and J.        ~Meichsner                     \orcit{0000-0002-9900-7864}\inst{\ref{inst:0016}}
\and S.        ~Messina                       \orcit{0000-0002-2851-2468}\inst{\ref{inst:0122}}
\and D.        ~Michalik                      \orcit{0000-0002-7618-6556}\inst{\ref{inst:0030}}
\and N.R.      ~Millar                        \inst{\ref{inst:0037}}
\and A.        ~Mints                         \orcit{0000-0002-8440-1455}\inst{\ref{inst:0146}}
\and D.        ~Molina                        \orcit{0000-0003-4814-0275}\inst{\ref{inst:0025},\ref{inst:0024},\ref{inst:0026}}
\and R.        ~Molinaro                      \orcit{0000-0003-3055-6002}\inst{\ref{inst:0327}}
\and L.        ~Moln\'{a}r                    \orcit{0000-0002-8159-1599}\inst{\ref{inst:0148},\ref{inst:0374},\ref{inst:0149}}
\and G.        ~Monari                        \orcit{0000-0002-6863-0661}\inst{\ref{inst:0126}}
\and M.        ~Mongui\'{o}                   \orcit{0000-0002-4519-6700}\inst{\ref{inst:0024},\ref{inst:0025},\ref{inst:0026}}
\and P.        ~Montegriffo                   \orcit{0000-0001-5013-5948}\inst{\ref{inst:0098}}
\and A.        ~Montero                       \inst{\ref{inst:0097}}
\and R.        ~Mor                           \orcit{0000-0002-8179-6527}\inst{\ref{inst:0382},\ref{inst:0025},\ref{inst:0026}}
\and A.        ~Mora                          \inst{\ref{inst:0040}}
\and R.        ~Morbidelli                    \orcit{0000-0001-7627-4946}\inst{\ref{inst:0010}}
\and T.        ~Morel                         \orcit{0000-0002-8176-4816}\inst{\ref{inst:0099}}
\and D.        ~Morris                        \orcit{0000-0002-1952-6251}\inst{\ref{inst:0101}}
\and N.        ~Mowlavi                       \orcit{0000-0003-1578-6993}\inst{\ref{inst:0038}}
\and D.        ~Munoz                         \inst{\ref{inst:0017}}
\and T.        ~Muraveva                      \orcit{0000-0002-0969-1915}\inst{\ref{inst:0098}}
\and C.P.      ~Murphy                        \inst{\ref{inst:0039}}
\and I.        ~Musella                       \orcit{0000-0001-5909-6615}\inst{\ref{inst:0327}}
\and Z.        ~Nagy                          \orcit{0000-0002-3632-1194}\inst{\ref{inst:0148}}
\and S.        ~Nieto                         \inst{\ref{inst:0104}}
\and L.        ~Noval                         \inst{\ref{inst:0017}}
\and A.        ~Ogden                         \inst{\ref{inst:0037}}
\and C.        ~Pagani                        \orcit{0000-0001-5477-4720}\inst{\ref{inst:0398}}
\and I.        ~Pagano                        \orcit{0000-0001-9573-4928}\inst{\ref{inst:0122}}
\and L.        ~Palaversa                     \orcit{0000-0003-3710-0331}\inst{\ref{inst:0261}}
\and P.A.      ~Palicio                       \orcit{0000-0002-7432-8709}\inst{\ref{inst:0002}}
\and L.        ~Pallas-Quintela               \orcit{0000-0001-9296-3100}\inst{\ref{inst:0105}}
\and A.        ~Panahi                        \orcit{0000-0001-5850-4373}\inst{\ref{inst:0174}}
\and S.        ~Payne-Wardenaar               \inst{\ref{inst:0034}}
\and L.        ~Pegoraro                      \inst{\ref{inst:0008}}
\and A.        ~Penttil\"{ a}                 \orcit{0000-0001-7403-1721}\inst{\ref{inst:0018}}
\and P.        ~Pesciullesi                   \inst{\ref{inst:0104}}
\and A.M.      ~Piersimoni                    \orcit{0000-0002-8019-3708}\inst{\ref{inst:0242}}
\and M.        ~Pinamonti                     \orcit{0000-0002-4445-1845}\inst{\ref{inst:0010}}
\and F.-X.     ~Pineau                        \orcit{0000-0002-2335-4499}\inst{\ref{inst:0126}}
\and E.        ~Plachy                        \orcit{0000-0002-5481-3352}\inst{\ref{inst:0148},\ref{inst:0374},\ref{inst:0149}}
\and G.        ~Plum                          \inst{\ref{inst:0032}}
\and E.        ~Poggio                        \orcit{0000-0003-3793-8505}\inst{\ref{inst:0002},\ref{inst:0010}}
\and D.        ~Pourbaix$^\dagger$            \orcit{0000-0002-3020-1837}\inst{\ref{inst:0141},\ref{inst:0289}}
\and A.        ~Pr\v{s}a                      \orcit{0000-0002-1913-0281}\inst{\ref{inst:0419}}
\and L.        ~Pulone                        \orcit{0000-0002-5285-998X}\inst{\ref{inst:0127}}
\and E.        ~Racero                        \orcit{0000-0002-6101-9050}\inst{\ref{inst:0097},\ref{inst:0422}}
\and M.        ~Rainer                        \orcit{0000-0002-8786-2572}\inst{\ref{inst:0013},\ref{inst:0424}}
\and C.M.      ~Raiteri                       \orcit{0000-0003-1784-2784}\inst{\ref{inst:0010}}
\and P.        ~Ramos                         \orcit{0000-0002-5080-7027}\inst{\ref{inst:0426},\ref{inst:0024},\ref{inst:0026}}
\and M.        ~Ramos-Lerate                  \orcit{0009-0005-4677-8031}\inst{\ref{inst:0089}}
\and M.        ~Ratajczak                     \orcit{0000-0002-3218-2684}\inst{\ref{inst:0306}}
\and P.        ~Re Fiorentin                  \orcit{0000-0002-4995-0475}\inst{\ref{inst:0010}}
\and S.        ~Regibo                        \orcit{0000-0001-7227-9563}\inst{\ref{inst:0151}}
\and C.        ~Reyl\'{e}                     \orcit{0000-0003-2258-2403}\inst{\ref{inst:0023}}
\and V.        ~Ripepi                        \orcit{0000-0003-1801-426X}\inst{\ref{inst:0327}}
\and A.        ~Riva                          \orcit{0000-0002-6928-8589}\inst{\ref{inst:0010}}
\and H.-W.     ~Rix                           \orcit{0000-0003-4996-9069}\inst{\ref{inst:0053}}
\and G.        ~Rixon                         \orcit{0000-0003-4399-6568}\inst{\ref{inst:0037}}
\and N.        ~Robichon                      \orcit{0000-0003-4545-7517}\inst{\ref{inst:0032}}
\and C.        ~Robin                         \inst{\ref{inst:0017}}
\and M.        ~Romero-G\'{o}mez              \orcit{0000-0003-3936-1025}\inst{\ref{inst:0024},\ref{inst:0025},\ref{inst:0026}}
\and N.        ~Rowell                        \orcit{0000-0003-3809-1895}\inst{\ref{inst:0101}}
\and F.        ~Royer                         \orcit{0000-0002-9374-8645}\inst{\ref{inst:0032}}
\and D.        ~Ruz Mieres                    \orcit{0000-0002-9455-157X}\inst{\ref{inst:0037}}
\and K.A.      ~Rybicki                       \orcit{0000-0002-9326-9329}\inst{\ref{inst:0446}}
\and G.        ~Sadowski                      \orcit{0000-0002-3411-1003}\inst{\ref{inst:0141}}
\and A.        ~S\'{a}ez N\'{u}\~{n}ez        \orcit{0009-0001-6078-0868}\inst{\ref{inst:0026},\ref{inst:0024},\ref{inst:0025}}
\and A.        ~Sagrist\`{a} Sell\'{e}s       \orcit{0000-0001-6191-2028}\inst{\ref{inst:0034}}
\and J.        ~Sahlmann                      \orcit{0000-0001-9525-3673}\inst{\ref{inst:0104}}
\and V.        ~Sanchez Gimenez               \orcit{0000-0003-1797-3557}\inst{\ref{inst:0026},\ref{inst:0024},\ref{inst:0025}}
\and N.        ~Sanna                         \orcit{0000-0001-9275-9492}\inst{\ref{inst:0013}}
\and R.        ~Santove\~{n}a                 \orcit{0000-0002-9257-2131}\inst{\ref{inst:0105}}
\and M.        ~Sarasso                       \orcit{0000-0001-5121-0727}\inst{\ref{inst:0010}}
\and C.        ~Sarrate Riera                 \inst{\ref{inst:0064},\ref{inst:0024},\ref{inst:0026}}
\and E.        ~Sciacca                       \orcit{0000-0002-5574-2787}\inst{\ref{inst:0122}}
\and J.C.      ~Segovia                       \inst{\ref{inst:0097}}
\and D.        ~S\'{e}gransan                 \orcit{0000-0003-2355-8034}\inst{\ref{inst:0038}}
\and S.        ~Shahaf                        \orcit{0000-0001-9298-8068}\inst{\ref{inst:0446}}
\and A.        ~Siebert                       \orcit{0000-0001-8059-2840}\inst{\ref{inst:0126},\ref{inst:0467}}
\and L.        ~Siltala                       \orcit{0000-0002-6938-794X}\inst{\ref{inst:0018}}
\and E.        ~Slezak                        \inst{\ref{inst:0002}}
\and R.L.      ~Smart                         \orcit{0000-0002-4424-4766}\inst{\ref{inst:0010},\ref{inst:0234}}
\and O.N.      ~Snaith                        \orcit{0000-0003-1414-1296}\inst{\ref{inst:0032},\ref{inst:0473}}
\and E.        ~Solano                        \orcit{0000-0003-1885-5130}\inst{\ref{inst:0474}}
\and F.        ~Solitro                       \inst{\ref{inst:0077}}
\and D.        ~Souami                        \orcit{0000-0003-4058-0815}\inst{\ref{inst:0324},\ref{inst:0477}}
\and J.        ~Souchay                       \inst{\ref{inst:0091}}
\and L.        ~Spina                         \orcit{0000-0002-9760-6249}\inst{\ref{inst:0029}}
\and E.        ~Spitoni                       \orcit{0000-0001-9715-5727}\inst{\ref{inst:0002},\ref{inst:0481}}
\and L.A.      ~Squillante                    \inst{\ref{inst:0077}}
\and I.A.      ~Steele                        \orcit{0000-0001-8397-5759}\inst{\ref{inst:0346}}
\and H.        ~Steidelm\"{ u}ller            \inst{\ref{inst:0016}}
\and J.        ~Surdej                        \orcit{0000-0002-7005-1976}\inst{\ref{inst:0099}}
\and L.        ~Szabados                      \orcit{0000-0002-2046-4131}\inst{\ref{inst:0148}}
\and F.        ~Taris                         \inst{\ref{inst:0091}}
\and M.B.      ~Taylor                        \orcit{0000-0002-4209-1479}\inst{\ref{inst:0488}}
\and R.        ~Teixeira                      \orcit{0000-0002-6806-6626}\inst{\ref{inst:0244}}
\and K.        ~Tisani\'{c}                   \orcit{0000-0001-6382-4937}\inst{\ref{inst:0261}}
\and L.        ~Tolomei                       \orcit{0000-0002-3541-3230}\inst{\ref{inst:0077}}
\and F.        ~Torra                         \orcit{0000-0002-8429-299X}\inst{\ref{inst:0064},\ref{inst:0024},\ref{inst:0026}}
\and G.        ~Torralba Elipe                \orcit{0000-0001-8738-194X}\inst{\ref{inst:0105},\ref{inst:0496},\ref{inst:0497}}
\and M.        ~Trabucchi                     \orcit{0000-0002-1429-2388}\inst{\ref{inst:0498},\ref{inst:0038}}
\and M.        ~Tsantaki                      \orcit{0000-0002-0552-2313}\inst{\ref{inst:0013}}
\and A.        ~Ulla                          \orcit{0000-0001-6424-5005}\inst{\ref{inst:0501},\ref{inst:0502}}
\and N.        ~Unger                         \orcit{0000-0003-3993-7127}\inst{\ref{inst:0038}}
\and O.        ~Vanel                         \orcit{0000-0002-7898-0454}\inst{\ref{inst:0032}}
\and A.        ~Vecchiato                     \orcit{0000-0003-1399-5556}\inst{\ref{inst:0010}}
\and D.        ~Vicente                       \orcit{0000-0002-1584-1182}\inst{\ref{inst:0282}}
\and S.        ~Voutsinas                     \inst{\ref{inst:0101}}
\and M.        ~Weiler                        \inst{\ref{inst:0026},\ref{inst:0024},\ref{inst:0025}}
\and \L{}.     ~Wyrzykowski                   \orcit{0000-0002-9658-6151}\inst{\ref{inst:0306}}
\and H.        ~Zhao                          \orcit{0000-0003-2645-6869}\inst{\ref{inst:0002},\ref{inst:0513}}
\and J.        ~Zorec                         \orcit{0000-0003-1257-6915}\inst{\ref{inst:0514}}
\and T.        ~Zwitter                       \orcit{0000-0002-2325-8763}\inst{\ref{inst:0515}}
\and L.        ~Balaguer-N\'{u}\~{n}ez      \orcit{0000-0001-9789-7069}\inst{\ref{inst:0026},\ref{inst:0024},\ref{inst:0025}}
\and N.        ~Leclerc                      \orcit{0009-0001-5569-6098}\inst{\ref{inst:0032}}
\and S.        ~Morgenthaler               \orcit{0009-0005-6349-3716}\inst{\ref{inst:0555}}
\and G.        ~Robert                         \inst{\ref{inst:0017}}
\and S.        ~Zucker                    \orcit{0000-0003-3173-3138}\inst{\ref{inst:0191}}
}
\institute{
     IMCCE, Observatoire de Paris, Universit\'{e} PSL, CNRS, Sorbonne Universit{\'e}, Univ. Lille, 77 av. Denfert-Rochereau, 75014 Paris, France\relax                                                                                                                                                                                                                                                               \label{inst:0001}
\and Universit\'{e} C\^{o}te d'Azur, Observatoire de la C\^{o}te d'Azur, CNRS, Laboratoire Lagrange, Bd de l'Observatoire, CS 34229, 06304 Nice Cedex 4, France\relax                                                                                                                                                                                                                                                \label{inst:0002}
\and Harvard-Smithsonian Center for Astrophysics, 60 Garden St., MS 15, Cambridge, MA 02138, USA\relax                                                                                                                                                                                                                                                                                                               \label{inst:0005}
\and Royal Observatory of Belgium, Ringlaan 3, 1180 Brussels, Belgium\relax                                                                                                                                                                                                                                                                                                                                          \label{inst:0007}
\and CNES Centre Spatial de Toulouse, 18 avenue Edouard Belin, 31401 Toulouse Cedex 9, France\relax                                                                                                                                                                                                                                                                                                                  \label{inst:0008}
\and INAF - Osservatorio Astrofisico di Torino, via Osservatorio 20, 10025 Pino Torinese (TO), Italy\relax                                                                                                                                                                                                                                                                                                           \label{inst:0010}
\and INAF - Osservatorio Astrofisico di Arcetri, Largo Enrico Fermi 5, 50125 Firenze, Italy\relax                                                                                                                                                                                                                                                                                                                    \label{inst:0013}
\and Telespazio for CNES Centre Spatial de Toulouse, 18 avenue Edouard Belin, 31401 Toulouse Cedex 9, France\relax                                                                                                                                                                                                                                                                                                   \label{inst:0014}
\and Lohrmann Observatory, Technische Universit\"{ a}t Dresden, Mommsenstra{\ss}e 13, 01062 Dresden, Germany\relax                                                                                                                                                                                                                                                                                                   \label{inst:0016}
\and Thales Services for CNES Centre Spatial de Toulouse, 18 avenue Edouard Belin, 31401 Toulouse Cedex 9, France\relax                                                                                                                                                                                                                                                                                              \label{inst:0017}
\and Department of Physics, University of Helsinki, P.O. Box 64, 00014 Helsinki, Finland\relax                                                                                                                                                                                                                                                                                                                       \label{inst:0018}
\and Finnish Geospatial Research Institute FGI, Vuorimiehentie 5, 02150 Espoo, Finland\relax                                                                                                                                                                                                                                                                                                                         \label{inst:0019}
\and Institut UTINAM CNRS UMR6213, Universit\'{e} de Franche-Comt\'{e}, OSU THETA Franche-Comt\'{e} Bourgogne, Observatoire de Besan\c{c}on, BP1615, 25010 Besan\c{c}on Cedex, France\relax                                                                                                                                                                                                                          \label{inst:0023}
\and Institut de Ci\`{e}ncies del Cosmos (ICCUB), Universitat  de  Barcelona  (UB), Mart\'{i} i  Franqu\`{e}s  1, 08028 Barcelona, Spain\relax                                                                                                                                                                                                                                                                       \label{inst:0024}
\and Departament de F\'{i}sica Qu\`{a}ntica i Astrof\'{i}sica (FQA), Universitat de Barcelona (UB), c. Mart\'{i} i Franqu\`{e}s 1, 08028 Barcelona, Spain\relax                                                                                                                                                                                                                                                      \label{inst:0025}
\and Institut d'Estudis Espacials de Catalunya (IEEC), c. Gran Capit\`{a}, 2-4, 08034 Barcelona, Spain\relax                                                                                                                                                                                                                                                                                                         \label{inst:0026}
\and Leiden Observatory, Leiden University, Niels Bohrweg 2, 2333 CA Leiden, The Netherlands\relax                                                                                                                                                                                                                                                                                                                   \label{inst:0027}
\and INAF - Osservatorio astronomico di Padova, Vicolo Osservatorio 5, 35122 Padova, Italy\relax                                                                                                                                                                                                                                                                                                                     \label{inst:0029}
\and European Space Agency (ESA), European Space Research and Technology Centre (ESTEC), Keplerlaan 1, 2201AZ, Noordwijk, The Netherlands\relax                                                                                                                                                                                                                                                                      \label{inst:0030}
\and GEPI, Observatoire de Paris, Universit\'{e} PSL, CNRS, 5 Place Jules Janssen, 92190 Meudon, France\relax                                                                                                                                                                                                                                                                                                        \label{inst:0032}
\and Univ. Grenoble Alpes, CNRS, IPAG, 38000 Grenoble, France\relax                                                                                                                                                                                                                                                                                                                                                  \label{inst:0033}
\and Astronomisches Rechen-Institut, Zentrum f\"{ u}r Astronomie der Universit\"{ a}t Heidelberg, M\"{ o}nchhofstr. 12-14, 69120 Heidelberg, Germany\relax                                                                                                                                                                                                                                                           \label{inst:0034}
\and Laboratoire d'astrophysique de Bordeaux, Univ. Bordeaux, CNRS, B18N, all{\'e}e Geoffroy Saint-Hilaire, 33615 Pessac, France\relax                                                                                                                                                                                                                                                                               \label{inst:0036}
\and Institute of Astronomy, University of Cambridge, Madingley Road, Cambridge CB3 0HA, United Kingdom\relax                                                                                                                                                                                                                                                                                                        \label{inst:0037}
\and Department of Astronomy, University of Geneva, Chemin Pegasi 51, 1290 Versoix, Switzerland\relax                                                                                                                                                                                                                                                                                                                \label{inst:0038}
\and European Space Agency (ESA), European Space Astronomy Centre (ESAC), Camino bajo del Castillo, s/n, Urbanizaci\'{o}n Villafranca del Castillo, Villanueva de la Ca\~{n}ada, 28692 Madrid, Spain\relax                                                                                                                                                                                                           \label{inst:0039}
\and Aurora Technology for European Space Agency (ESA), Camino bajo del Castillo, s/n, Urbanizaci\'{o}n Villafranca del Castillo, Villanueva de la Ca\~{n}ada, 28692 Madrid, Spain\relax                                                                                                                                                                                                                             \label{inst:0040}
\and Lund Observatory, Division of Astrophysics, Department of Physics, Lund University, Box 43, 22100 Lund, Sweden\relax                                                                                                                                                                                                                                                                                            \label{inst:0045}
\and Nicolaus Copernicus Astronomical Center, Polish Academy of Sciences, ul. Bartycka 18, 00-716 Warsaw, Poland\relax                                                                                                                                                                                                                                                                                               \label{inst:0051}
\and Max Planck Institute for Astronomy, K\"{ o}nigstuhl 17, 69117 Heidelberg, Germany\relax                                                                                                                                                                                                                                                                                                                         \label{inst:0053}
\and Mullard Space Science Laboratory, University College London, Holmbury St Mary, Dorking, Surrey RH5 6NT, United Kingdom\relax                                                                                                                                                                                                                                                                                    \label{inst:0055}
\and Department of Astronomy, University of Geneva, Chemin d'Ecogia 16, 1290 Versoix, Switzerland\relax                                                                                                                                                                                                                                                                                                              \label{inst:0061}
\and DAPCOM Data Services, c. dels Vilabella, 5-7, 80500 Vic, Barcelona, Spain\relax                                                                                                                                                                                                                                                                                                                                 \label{inst:0064}
\and ALTEC S.p.a, Corso Marche, 79,10146 Torino, Italy\relax                                                                                                                                                                                                                                                                                                                                                         \label{inst:0077}
\and Sednai S\`{a}rl, Geneva, Switzerland\relax                                                                                                                                                                                                                                                                                                                                                                      \label{inst:0079}
\and Gaia DPAC Project Office, ESAC, Camino bajo del Castillo, s/n, Urbanizaci\'{o}n Villafranca del Castillo, Villanueva de la Ca\~{n}ada, 28692 Madrid, Spain\relax                                                                                                                                                                                                                                                \label{inst:0087}
\and Telespazio UK S.L. for European Space Agency (ESA), Camino bajo del Castillo, s/n, Urbanizaci\'{o}n Villafranca del Castillo, Villanueva de la Ca\~{n}ada, 28692 Madrid, Spain\relax                                                                                                                                                                                                                            \label{inst:0089}
\and SYRTE, Observatoire de Paris, Universit\'{e} PSL, CNRS, Sorbonne Universit\'{e}, LNE, 61 avenue de l'Observatoire 75014 Paris, France\relax                                                                                                                                                                                                                                                                     \label{inst:0091}
\and Serco Gesti\'{o}n de Negocios for European Space Agency (ESA), Camino bajo del Castillo, s/n, Urbanizaci\'{o}n Villafranca del Castillo, Villanueva de la Ca\~{n}ada, 28692 Madrid, Spain\relax                                                                                                                                                                                                                 \label{inst:0097}
\and INAF - Osservatorio di Astrofisica e Scienza dello Spazio di Bologna, via Piero Gobetti 93/3, 40129 Bologna, Italy\relax                                                                                                                                                                                                                                                                                        \label{inst:0098}
\and Institut d'Astrophysique et de G\'{e}ophysique, Universit\'{e} de Li\`{e}ge, 19c, All\'{e}e du 6 Ao\^{u}t, B-4000 Li\`{e}ge, Belgium\relax                                                                                                                                                                                                                                                                      \label{inst:0099}
\and CRAAG - Centre de Recherche en Astronomie, Astrophysique et G\'{e}ophysique, Route de l'Observatoire Bp 63 Bouzareah 16340 Algiers, Algeria\relax                                                                                                                                                                                                                                                               \label{inst:0100}
\and Institute for Astronomy, University of Edinburgh, Royal Observatory, Blackford Hill, Edinburgh EH9 3HJ, United Kingdom\relax                                                                                                                                                                                                                                                                                    \label{inst:0101}
\and RHEA for European Space Agency (ESA), Camino bajo del Castillo, s/n, Urbanizaci\'{o}n Villafranca del Castillo, Villanueva de la Ca\~{n}ada, 28692 Madrid, Spain\relax                                                                                                                                                                                                                                          \label{inst:0104}
\and CIGUS CITIC - Department of Computer Science and Information Technologies, University of A Coru\~{n}a, Campus de Elvi\~{n}a s/n, A Coru\~{n}a, 15071, Spain\relax                                                                                                                                                                                                                                               \label{inst:0105}
\and ATG Europe for European Space Agency (ESA), Camino bajo del Castillo, s/n, Urbanizaci\'{o}n Villafranca del Castillo, Villanueva de la Ca\~{n}ada, 28692 Madrid, Spain\relax                                                                                                                                                                                                                                    \label{inst:0107}
\and Kavli Institute for Cosmology Cambridge, Institute of Astronomy, Madingley Road, Cambridge, CB3 0HA\relax                                                                                                                                                                                                                                                                                                       \label{inst:0112}
\and Department of Astrophysics, Astronomy and Mechanics, National and Kapodistrian University of Athens, Panepistimiopolis, Zografos, 15783 Athens, Greece\relax                                                                                                                                                                                                                                                    \label{inst:0113}
\and Donald Bren School of Information and Computer Sciences, University of California, Irvine, CA 92697, USA\relax                                                                                                                                                                                                                                                                                                  \label{inst:0120}
\and CENTRA, Faculdade de Ci\^{e}ncias, Universidade de Lisboa, Edif. C8, Campo Grande, 1749-016 Lisboa, Portugal\relax                                                                                                                                                                                                                                                                                              \label{inst:0121}
\and INAF - Osservatorio Astrofisico di Catania, via S. Sofia 78, 95123 Catania, Italy\relax                                                                                                                                                                                                                                                                                                                         \label{inst:0122}
\and Dipartimento di Fisica e Astronomia ""Ettore Majorana"", Universit\`{a} di Catania, Via S. Sofia 64, 95123 Catania, Italy\relax                                                                                                                                                                                                                                                                                 \label{inst:0123}
\and Universit\'{e} de Strasbourg, CNRS, Observatoire astronomique de Strasbourg, UMR 7550,  11 rue de l'Universit\'{e}, 67000 Strasbourg, France\relax                                                                                                                                                                                                                                                              \label{inst:0126}
\and INAF - Osservatorio Astronomico di Roma, Via Frascati 33, 00078 Monte Porzio Catone (Roma), Italy\relax                                                                                                                                                                                                                                                                                                         \label{inst:0127}
\and Space Science Data Center - ASI, Via del Politecnico SNC, 00133 Roma, Italy\relax                                                                                                                                                                                                                                                                                                                               \label{inst:0128}
\and HE Space Operations BV for European Space Agency (ESA), Keplerlaan 1, 2201AZ, Noordwijk, The Netherlands\relax                                                                                                                                                                                                                                                                                                  \label{inst:0138}
\and Dpto. de Inteligencia Artificial, UNED, c/ Juan del Rosal 16, 28040 Madrid, Spain\relax                                                                                                                                                                                                                                                                                                                         \label{inst:0139}
\and Institut d'Astronomie et d'Astrophysique, Universit\'{e} Libre de Bruxelles CP 226, Boulevard du Triomphe, 1050 Brussels, Belgium\relax                                                                                                                                                                                                                                                                         \label{inst:0141}
\and Leibniz Institute for Astrophysics Potsdam (AIP), An der Sternwarte 16, 14482 Potsdam, Germany\relax                                                                                                                                                                                                                                                                                                            \label{inst:0146}
\and Konkoly Observatory, Research Centre for Astronomy and Earth Sciences, E\"{ o}tv\"{ o}s Lor{\'a}nd Research Network (ELKH), MTA Centre of Excellence, Konkoly Thege Mikl\'{o}s \'{u}t 15-17, 1121 Budapest, Hungary\relax                                                                                                                                                                                       \label{inst:0148}
\and ELTE E\"{ o}tv\"{ o}s Lor\'{a}nd University, Institute of Physics, 1117, P\'{a}zm\'{a}ny P\'{e}ter s\'{e}t\'{a}ny 1A, Budapest, Hungary\relax                                                                                                                                                                                                                                                                   \label{inst:0149}
\and Instituut voor Sterrenkunde, KU Leuven, Celestijnenlaan 200D, 3001 Leuven, Belgium\relax                                                                                                                                                                                                                                                                                                                        \label{inst:0151}
\and Department of Astrophysics/IMAPP, Radboud University, P.O.Box 9010, 6500 GL Nijmegen, The Netherlands\relax                                                                                                                                                                                                                                                                                                     \label{inst:0152}
\and University of Vienna, Department of Astrophysics, T\"{ u}rkenschanzstra{\ss}e 17, A1180 Vienna, Austria\relax                                                                                                                                                                                                                                                                                                   \label{inst:0157}
\and Institute of Physics, Ecole Polytechnique F\'ed\'erale de Lausanne (EPFL), Observatoire de Sauverny, 1290 Versoix, Switzerland\relax                                                                                                                                                                                                                                                                            \label{inst:0158}
\and Quasar Science Resources for European Space Agency (ESA), Camino bajo del Castillo, s/n, Urbanizaci\'{o}n Villafranca del Castillo, Villanueva de la Ca\~{n}ada, 28692 Madrid, Spain\relax                                                                                                                                                                                                                      \label{inst:0162}
\and LASIGE, Faculdade de Ci\^{e}ncias, Universidade de Lisboa, Edif. C6, Campo Grande, 1749-016 Lisboa, Portugal\relax                                                                                                                                                                                                                                                                                              \label{inst:0169}
\and School of Physics and Astronomy , University of Leicester, University Road, Leicester LE1 7RH, United Kingdom\relax                                                                                                                                                                                                                                                                                             \label{inst:0170}
\and School of Physics and Astronomy, Tel Aviv University, Tel Aviv 6997801, Israel\relax                                                                                                                                                                                                                                                                                                                            \label{inst:0174}
\and Cavendish Laboratory, JJ Thomson Avenue, Cambridge CB3 0HE, United Kingdom\relax                                                                                                                                                                                                                                                                                                                                \label{inst:0175}
\and National Observatory of Athens, I. Metaxa and Vas. Pavlou, Palaia Penteli, 15236 Athens, Greece\relax                                                                                                                                                                                                                                                                                                           \label{inst:0180}
\and University of Turin, Department of Physics, Via Pietro Giuria 1, 10125 Torino, Italy\relax                                                                                                                                                                                                                                                                                                                      \label{inst:0183}
\and Depto. Estad\'istica e Investigaci\'on Operativa. Universidad de C\'adiz, Avda. Rep\'ublica Saharaui s/n, 11510 Puerto Real, C\'adiz, Spain\relax                                                                                                                                                                                                                                                               \label{inst:0184}
\and EURIX S.r.l., Corso Vittorio Emanuele II 61, 10128, Torino, Italy\relax                                                                                                                                                                                                                                                                                                                                         \label{inst:0190}
\and Porter School of the Environment and Earth Sciences, Tel Aviv University, Tel Aviv 6997801, Israel\relax                                                                                                                                                                                                                                                                                                        \label{inst:0191}
\and ATOS for CNES Centre Spatial de Toulouse, 18 avenue Edouard Belin, 31401 Toulouse Cedex 9, France\relax                                                                                                                                                                                                                                                                                                         \label{inst:0192}
\and HE Space Operations BV for European Space Agency (ESA), Camino bajo del Castillo, s/n, Urbanizaci\'{o}n Villafranca del Castillo, Villanueva de la Ca\~{n}ada, 28692 Madrid, Spain\relax                                                                                                                                                                                                                        \label{inst:0194}
\and LFCA/DAS,Universidad de Chile,CNRS,Casilla 36-D, Santiago, Chile\relax                                                                                                                                                                                                                                                                                                                                          \label{inst:0196}
\and SISSA - Scuola Internazionale Superiore di Studi Avanzati, via Bonomea 265, 34136 Trieste, Italy\relax                                                                                                                                                                                                                                                                                                          \label{inst:0201}
\and University of Turin, Department of Computer Sciences, Corso Svizzera 185, 10149 Torino, Italy\relax                                                                                                                                                                                                                                                                                                             \label{inst:0209}
\and Dpto. de Matem\'{a}tica Aplicada y Ciencias de la Computaci\'{o}n, Univ. de Cantabria, ETS Ingenieros de Caminos, Canales y Puertos, Avda. de los Castros s/n, 39005 Santander, Spain\relax                                                                                                                                                                                                                     \label{inst:0212}
\and Institut de F\'{i}sica d'Altes Energies (IFAE), The Barcelona Institute of Science and Technology, Campus UAB, 08193 Bellaterra (Barcelona), Spain\relax                                                                                                                                                                                                                                                        \label{inst:0218}
\and Port d'Informaci\'{o} Cient\'{i}fica (PIC), Campus UAB, C. Albareda s/n, 08193 Bellaterra (Barcelona), Spain\relax                                                                                                                                                                                                                                                                                              \label{inst:0219}
\and Instituto de Astrof\'{i}sica, Universidad Andres Bello, Fernandez Concha 700, Las Condes, Santiago RM, Chile\relax                                                                                                                                                                                                                                                                                              \label{inst:0227}
\and Centre for Astrophysics Research, University of Hertfordshire, College Lane, AL10 9AB, Hatfield, United Kingdom\relax                                                                                                                                                                                                                                                                                           \label{inst:0234}
\and University of Turin, Mathematical Department ""G.Peano"", Via Carlo Alberto 10, 10123 Torino, Italy\relax                                                                                                                                                                                                                                                                                                       \label{inst:0238}
\and INAF - Osservatorio Astronomico d'Abruzzo, Via Mentore Maggini, 64100 Teramo, Italy\relax                                                                                                                                                                                                                                                                                                                       \label{inst:0242}
\and Instituto de Astronomia, Geof\`{i}sica e Ci\^{e}ncias Atmosf\'{e}ricas, Universidade de S\~{a}o Paulo, Rua do Mat\~{a}o, 1226, Cidade Universitaria, 05508-900 S\~{a}o Paulo, SP, Brazil\relax                                                                                                                                                                                                                  \label{inst:0244}
\and M\'{e}socentre de calcul de Franche-Comt\'{e}, Universit\'{e} de Franche-Comt\'{e}, 16 route de Gray, 25030 Besan\c{c}on Cedex, France\relax                                                                                                                                                                                                                                                                    \label{inst:0249}
\and Ru{\dj}er Bo\v{s}kovi\'{c} Institute, Bijeni\v{c}ka cesta 54, 10000 Zagreb, Croatia\relax                                                                                                                                                                                                                                                                                                                       \label{inst:0261}
\and Astrophysics Research Centre, School of Mathematics and Physics, Queen's University Belfast, Belfast BT7 1NN, UK\relax                                                                                                                                                                                                                                                                                          \label{inst:0263}
\and Data Science and Big Data Lab, Pablo de Olavide University, 41013, Seville, Spain\relax                                                                                                                                                                                                                                                                                                                         \label{inst:0275}
\and Institute of Astrophysics, FORTH, Crete, Greece\relax                                                                                                                                                                                                                                                                                                                                                           \label{inst:0281}
\and Barcelona Supercomputing Center (BSC), Pla\c{c}a Eusebi G\"{ u}ell 1-3, 08034-Barcelona, Spain\relax                                                                                                                                                                                                                                                                                                            \label{inst:0282}
\and ETSE Telecomunicaci\'{o}n, Universidade de Vigo, Campus Lagoas-Marcosende, 36310 Vigo, Galicia, Spain\relax                                                                                                                                                                                                                                                                                                     \label{inst:0286}
\and F.R.S.-FNRS, Rue d'Egmont 5, 1000 Brussels, Belgium\relax                                                                                                                                                                                                                                                                                                                                                       \label{inst:0289}
\and Asteroid Engineering Laboratory, Lule\aa{} University of Technology, Box 848, S-981 28 Kiruna, Sweden\relax                                                                                                                                                                                                                                                                                                     \label{inst:0291}
\and Kapteyn Astronomical Institute, University of Groningen, Landleven 12, 9747 AD Groningen, The Netherlands\relax                                                                                                                                                                                                                                                                                                 \label{inst:0298}
\and IAC - Instituto de Astrofisica de Canarias, Via L\'{a}ctea s/n, 38200 La Laguna S.C., Tenerife, Spain\relax                                                                                                                                                                                                                                                                                                     \label{inst:0300}
\and Department of Astrophysics, University of La Laguna, Via L\'{a}ctea s/n, 38200 La Laguna S.C., Tenerife, Spain\relax                                                                                                                                                                                                                                                                                            \label{inst:0301}
\and Astronomical Observatory, University of Warsaw,  Al. Ujazdowskie 4, 00-478 Warszawa, Poland\relax                                                                                                                                                                                                                                                                                                               \label{inst:0306}
\and Research School of Astronomy \and Astrophysics, Australian National University, Cotter Road, Weston, ACT 2611, Australia\relax                                                                                                                                                                                                                                                                                     \label{inst:0307}
\and European Space Agency (ESA, retired), European Space Research and Technology Centre (ESTEC), Keplerlaan 1, 2201AZ, Noordwijk, The Netherlands\relax                                                                                                                                                                                                                                                             \label{inst:0308}
\and LESIA, Observatoire de Paris, Universit\'{e} PSL, CNRS, Sorbonne Universit\'{e}, Universit\'{e} de Paris, 5 Place Jules Janssen, 92190 Meudon, France\relax                                                                                                                                                                                                                                                     \label{inst:0324}
\and Universit\'{e} Rennes, CNRS, IPR (Institut de Physique de Rennes) - UMR 6251, 35000 Rennes, France\relax                                                                                                                                                                                                                                                                                                        \label{inst:0325}
\and INAF - Osservatorio Astronomico di Capodimonte, Via Moiariello 16, 80131, Napoli, Italy\relax                                                                                                                                                                                                                                                                                                                   \label{inst:0327}
\and Shanghai Astronomical Observatory, Chinese Academy of Sciences, 80 Nandan Road, Shanghai 200030, People's Republic of China\relax                                                                                                                                                                                                                                                                               \label{inst:0330}
\and University of Chinese Academy of Sciences, No.19(A) Yuquan Road, Shijingshan District, Beijing 100049, People's Republic of China\relax                                                                                                                                                                                                                                                                         \label{inst:0332}
\and S\~{a}o Paulo State University, Grupo de Din\^{a}mica Orbital e Planetologia, CEP 12516-410, Guaratinguet\'{a}, SP, Brazil\relax                                                                                                                                                                                                                                                                                \label{inst:0334}
\and Niels Bohr Institute, University of Copenhagen, Juliane Maries Vej 30, 2100 Copenhagen {\O}, Denmark\relax                                                                                                                                                                                                                                                                                                      \label{inst:0337}
\and DXC Technology, Retortvej 8, 2500 Valby, Denmark\relax                                                                                                                                                                                                                                                                                                                                                          \label{inst:0338}
\and Las Cumbres Observatory, 6740 Cortona Drive Suite 102, Goleta, CA 93117, USA\relax                                                                                                                                                                                                                                                                                                                              \label{inst:0339}
\and CIGUS CITIC, Department of Nautical Sciences and Marine Engineering, University of A Coru\~{n}a, Paseo de Ronda 51, 15071, A Coru\~{n}a, Spain\relax                                                                                                                                                                                                                                                            \label{inst:0345}
\and Astrophysics Research Institute, Liverpool John Moores University, 146 Brownlow Hill, Liverpool L3 5RF, United Kingdom\relax                                                                                                                                                                                                                                                                                    \label{inst:0346}
\and IRAP, Universit\'{e} de Toulouse, CNRS, UPS, CNES, 9 Av. colonel Roche, BP 44346, 31028 Toulouse Cedex 4, France\relax                                                                                                                                                                                                                                                                                          \label{inst:0353}
\and MTA CSFK Lend\"{ u}let Near-Field Cosmology Research Group, Konkoly Observatory, MTA Research Centre for Astronomy and Earth Sciences, Konkoly Thege Mikl\'{o}s \'{u}t 15-17, 1121 Budapest, Hungary\relax                                                                                                                                                                                                      \label{inst:0374}
\and Pervasive Technologies s.l., c. Saragossa 118, 08006 Barcelona, Spain\relax                                                                                                                                                                                                                                                                                                                                     \label{inst:0382}
\and School of Physics and Astronomy, University of Leicester, University Road, Leicester LE1 7RH, United Kingdom\relax                                                                                                                                                                                                                                                                                              \label{inst:0398}
\and Villanova University, Department of Astrophysics and Planetary Science, 800 E Lancaster Avenue, Villanova PA 19085, USA\relax                                                                                                                                                                                                                                                                                   \label{inst:0419}
\and Departmento de F\'{i}sica de la Tierra y Astrof\'{i}sica, Universidad Complutense de Madrid, 28040 Madrid, Spain\relax                                                                                                                                                                                                                                                                                          \label{inst:0422}
\and INAF - Osservatorio Astronomico di Brera, via E. Bianchi, 46, 23807 Merate (LC), Italy\relax                                                                                                                                                                                                                                                                                                                    \label{inst:0424}
\and National Astronomical Observatory of Japan, 2-21-1 Osawa, Mitaka, Tokyo 181-8588, Japan\relax                                                                                                                                                                                                                                                                                                                   \label{inst:0426}
\and Department of Particle Physics and Astrophysics, Weizmann Institute of Science, Rehovot 7610001, Israel\relax                                                                                                                                                                                                                                                                                                   \label{inst:0446}
\and Centre de Donn\'{e}es Astronomique de Strasbourg, Strasbourg, France\relax                                                                                                                                                                                                                                                                                                                                      \label{inst:0467}
\and University of Exeter, School of Physics and Astronomy, Stocker Road, Exeter, EX2 7SJ, United Kingdom\relax                                                                                                                                                                                                                                                                                                      \label{inst:0473}
\and Departamento de Astrof\'{i}sica, Centro de Astrobiolog\'{i}a (CSIC-INTA), ESA-ESAC. Camino Bajo del Castillo s/n. 28692 Villanueva de la Ca\~{n}ada, Madrid, Spain\relax                                                                                                                                                                                                                                        \label{inst:0474}
\and naXys, Department of Mathematics, University of Namur, Rue de Bruxelles 61, 5000 Namur, Belgium\relax                                                                                                                                                                                                                                                                                                           \label{inst:0477}
\and INAF. Osservatorio Astronomico di Trieste, via G.B. Tiepolo 11, 34131, Trieste, Italy\relax                                                                                                                                                                                                                                                                                                                     \label{inst:0481}
\and H H Wills Physics Laboratory, University of Bristol, Tyndall Avenue, Bristol BS8 1TL, United Kingdom\relax                                                                                                                                                                                                                                                                                                      \label{inst:0488}
\and Escuela de Arquitectura y Polit\'{e}cnica - Universidad Europea de Valencia, Spain\relax                                                                                                                                                                                                                                                                                                                        \label{inst:0496}
\and Escuela Superior de Ingenier\'{i}a y Tecnolog\'{i}a - Universidad Internacional de la Rioja, Spain\relax                                                                                                                                                                                                                                                                                                        \label{inst:0497}
\and Department of Physics and Astronomy G. Galilei, University of Padova, Vicolo dell'Osservatorio 3, 35122, Padova, Italy\relax                                                                                                                                                                                                                                                                                    \label{inst:0498}
\and Applied Physics Department, Universidade de Vigo, 36310 Vigo, Spain\relax                                                                                                                                                                                                                                                                                                                                       \label{inst:0501}
\and Instituto de F{'i}sica e Ciencias Aeroespaciais (IFCAE), Universidade de Vigo‚ \'{A} Campus de As Lagoas, 32004 Ourense, Spain\relax                                                                                                                                                                                                                                                                          \label{inst:0502}
\and Purple Mountain Observatory, Chinese Academy of Sciences, Nanjing 210023, China\relax                                                                                                                                                                                                                                                                                                                           \label{inst:0513}
\and Sorbonne Universit\'{e}, CNRS, UMR7095, Institut d'Astrophysique de Paris, 98bis bd. Arago, 75014 Paris, France\relax                                                                                                                                                                                                                                                                                           \label{inst:0514}
\and Faculty of Mathematics and Physics, University of Ljubljana, Jadranska ulica 19, 1000 Ljubljana, Slovenia\relax                                                                                                                                                                                                                                                                                                 \label{inst:0515}
\and Institute of Mathematics, Ecole Polytechnique F\'ed\'erale de Lausanne (EPFL), Switzerland\relax
\label{inst:0555}
}

\date{Received / Accepted}

  \abstract
   {We report the exploitation of a sample of Solar System observations based on data from the third Gaia Data Release  (\gdrthree) of nearly $157\,000$ asteroids that extends the epoch astrometric solution over the time coverage planned for the complete \gdrfour, which is not expected  before the end of 2025. This extended data set covers more than one full orbital period for the vast majority of these asteroids. The orbital solutions are derived from the \gaia data alone over a relatively short arc compared to the observation history of many of these asteroids. } 
   {The work aims to produce orbital elements for a large set of  asteroids based on 66 months of  accurate astrometry provided by Gaia and to assess the accuracy of these orbital solutions with a comparison to the best available orbits estimated from independent observations. A second validation is performed with accurate occultation timings.}
   {We processed the raw astrometric measurements of Gaia to obtain astrometric positions of moving objects with 1D sub-mas accuracy at the bright end. For each asteroid that we matched to the data,  an orbit fitting was attempted in the form of the best fit of the initial conditions (state vector) at the median epoch. The force model included Newtonian and relativistic accelerations to derive the observation equations, which were solved with a linear least-squares fit. }
   {Orbits are provided in the form of state vectors in the International Celestial Reference Frame for $156\,764$ asteroids, including near-Earth objects, main-belt asteroids, and Trojans. For the asteroids with the best observations, the (formal) relative accuracy measured by $\sigma_a/a$ is better than $10^{-10}$. Results are compared to orbits available from the Jet Propulsion Laboratory and MPC. Their orbits are based on much longer data arcs, but the positions are of lower quality. The relative differences in semi-major axes have a mean of $5\times 10^{-10}$ and a scatter of $5\times 10^{-9}$.}
   {}

   \maketitle
%



\section{Introduction} \label{sect:intro}
In the course of its systematic scan of the sky, the ESA \gaia spacecraft has detected a wide variety of celestial sources. It will continue operating until the end of its operational life sometime in early 2025. The celestial sources are  primarily stars from the Milky Way, which form the bulk of the massive Gaia catalogues that were released between 2016 and 2022 in data releases \gdrone, \gdrtwo, \gedrthree, and  \gdrthree proper (\cite{DR1-DPACP-8}, \cite{DR1-DPACP-18},  \cite{2023A&A...674A...1G}). However, sources closer to Earth are also caught in the \gaia net, with representatives of every category of small Solar System bodies (SSSBs), 
such as eear-Earth objects (NEOs), main-belt asteroids (MBA), 
Trojans, and a few more distant objects from the Kuiper belt and from the trans-Neptunian region. In addition, \gaia data include the largest planetary moons and a sample of periodic comets that were observed not too far from their perihelion passage. 

In this Gaia-coordinated Focus Product Release (\gfpr), the astrometry of 156,792 minor planets and 31 natural satellites is provided. However, this paper focuses only on the orbit computation of minor planets, and only this category of Solar System bodies is investigated. Likewise, photometric and spectral data are also beyond the scope of this work and have been largely presented with the \gdrthree results in \cite{2023A&A...674A..12T}.

\gdrtwo has shown the quality of \gaia observations and their huge value for asteroids. Based on the high-precision astrometry and because the time span covered by the observations was long enough (at least $\simeq$  1000 days), it was possible to derive accurate orbits for the observed asteroids, as was shown in \gdrthree \citep{2023A&A...674A..12T},  who reported a summary of the observational technique and the epoch astrometric accuracy. More precisely,  \gdrthree  was built upon 34 months of \gaia operations extending from July 2014 to May 2017. A well-observed Solar System body came into the \gaia fields of view about 25-30 times on average,  resulting in up to nine accurate astrometric positions at every passage. \footnote{We adopted the following terminology for the data sequencing. A passage or a transit is the crossing of a \gaia field of view. During a passage, up to nine position measurements at the CCD level are possible (primarily in the along-scan direction), which are referred to as observations, or elementary observations  to remove ambiguity. An epoch or a visibility period is a set of a few successive transits within one or two days, followed by a gap of several weeks before the start of a new period.} \cite{2023A&A...674A..12T} ranked the orbits from excellent to very approximate, with a clear difference in reliability as a function of the fraction of the orbital period covered over 34 months.

For the majority of the MBAs, this time was too short to remove strong constraints between  orbital elements. In the processing reported in this paper, which yielded the Focused Product Release, 
the data arc can be as long as 66 months. This typically is a little longer than the orbital period of an average asteroid that orbits in the main belt between Mars and Jupiter. Therefore, the situation  considerably improved as soon as we had a few pairs of data points that lie one orbital period apart from each other. This is sufficient to remove the partial degeneracy that is observed when the arc was too short. Although we have almost twice as many observations as in  \gdrthree, the main source of improvement remains the complete coverage of the orbit in the data set, which is in sharp contrast with the previous solution in \gdrthree, which was based on only 34 months of data. The length of arc that is used is the primary factor in the improved uncertainty of the orbit.

In this paper, we aim to present and discuss
the data that are released in the \gfpr, together with the results of the orbital fitting obtained in an extended time span, using a refined version of the fitting algorithm, with better weighting and an improved outlier rejection. Several shortcomings in the code were found and corrected in this new exploitation, and the orbits we 
provide now are the best we can achieve at the moment with the extremely high 
astrometric precision of \gaia. It is important to mention at this stage that the orbital solutions are derived from the \gaia data alone, meaning that the arc is still relatively short in comparison with the fits performed at MPC or JPL using all the available observations that so conveniently archived by the MPC, which were collected over decades or even centuries. As shown by the results, the relatively short time-span is somewhat balanced by the homogeneity of the data and the unparalleled single-observation accuracy and precision.

As mentioned above, no new photometric data in G band or reflectance spectra are released in the FPR. To associate photometry with FPR data, the G fluxes and magnitudes published in DR3 can be adopted for the corresponding transits in common. New spectrophotometric processing will be available in DR4, which is currently scheduled for the last term of 2025. Likewise, new astrometry is published in the FPR for planetary satellites, but their orbits have not yet been adjusted on these new positions.

The paper is broken down into a few sections. In \secrefalt{sect:dataset} we discuss the observation material used in the paper, and then we very briefly recall the relevant properties of the astrometric solution in \secrefalt{sect:astrometry}. In \secrefalt{sect:orbfit} we present the principle of the orbit fitting that produced the set of state vectors at a median epoch for each asteroid. The validation and discussion of the results are taken up in detail in \secrefalt{sect:orbeval}, along with some recommendation for users.

\section{Properties of the data set}\label{sect:dataset}

Gaia scans the sky throughout the year in a continuous way, without virtually any significant interruption beyond unavoidable but short dead times. The principles have been described in several papers from the Gaia Collaboration, and we refer to the reliable information therein \citep{DR1-DPACP-18}. The relevant facts for the orbital solutions we use here are the repeated observations of the same source, with an interval of 6 to 10 weeks between two sequences. A typical sequence consists of a small set of successive passages in the Preceding Field of View (pfov) and Following Field of View (ffov). The most common sequence is just a simple sequence of two passages pfov to ffov (PF) or ffov to pfov  (FP) within 6 hours, and then a few weeks before a new similar sequence. However,  every asteroid during the lifetime of Gaia enjoyed one or two much longer sequences, such as PFPF... or FPFP... and so on. In some cases, there were more than 15 consecutive transits within 4 or 5 days in a row. For the well-observed asteroids, typically, 40 to 90 passages (field crossings) were made in 66 months, and the number of visibility periods (closely packed repeating sequences at distinct epochs) is $30 \pm10$ on average. This has the strongest effect for the orbit determination because it drives the sampling in orbital phases.

\begin{figure}
   \centering
\includegraphics[width=0.95\hsize]{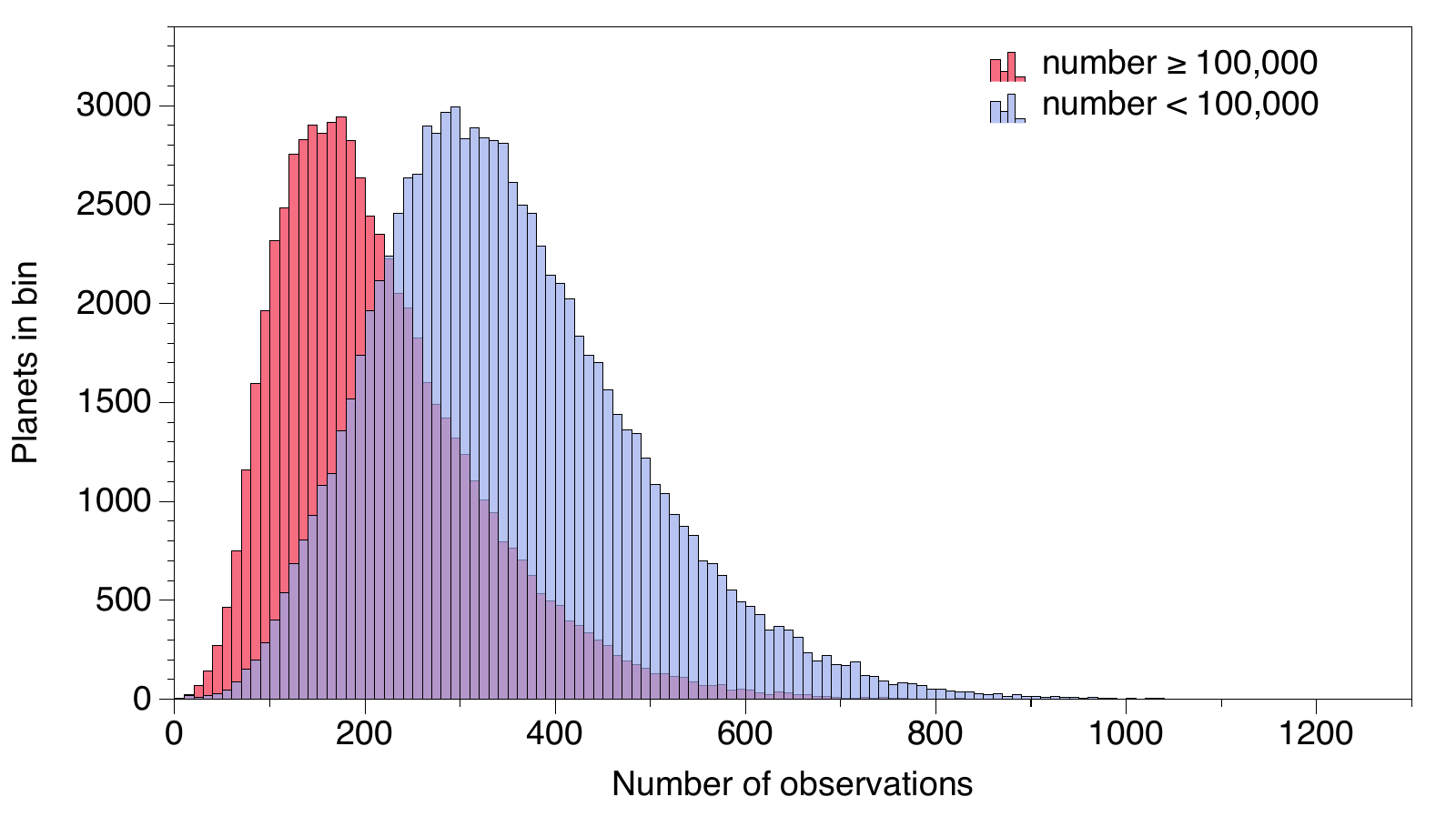} 
      \caption{Distribution of the number of  observations used in the orbital solution. Blue shows the first $100\,000$ numbered asteroids with a median of 330 individual 
      observations ($91\,785$ orbits), and red shows the numbered asteroids with numbers $\ge 100\,000$ and a median of 190 observations ($64\,979$ orbits).
              }
         \label{fig:numobs_all}
   \end{figure}

During each passage,  a maximum of nine astrometric positions is  obtained.
These make up the  observations at the CCD level.  The number of observations that were used (after rejections in the iterative model fitting) is shown in \autoref{fig:numobs_all} for the asteroids that were observed best with numbers $< 100\,000,$ and for the fainter and less frequently observed asteroids with numbers $\ge 100\,000$, typically with half as many transits ending with a successful Gaia detection. The transits were missed because the magnitude was too faint in some part of the orbits or because problems were found in the astrometric solution, which led to a rejection. The transition at $100,000$ between the two groups is just a convenient number to obtain balanced distributions. It has no deeper significance.

\begin{figure}
   \centering
\includegraphics[width=0.95\hsize]{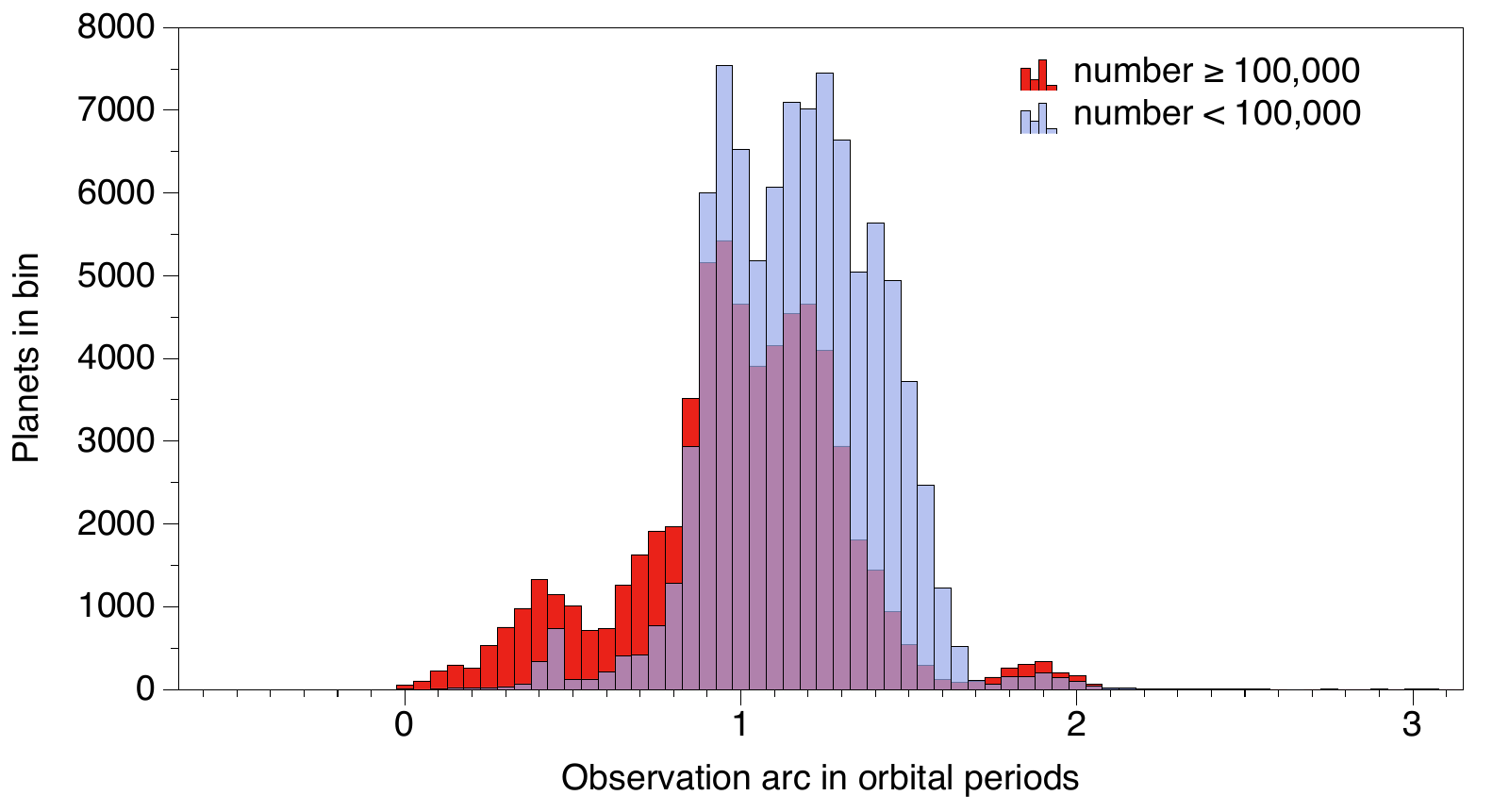} 
      \caption{Observation coverage expressed in orbital periods. Asteroid numbers $<100\,000$ are plotted in blue, and  the complementary set is shown in red.}
         \label{fig:extent_periods}
   \end{figure}

Beyond the number of transits, a key parameter for the orbital solution is the time coverage, or in this context, more precisely, the fraction ($< 1 \text{ or }> 1$) of the orbit that is covered between the first and last observation. That is to say, the arc length measured in orbital periods.  This is shown in blue in \autoref{fig:extent_periods} for the asteroids that were observed best with  numbers $ < 100,000,$  and the fainter and smaller asteroids with numbers $ \ge 100,000$ are superimposed in red. The typical coverage is about $1.2$ orbital periods for the first group, and it is about $0.5$ orbital periods for the Trojans. The typical coverage is shorter at about one orbital period for the fainter asteroids, with a marked tail of short arcs ($\text{less than}\text{}$ one orbital period) that is caused not by the periods themselves, but by a systematic of the Gaia sensitivity threshold: The faint asteroids are not observed at all elongations that are theoretically allowed by the Gaia scan, but only when their distance to the Earth is small enough to have an apparent magnitude $<20.7$ in the Gaia white band. This on-board brightness cut is imposed by the instrument sensitivity and the ground-link data rate. An orbit is better determined when the observation arc covers a complete orbit or more, and the solution is less constrained for short arcs. This qualitative insight is confirmed in \secrefalt{sect:orbeval}, in which we analyse the solutions.

\section{Astrometric solution} \label{sect:astrometry}
The astrometric processing of Solar System objects has been presented  in \citet[Sect. 3.4]{2023A&A...674A..12T} and was described with more technical details in the \gdrthree documentation in  \citet[Sect. 8.3.3]{2022gdr3.reptE...8M}. We refer to these papers for more detailed information.  Here we just recall the key points needed for the readability of this paper and draw attention to the significant departures from the classical ground-based CCD or traditional photographic plate astrometry. 
As a mission destined to survey the sky several times, a scanning strategy was imposed early during the mission design. The \gaia time sampling is a direct consequence of the parameters selected to optimise this intricate scanning law. Because of the \gaia core science, the free parameters were not optimised with Solar System sources as the main targets, but for the best science return in stellar and fundamental physics. The Solar System objects are well sampled, however, just below the sky average, because their ecliptic latitude is predominantly low. This sky area is less frequently visited by \gaia than the mid latitudes.

Each passage of a source in the field of view nominally includes nine crossings of the astrometric CCDs either in the preceding
 or the following field of view. This provides as many 1D astrometric locations in the instrument frame (pixel number) that are perfectly time tagged. These repeated measurements form the basic astrometric data from Gaia. Based on the instrument calibration and the orientation parameters (attitude), these instrument-tied positions can be transformed into nine astrometric positions on the sky, with an excellent accuracy ($\approx \text{ sub-milliarcsecon to milliarcsecond level}$) in the scan direction and much poorer accuracy (up to $0.5 \arcsec$) in the transverse direction. When it is transformed into right ascension and declination, the uncertainty appears degraded in both coordinates, but the true statistical information is kept in the covariance matrix, where correlations are close to 1. This encodes the extreme accuracy in the along-scan direction. Therefore, a standard usage of the RA, DEC positions that does not take the correlation coefficient into account would completely fail to bring the true Gaia accuracy.

The positions used in this paper to fit the orbits are at the CCD level, meaning that the number of observations is always the number of validated CCD crossings. Each field transit has a maximum of nine crossings, but  the actual average is between seven and eight crossings, so that there are about seven to eight times as many observations as field transits. These are also called observations in other contexts of the Gaia data processing. Consistency checks between the multiple 1D positions in the same transit were used to accept or reject observations for inclusion in the data release, taking the expected linear motion of the Solar System object during the $\approx 40$\,s  of the field crossing into account. All these published observations are considered in the subsequent fits, and normal points for position and velocity were not computed.

We implemented an updated version for the FPR reported in this paper for the astrometric calibration used in DR3. The global transformation that produces astrometric positions is also more accurate because it is now based on the full duration of the nominal Gaia mission of 66 months.  Moreover, the FPR processing is mostly independent of the previous data releases. Even in the overlapping time between DR3 and the FPR,  observations can be found in one data set but not in the other. This is a result of a different selection that was used by the internal validation procedures or residuals that are now below or above the acceptance threshold with the new orbit. These are rare occurrences, and they arise as a natural consequence of the reprocessing of the whole data set.

\begin{figure}[htbp]
\centering
\includegraphics[width=0.95\hsize]{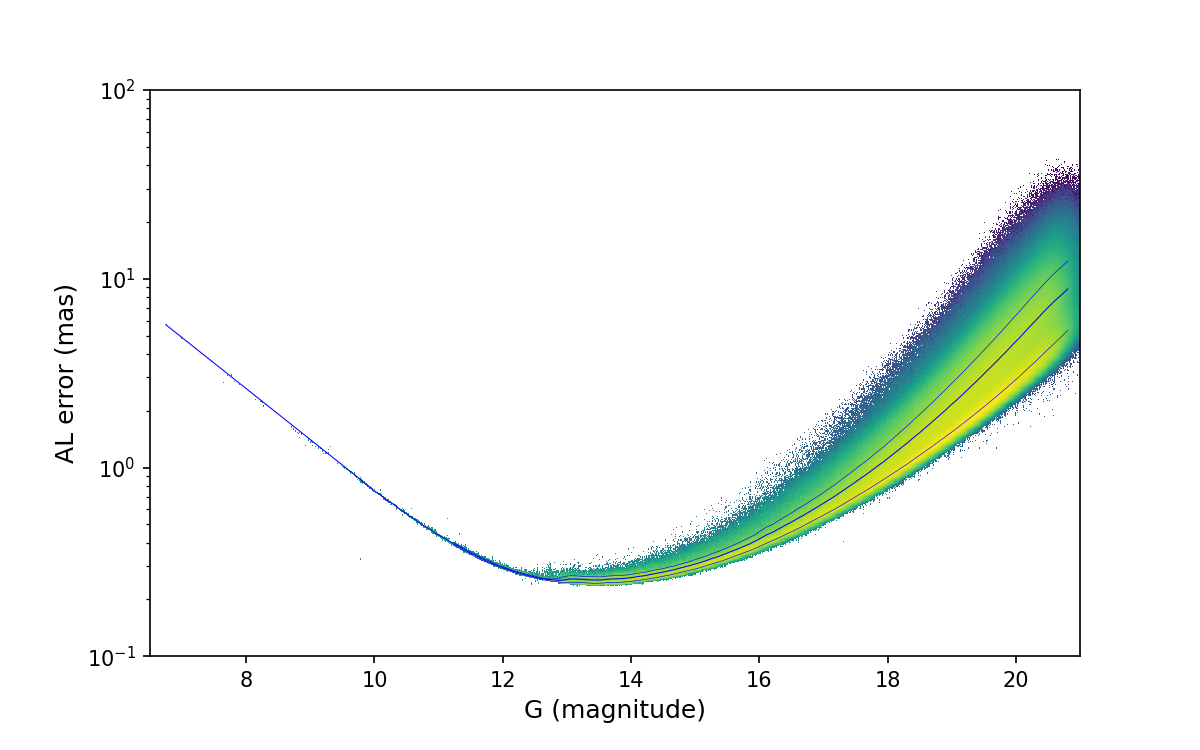}
\caption{Error model in the along-scan direction for the SSO astrometry in \gdrthree as a function of the G magnitude. The total error is represented as given by the squared sum of the random and the systematic component. The colour represents the data density (yellow or light means a higher density). The thick line and the two thin lines on each side are the quantiles that correspond to the mean and the 1$\sigma$ level.}
\label{fig:al_errormodel}
\end{figure}

The precision of the astrometric measurement in the along-scan direction  has been much discussed in \cite{2023A&A...674A..12T}. The precision that is relevant here for the orbit fitting is the combination of a random and systematic part. Its variation with G magnitude is shown in \autoref{fig:al_errormodel}, where it is reproduced in the same way as in Sect. 3.4.2 of the \gdrthree paper The elementary measurement at the CCD level is better than 1 mas for $G < 18$\,mag and grows to reach about 10~mas at $G \simeq 20$\,mag. This makes \gaia data outstanding and unusual in comparison with the core of the historical observations: They are extremely accurate, rather dense, homogeneous, but over a limited time range that is not too short to preclude orbit fitting. 

The orbit can be fit either with observation equations expressed in the along-scan and across-scan directions, or in the more usual way in RA, DEC coordinates. The use of either coordinate set is rigorously equivalent theoretically in any case provided that the $2\times 2$ variance or covariance matrix is well implemented in the global weight matrix. Tests have shown that identical results are obtained with both approaches, as expected, but great care must be exercised in using the statistical information in the covariance matrix in the right manner. 

\section{Orbit fitting} \label{sect:orbfit}
\subsection{Presentation}
We only consider a subset of known asteroids here that have known (elliptical) orbits, as in \gdrthree. We determined a least-squares orbit solution for them that represents the best fit to Gaia data. Detection, follow-up, and initial orbit determination of newly discovered asteroids are made within another pipeline that was described in \cite{carryPotentialAsteroidDiscoveries2021}. 

The overall principle of the orbit computation from the Gaia astrometry was summarised in \cite{2023A&A...674A..12T}. As a rule, the general procedure used in the \gdrthree release is also used here, but the code has been updated to correct minor bugs and improve the overall consistency between the fundamental constants in their form in astronomical or metric units. 
\subsection{Units of time and length} \label{sect:unit_timelength}
\gaia time tagging is the Barycentric coordinate time (TCB)  internally for the data collection and processing, and externally in the public releases. This means that, for example, the Solar System ephemerides have TCB as an independent variable and that  all epoch data are given with a timing in TCB. For stellar astrometry, and even for epoch photometry, the difference between TCB and TDB (or TT), which is about 20 s, is not an issue for the user. This is no longer true for Solar System objects moving at speed of several 10 mas per second, however. This has consequences for the orbit fitting. Observations are timed in TCB, and the equations of motions must be consistent with this timescale. More specifically, the solar mass parameter must be TCB compatible.

Unfortunately, a small glitch has remained in the software between the TCB and our using of an outdated Solar System $GM_\odot$  equal to the square of the Gauss constant $k = 0.017\, 202\, 098\, 95$. This is equivalent to our using a unit of length that is not strictly equal to the IAU astronomical unit. Because of the constraints of the complex integrated processing of the \gaia data and the huge validation pipeline, it was not possible to correct for the scale factor in the FPR results. It seems to be a benign correction, and it is, but nothing is simple in a complex system. Fundamentally, this is a mistake, and this issue has been fixed in the software for the planned \gdrfour. The positive side of this unfortunate bug  is that noticing a systematic effect as small as $0.5\times 10^{-8}$ in the results already means something good about the quality of the solution.

Mixing the Gauss constant and the TCB timescale results in an $\text{au}_\text{FPR}= 149\,597\,871\,473.216\, \text{m}$ instead of the IAU fundamental constant for the scale length in SI equal to $149\,597\,870\,700.0\,\text{m}$. The consequence is that the Gaia published state vectors of the asteroids must be scaled to scale them properly in TCB and au, as it should have been in the first place. Afterwards, the transformation to osculating elements provides a TCB-compatible semi-major axis. More explicitly, we have\begin{align*}\label{eq:scaling_axis}
\rho & =  \text{au}_\text{FPR}/\text{au}\\
 &=149597871473.216/149597870700\\
 &=1.0000000051686297,
\end{align*}
which with $ k^2 \approx (GM_\odot)_\text{TDB} $ , is essentially
\begin{equation*}
    \left(\frac{ (GM_\odot)_\text{TCB}}{(GM_\odot)_\text{TDB}}\right)^{1/3.}
\end{equation*}
 Now the TCB compatible values in au and au/day can be computed from the published components $\mathbf{X}_\text{FPR}, \mathbf{V}_\text{FPR}$ of the state vector with
\begin{align*}
    \mathbf{X}_\text{TCB} &= \rho\times  \mathbf{X}_\text{FPR}\\
    \mathbf{V}_\text{TCB} &= \rho\times  \mathbf{V}_\text{FPR.}
\end{align*}  
It can also be added  that when the FPR state vector in osculating elements is transformed before the scaling, only a single scale transformation has to be applied to the semi-major axis,
\begin{equation*}
    a_\text{TCB} = \rho\times a_\text{FPR.}
\end{equation*}  
We show in \secrefalt{sect:JPL_compar} how the Gaia TCB compatible orbits and the equivalent from JPL TDB compatible orbits are handled.
\subsection{Principles and algorithm} \label{sect:orbit_principles}
The \gaia-centric astrometric data in right ascension and declination at CCD-level are adjusted using a standard weighted linear least-squares (LLS) fit. The error model with non-diagonal covariance matrix as derived in \cite{2023A&A...674A..12T} provides the weights to be applied for each observation equation. The same set of asteroids as in \gdrthree is considered for this new release, but over an observational time span that is roughly twice as long.

For the LLS method, the theoretical expectation  is required. To spell out this function, the equations of motion are integrated simultaneously with the variational equations.  In the ICRF reference frame and on the TCB timescale, the heliocentric equation of motion is expressed as
\begin{equation}
        \ddot{\mathbf r}_i = -(GM_{\odot}+Gm_i) {\frac {\mathbf r_i}{r_i^3}} + \sum_p \mathbf f_p
\label{E:motion}
,\end{equation}
where $r_{i} = |\mathbf r_i|$ and $\mathbf r_i$ is the heliocentric position vector of asteroid $i$, $GM_{\odot}$ is the mass parameter of the Sun, and $Gm_i$ is the mass parameter of asteroid $i,$ which can generally be neglected.
The additional perturbing accelerations, added linearly, correspond to
\begin{equation}
        \sum_p \mathbf f_{p} = \mathbf f_{p{|Grav}} + \mathbf f_{p{|relat}} + \cdots  
\label{E:motionPert}
\end{equation}
They account as in the \gdrthree orbital fitting for the planetary perturbations and the relativistic terms, respectively. 
The part due to the gravitational perturbations (by a total of $n_{p}$ planets, and potentially other perturbing asteroids) is given by
\begin{equation}
        \mathbf{f}_{p{|Grav}} =- \sum_{j\ne i,j\ge1}^{n_p} G m_j \left\{\frac{\mathbf {r}_i-\mathbf {r}_j}  {r^3_{ij} } + \frac{\mathbf {r}_j }{r_{j}^3}\right\}
\label{E:motionPGrav}
.\end{equation}
The relativistic contribution to the equations of motion can be approximated by
\begin{equation}
\mathbf {f}_{p{|relat}} \simeq {\frac{{G}M_{\odot}}{c^2r_i^3}} 
\left\{ \,\left[ \frac{GM_{\odot}}{r_i}\mathbf{\dot{r}_i^2}\right] \,\mathbf r_i +4(\mathbf{r}_i\cdot\mathbf{\dot{r}}_i) \mathbf {\dot{r}}_i \,\right\}
\label{E:motionPrelat}
.\end{equation}
The variational equations are generated by the derivation of Eq.~(\ref{E:motion}) with respect to the parameter to be corrected. \citet{Beutler2005} and \citet{Pontriaguine1969} give a very thorough exposition of how the variational equations can be obtained from a general point of view. 

The heliocentric positions of the asteroids are computed from numerical integration\footnote{We integrated the equations using the implementation of the Gragg--Bulirsch--Stoer integrator provided by the Apache Commons software available at \url{http://apache.commons.org/}} of Eq.~(\ref{E:motion}). In \gdrthree, the Solar System model is the planetary solution INPOP10e\footnote{The INPOP models are developed and made available by the Institut de Mécanique Céleste et de Calcul des Ephémérides, Paris Observatory, France. Visit \url{https://www.imcce.fr/}}. In the current solution, the Solar System model has been updated to INPOP19a, which presents some important differences with INPOP10e. As with INPOP10e, the Earth-Moon barycentre with the seven other planets, dwarf planet Pluto, and a selection of 343 asteroids are included in the Solar System model \citep{2013arXiv1301.1510F, arXiv:2203.01586}. For INPOP19a, the ten most massive trans--neptunian objects (TNO) have additionally been included as perturbing bodies. Furthermore, in INPOP19a, influences by the less massive TNOs are modelled by a ring. These added perturbations in INPOP19a cause a shift in the Solar System barycentre that affects the barycentric position of the Gaia satellite \citep{2019NSTIM.109.....F}.

No perturbation of asteroids is included yet in Eq.~(\ref{E:motionPGrav}). This is still a limitation compared to other databases of asteroid orbits, which consider systematic perturbations by several massive asteroids (e.g. Ceres, Pallas, or Vesta) together with the major planets, or for consistency with INPOP19a. This approximation on the dynamical model introduces a small bias on the orbital solution (quantified in internal validations; see Sect.~\ref{sect:JPL_compar}), but almost no  effect on the uncertainty of the orbits themselves. Thus we will compare and analyse the formal uncertainties of the orbital solution, not its {\em \textup{accuracy}}.  These systematic perturbations will be included in subsequent data releases (starting from DR4), together with other mutual perturbations, that should also allow the mass determination from a global inversion. 

Because mutual interactions between asteroids are neglected, equations~(\ref{E:motion}) and~(\ref{E:motionPGrav}) can be integrated simultaneously with the variational equations for each asteroid separately. The small corrections relative to the reference orbit are then computed at a given reference epoch by an LLS method that minimises the residuals on the astrometric position. The correction is determined for the six-dimension initial state-vector in Cartesian coordinates taken from an auxiliary file of orbital elements. In our case, it was derived from the {\tt astorb} database\footnote{Information on the astorb data base can be found at https://asteroid.lowell.edu/astorb/} and was propagated to the reference epoch. The reference epoch for the orbital fit was taken to be the mid-point of the total time span of the observations available for the given asteroid as a proxy for the weighted mean time of the observational arc. This reference epoch is thus not common to all asteroids in the orbital data base.

To be more concise, we must solve the following linear system of equations for each asteroid source:

\begin{equation}
\sqrt{\mathbf W} \, \textnormal{d} {\boldsymbol\lambda} = 
\sqrt{\mathbf W} \, \left(\begin{array}{l}
\Delta\alpha_0\cos(\delta_0)     \\
\Delta\delta_0 \\
\Delta\alpha_1\cos(\delta_1)     \\
\Delta\delta_1 \\
\vdots \\
\Delta\alpha_{N-1}\cos(\delta_{N-1})     \\
\Delta\delta_{N-1}
 \end{array}\right)\, = \sqrt{\mathbf W} \, \mathbf{A}. \textnormal{d}\mathbf {q}
\label{E:blocmatrix}
,\end{equation}
where the vector $\text d \boldsymbol\lambda(t)=\mathbf{(O-C)}$  represents the value of the difference between the observed and computed values of the measured quantity, in this case, the right ascension and declination for each of the $N$ observations for a given asteroid. $\text d \mathbf{q}$ is the differential correction to the parameters, here the components of the initial state-vector for the asteroid under consideration, and $\mathbf{A}$ is the matrix formed with the partial derivatives with respect to these parameters, projected to right ascension and declination. Furthermore, all equations are weighted by the matrix $\sqrt{\mathbf W}$ obtained through the corresponding variance or covariance matrix for the right ascension and declination of the observations. $\sqrt{\mathbf W}$ is computed as described in \cite{doi:10.1080/0025570X.1980.11976858}. We accepted only real values and used only the positive square roots where these are needed in the formulae. Moreover, because the variance or covariance matrix for right ascension and declination of the observations is positive semi-definite and symmetric, $\sqrt{\mathbf W}$ verifies ${\sqrt{\mathbf W}}^T\sqrt{\mathbf W} = {\mathbf W}$. We refer to \cite{2023A&A...674A..12T} for a detailed explanation of the uncertainties involved in the data reduction. As explained therein, each observation of an asteroid at a given epoch, that is, a right ascension $\alpha$ and declination $\delta$, has an associated variance or co--variance matrix of the form
\begin{equation}
\boldsymbol\gamma =
\left(\begin{array}{cc}
\textnormal{var}(\alpha^*)       & \textnormal{covar}(\alpha^*, \delta) \\
\textnormal{covar}(\alpha^*, \delta) & \textnormal{var}(\delta) 
 \end{array}\right) 
 =
\left(\begin{array}{cc}
\sigma_{\alpha^*\alpha^*} & \sigma_{\alpha^*\delta} \\
\sigma_{\alpha^*\delta} & \sigma_{\delta\delta} 
 \end{array}\right)
\label{E:variance_covariance_per_observation}
,\end{equation}
with $\alpha^* = \alpha\cos{(\delta)}$ in differential quantities. Then the weights are obtained as ${\mathbf W} = {\boldsymbol\gamma_{k}^{-1}}$ for the $kth^\text{}$ 
asteroid. We recall that the off-diagonal elements play a prominent role in the uncertainty deduced for the sky coordinates, as already stressed in \secrefalt{sect:astrometry}.
 
Planetary aberration  as well as the gravitational light bending are then added to the computed coordinates as so-called corrections to the observations to construct the (O$-$C) vector in Eq.(\ref{E:blocmatrix}). All corrections are computed in the general relativistic framework of Gaia data reduction, considering the source at finite distance \citep{2003AJ....125.1580K}. While the relativistic light deflection effect, which can reach several milliarcseconds at moderate solar elongations, is not included in ground-based observations, it is mandatory in the treatment of Gaia data due to the sub-milliarcsecond astrometry in the position of the asteroids. We recall that the  published right ascension and declination are corrected for the annual aberration, that is, the equivalent to the stellar aberration due only to the barycentric motion of the observer, without any contribution from the asteroid motion.

The LLS is based on the optimisation of a target function, which in our case is defined by
\begin{equation}
    Q = \frac{1}{N_i} {\sum_{j=0}^{N_i-1} (\mathbf {O - C})_j^\text{T}\,{\mathbf W}_j\,(\mathbf {O - C})_j}
\label{E:target_function}
,\end{equation}
that is, the sum of the squares of the weighted residuals \citep{Milani2010} of the superscript T meaning the transpose of the vector $(\mathbf {O - C})_j$. As in the \gdrthree release, optimisation is 
performed for each asteroid individually, considering all astrometric observations $N$. There are no more general parameters that would couple observations from different planets in the fit. Thus, the LLS solution of the system  in 
${\textnormal d} \mathbf{\bar q^{\null}} = 
(\mathbf{A}^{\textnormal T}\mathbf{W}\mathbf{A})^{-1}\!\mathbf{A}^{\textnormal T} \mathbf{W}
{\textnormal d}{\boldsymbol\lambda^{\null}}$
with corresponding variance/covariance matrix $(\mathbf{A}^{\textnormal T}\mathbf{W} \mathbf{A})^{-1}$ where $\mathbf {A}^{\textnormal T}$ indicates the transpose of the matrix $\mathbf A$.

\subsection{Iterations and convergence} \label{sect:orbit_iterations}
The procedure of orbit correction is iterated until convergence is reached with a given tolerance $p$  following  \citealp[sec. 5.2]{Milani2010}; initially this parameter has a value of $10^{-8}$ however, it can be increased in some cases. A preliminary outlier filtering is performed during the astrometric reduction (section \ref{sect:astrometry}), further outlier rejection at the observation level has been implemented and follows \citet{Carpinoetal2003}. 
 
The procedure is built around a post--fit $\chi^2$ computed as 
$\chi^2 = \boldsymbol\xi_j \, \boldsymbol\gamma_{\xi_j}^{-1} \, \boldsymbol\xi_j^{\textnormal T}$; 
with $\boldsymbol\xi$ the residuals of all observations evaluated with respect to the corresponding orbit and $\boldsymbol\gamma_{\xi_j}$ their expected covariance  matrix. The threshold for $\boldsymbol\xi$ is a parameter of the fitting process. The tolerance test for stopping the computation is evaluated between iteration $k$ and $k+1$ as follows:
\begin{equation}
c_1 = \frac{{\lvert Q_{k+1} - Q_k \rvert} }{ Q_{k+1}} < p
\label{E:tolerance_p}
,\end{equation}
where $p$ is a tolerance threshold, and $Q_k$ is given by Eq.(\ref{E:target_function}) for the $\textnormal{kth}^{}$ iteration. 

If the next iteration is unlikely to improve the corrections, the processing can also be stopped; this can be assessed by considering the size of the last correction using an appropriate norm (see \citet{Milani2010}), for example,
\begin{equation}
        c_2 = \lVert{\Delta \mathbf{X}}\rVert = \sqrt{\frac{\Delta \mathbf X^{\text T} \, \mathbf{M}\, \Delta \mathbf{X}} { N}} < p
\label{E:tolerance_norm}
,\end{equation}
$\mathbf{\text{where }M} = \mathbf{A}^T \mathbf{W\, A}$ is the normal matrix, $\Delta \mathbf{X}$ is the correction vector, and $N$ is the number of observations. In practice, this criterion is used far less frequently. In addition to the two criteria mentioned above, an absolute maximum number of iterations is also denoted $n_{max}$.

The iteration starts with $p = 10^{-8}$ , and the maximum number of iterations is set to $n_{max} = 15$. The value of $p = 10^{-8}$ is somewhat arbitrary, and different values could be imposed (or even a different tolerance for each condition $c_1$ and $c_2$). However, according to our tests, it is unlikely that this would significantly affect the overall statistical outcome of the orbit adjustment because typically, only a few iterations are required. When neither $c_1$ nor $c_2$ is lower than $p$ after 15 iterations, the iteration is continued and $p$ is replaced by $10\,p$. When another 15 iterations do not return a $c_1$ or $c_2$ that is lower than $10\,p$ , this is repeated one more time for a last trial. In general, convergence is reached after a few iterations, and in practice, the value of $p$ had to be changed very rarely.

Three situations are considered as a failure in the attempt to fit an orbit. First when the maximum number of iterations is reached, second if the orbital solution is not an ellipse and is rejected and third when all observations have finally been rejected by the rejection algorithm. 

A total of 59 sources were rejected, but their astrometry still appears in the FPR data; these are identified with a {\tt NULL} state vector in the tables. Interestingly, 31 of these 59 sources are moons of Mars, Jupiter, Saturn, Uranus, or Neptune. An attempt at orbit improvement was not even performed for these objects. Twenty-four of the remaining 28 asteroids are main-belt asteroids,  three are near-Earth objects ((433) Eros, 2001~XV$10$, and 2003~YV$37$), and  the last source is a Mars crosser (2006~PG$1$). For two of the main-belt objects, 2005~UB160 and 2001~QG115, the high rejection rate that is probably due to a larger fraction of outliers makes convergence unstable or the orbit computation based on overly short arcs unreliable. No convergence was reached. A dedicated investigation of this subset of asteroids, which consists of only $0.2\text{\textperthousand}$ of the orbital catalogue, is needed to fully understand why convergence has not been reached. This is beyond our concerns in this FPR, however, because  (433)~Eros is a difficult case in Gaia DR3 and requires further investigations. 

\subsection{Post-fit residuals}
The full processing resulted in 156\,762 orbits for the FPR. 
In contrast to \gdrthree, there is no post-processing of the FPR orbit determination. In the previous catalogue, a further filtering of solutions was applied based on the relative uncertainty in the semi--major axis $\sigma_a/a$ and arc length. We refer to \citealp[Sect.~3.5]{2023A&A...674A..12T} for an explanation. 

\begin{figure}
\centering
\includegraphics[width=0.85\hsize]{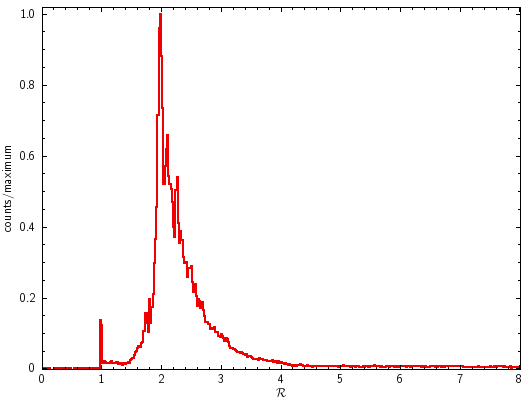}
\caption{ Ratio of the arc length of the FPR sources to that of the \gdrthree  ($\mathcal{R}$). The peak at $\simeq 2$ is marked, and the excursion about the mode is moderate.}  
\label{fig:ratioArcFPRtoDR3Density}
\end{figure}

\begin{figure}
\centering
\includegraphics[width=0.95\hsize]{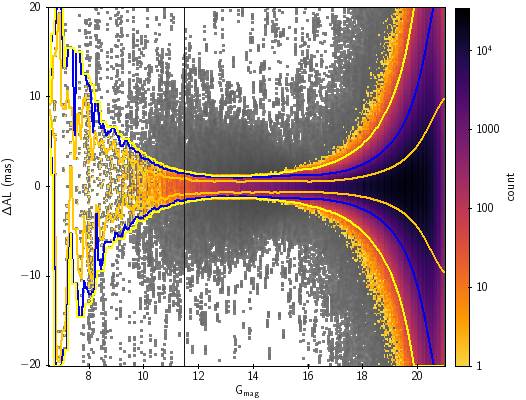}
\caption{Along-scan residuals against G magnitudes. The grey points represent all observations. The coloured points are the observations that were accepted by the orbit adjustment procedure after outlier rejection chosen at $|O-C|>5\,\sigma$. For $G_{mag} < 11.5,$ the distribution is not Gaussian. This is clearer in \autoref{fig:histALGmag}. For $G_{mag} \ge 11.5,$ the distribution is Gaussian. The orange lines in the plot are $\pm1\,\sigma$, the blue lines are $\pm2\,\sigma$, and the yellow lines are $\pm3\,\sigma$. The number density is given by the scale on the right.}
\label{fig:ALresVsGmag}
\end{figure}

\begin{figure}
\centering
\includegraphics[width=0.95\hsize]{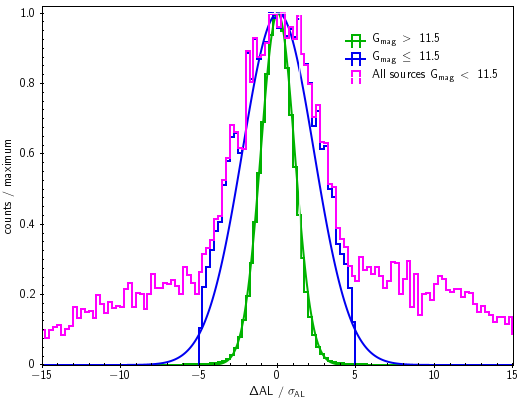}
\caption{Histograms of the along-scan residuals normalised to the formal uncertainties for the accepted solutions with $G_{mag} \le 11.5$ shown in blue or $G_{mag} > 11.5$  shown in green. The histogram in magenta shows all sources (accepted or rejected) with $G_{mag} \le 11.5$. A Gaussian fit to the data is shown as well. We note that for sources with $G_{mag}>11.5,$ the distribution is very well represented by a Gaussian ($\mu = 0.018$, $\sigma=1.08$), but for $G_{mag}\le 11.5,$ the distribution is clearly not Gaussian.}
\label{fig:histALGmag}
\end{figure}

The time span covered by the FPR observations is roughly twice that of \gdrthree. Although here the astrometric processing is more accurate (see  \secrefalt{sect:astrometry}), it was a reasonable guess to expect that the arc covered by any object would be at least as long as it was in the \gdrthree release. This is true for an overwhelming majority of the sources ($99 \%$) and even $76\%$ having the ratio $\mathcal{R}$ of the total arc length of the FPR sample to that of the \gdrthree at least equal to 2.  Figure \ref{fig:ratioArcFPRtoDR3Density} shows the normalised distribution of $\mathcal R$ where a sharp peak is seen around $\mathcal R = 2$ followed by a steep decline. A substantial fraction of the sources have a ratio $\mathcal R > 2$, while the global time coverage from Gaia has changed from $34$ to $66$ months. These happen to be sources with actual time coverage, between first and last observation, much below $34$ months in \gdrthree, thus increasing automatically the ratio in FPR. One can see also that there are very few sources with 
a ratio less than one, i.e. an unexpected longer arc in the \gdrthree data than in the FPR sample. This occurs when the orbit adjustment rejects a large part of the observations, resulting into a smaller orbital arc; only eight sources (over approx. 157\,000) belong to this group.
Figure \ref{fig:ALresVsGmag} displays the along scan residuals against the G magnitude, $G_{mag}$. The $5 \sigma$ cut--off value for rejection is visible; it is the envelope of the accepted observations, i.e. the limit between the grey and coloured points. Figure \ref{fig:histALGmag} shows the corresponding histograms for normalised residuals $\Delta \text{AL}/\sigma_\text{AL}$ for sources with a $G_{mag}$ less than or equal to $11.5$, or greater than $11.5$, respectively. For accepted observations with $G_{mag}$ greater than $11.5$, the normalised distribution is Gaussian with a mean of $0.018$ and standard deviation of $1.08$; for these sources the orbit adjustment is acceptable. Sources with $G_{mag}$ less than $11.5$ do not show a Gaussian distribution, even if the rejected solutions are included; in this magnitude range the model is not entirely satisfactory, but can be improved by including some other effects, as the photocentre offset for example. For $G_{mag}>11.5$ the symmetry is striking.

\begin{figure}
\centering
\includegraphics[width=0.95\hsize]{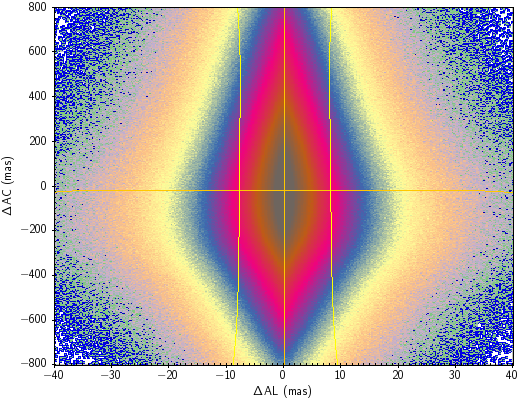}
\includegraphics[width=0.825\hsize]{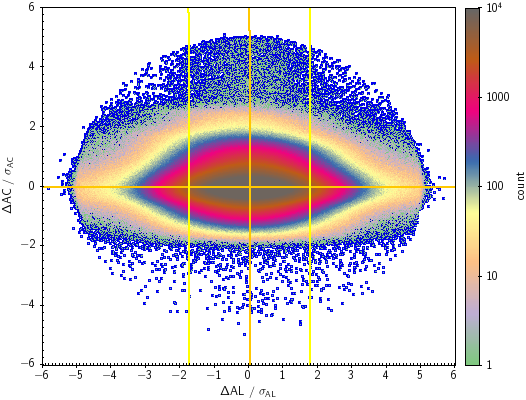}
\caption{Post-fit residuals of the across-scan direction  vs. the along scan direction. Top: True angular values. Bottom: Normalised quantities.  The orange lines  show the median, and the yellow lines show the $5$th (left) and $95$th percentile (right).  The density is given in log-scale, so that only the core of the plots is populated.}
\label{fig:ALvsAC}
\end{figure}

Figure \ref{fig:ALvsAC} displays the residuals in along and across scan. There is a small offset from zero, $(\Delta\,AL, \Delta\,AC) \approx (0, -15)\text{\,mas}$ or $(\Delta\,AL/\sigma_{AL}, \Delta\,AC/\sigma_{AC}) \approx (0.05, -0.028)$, this is also the true in \gdrthree. In the $\Delta\,AC/\sigma_{AC}$ vs $\Delta\,AL/\sigma_{AL}$ plot the distribution is clearly asymmetric about $\Delta\,AC/\sigma_{AC} = 0$ but appears to be much more symmetric about $\Delta\,AL/\sigma_{AL} = 0$. Again, the same is seen in the \gdrthree sample and is probably due to the greater uncertainties in the across scan direction. For $90\%$ of the sources $\Delta\,AL$ is in the interval $[-8, +8]$~mas, $\Delta\,AL/\sigma_{AL} \in [-1.8, +1.8]$.

\begin{figure}
\centering
\includegraphics[width=0.95\hsize]{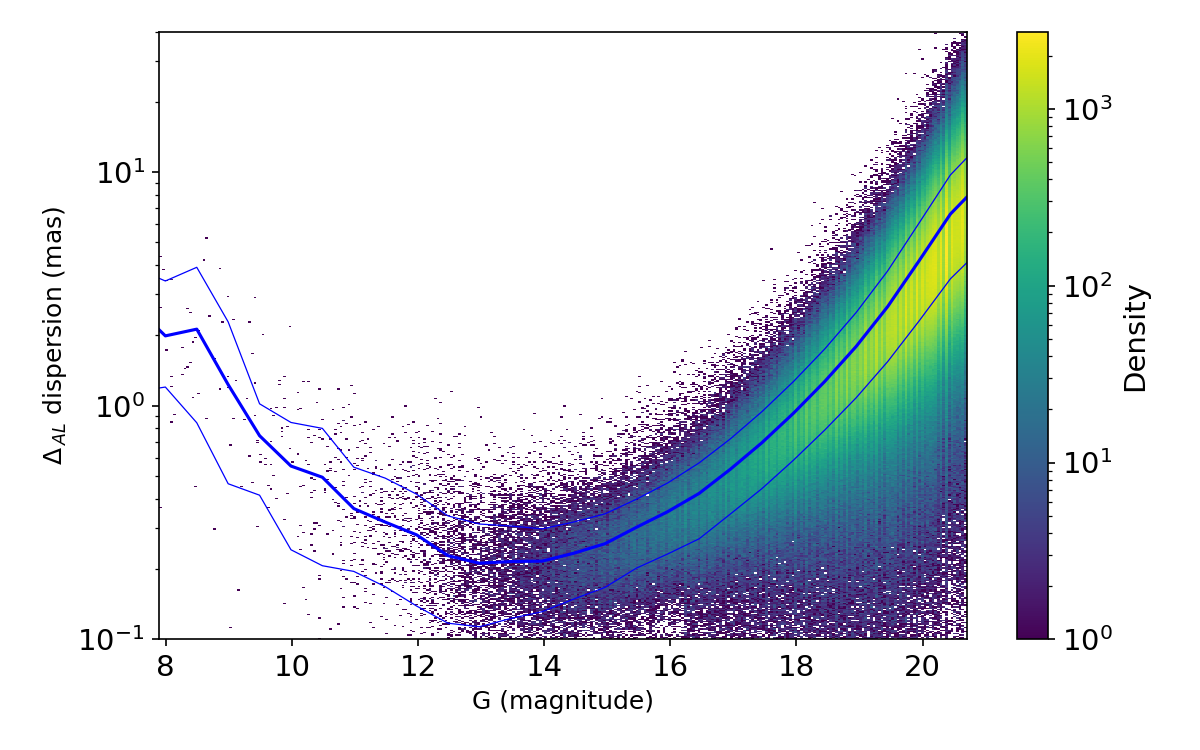}
\caption{Standard deviation of the values of the along-scan residuals for each CCD, computed over each transit. The colour corresponds to the point density. The lines represent the smoothed average, and the quantiles at the 1$\sigma$ level.}
\label{fig:res_std}
\end{figure}

A representation of the astrometry quality over a single transit can finally be obtained by considering the dispersion of the measurements that were recorded during each passage of a source in the focal plane in the along-scan direction (Fig.~\ref{fig:res_std}). This figure can be compared to Fig. 12 in \citet{2023A&A...674A..12T} for \gdrthree. The figures appear nearly equivalent, with a slight improvement around the G-magnitude interval of the best performance (G$_\text{mag} \sim$12-13).  


\subsection{Orbital elements} \label{sect:orbelem}
The output of the orbit fitting for a particular asteroid consists of a 6D vector of the initial conditions at a particular reference epoch, the midpoint of the observation times, and is given in TCB. This vector combines a position and a velocity vector and is heliocentric. It is given on the ICRF axes. The transformation from the state vector to the osculating elements is in principle routine work in celestial mechanics, and fully tested software is widely available. However, here we work with sub-milliarcsecond astrometry, or with orbits at the $10^{-9} - 10^{-10}$ accuracy level  ( as $\sigma(a)/a$), and care must be taken everywhere to ensure that no loss of accuracy is associated with the transformation and
that the underlying conventions are well understood and implemented. This is the object of this section, to explain all the conventions we used in our solution and describe where they may differ from other usages.

The transformation from the state vectors to the osculating elliptical elements is done in two steps: first, the transformation from heliocentric ICRF frame heliocentric ecliptic frame for the Cartesian coordinates and then the transformation in the ecliptic frame from  state vector to osculating elements.
In addition, the TCB is used consistently for Gaia while all other sources of osculating elements used TDB in their publications and this impacts the scale length.

The transformation of the state vector to the ecliptic frame depends on just one rotation matrix to connect the two frames. The main part is the rotation $\epsilon$ around the $x$-axis, where $\epsilon$ is the obliquity of the ecliptic. The devil lies in the details, as so often is the case.  The intersection of the dynamical ecliptic with the ICRF fundamental plane has no reason to coincide with the ICRF $x$-axis, and this is not the case, either. In addition, the obliquity of the ecliptic relates the ecliptic (taken as a dynamical plane with a strict definition) to the celestial equator. The latter is again different from the ICRF reference plane, and the angle between the two planes of interest is not exactly the physical obliquity of the ecliptic. It is very similar, but not equal. As a consequence, and for lack of a well-agreed convention, several definitions are in use. From a single state vector, we can derive six sets of osculating elements that differ by several tens of milliarcseconds in the longitude of the ascending node or the orbital inclination. This motivated our choice to publish Gaia orbits in the form of a heliocentric state vector in the ICRF axes instead of the usual osculating elements. The former is unambiguous, and from this vector, the orbital elements can be computed so that they agree with the different choices for the ecliptic. They differ slightly with the choice of the ecliptic, although they represent the same orbit.

\begin{figure}
    \centering
    \includegraphics[width=0.95\hsize]{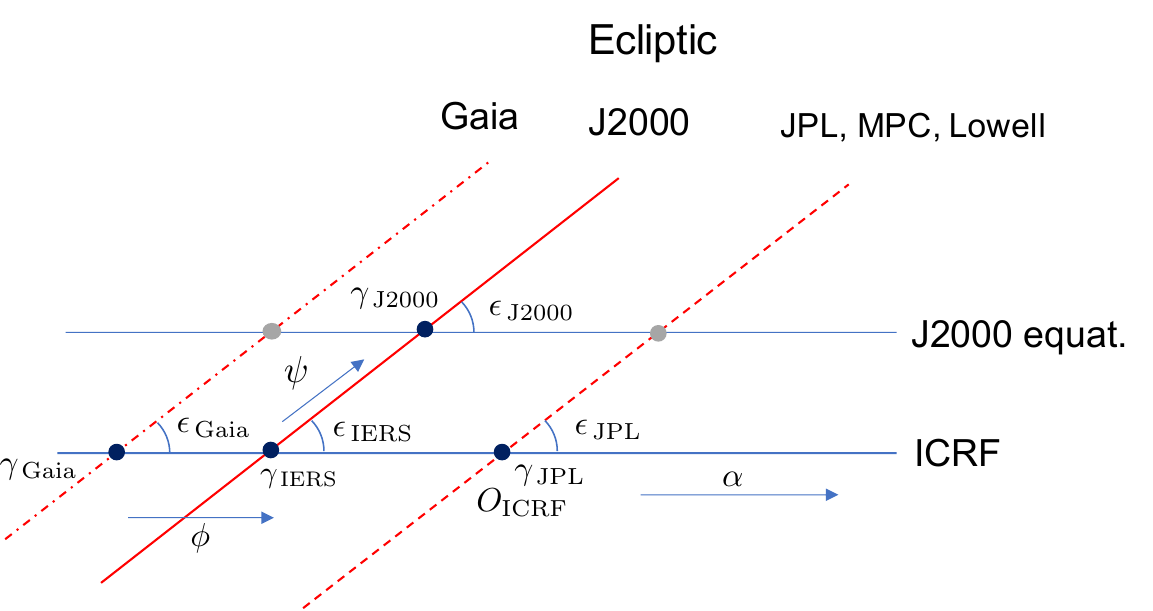} 
    \caption{Relative positions of the ICRF and J2000 equators and of the ecliptic with the different conventions found for the ecliptic frame. The origin of the ICRF ($O_\text{ICRF}$) is offset from the inertial equinox $\gamma_\text{J2000}$. The relevant numbers are given in \autoref{tab:icrf_ecliptic}.}
    \label{fig:ecliptics}
\end{figure}

The issue extends beyond the published osculating elements to the users of this information. They may start from osculating elements of an asteroid to propagate to another time and compute local coordinates using the true equator of the date. Then they will call a precession or nutation package, relating the J2000 celestial equator to another epoch, or the ICRF plane to the celestial equator by using the frame bias above the precession. Therefore, the reference plane and the origin of longitudes for the osculating element are key ingredients for high-precision computation. In this context, high precision means better than 50 mas,  which is the magnitude of the small angles in \autoref{tab:icrf_ecliptic} or of their difference between institutes.

We have attempted to collate the various choices made for the reference plane and the origin of the longitude of node in this plane by the most relevant sources of osculating elements. \autoref{fig:ecliptics} shows the different equatorial planes and ecliptics, together with their associated obliquity. The angle $\phi$ is the value of $\gamma_\text{xx} O_\text{ICRF}$ (xx: ICRF, Gaia, JPL) measured in the ICRF equator and taken positive, like right ascension from the local $\gamma_\text{xx}$ (the sign convention is opposite to \cite{2002A&A...387..700C} and similar to \cite{2004A&A...413..765H}). The so-called IERS values are taken from the SOFA software and come from the precession formulae derived in \cite{2003AJ....126..494F}. The angle $\psi$ is the angular distance between the intersection of the ecliptic with the ICRF plane and the intersection with the J2000 equator, positive like an ecliptic longitude. The numerical values of the angles needed to make the transformation from the ICRF to the selected ecliptic are given in \autoref{tab:icrf_ecliptic}. The Gaia offset was taken from \cite{2002A&A...387..700C} before the adoption of the Fukushima precession by IERS and SOFA, and has not yet been changed for continuity reasons. The ecliptic plane and origin in the ICRF equator are extremely similar (within 3 mas) to the VLBI-derived ecliptic adopted in SOFA. However, the origin of the longitude of node is not on the J2000 equator, but in the ICRF equator.

\begin{table}[htbp]
\centering
\caption{Relevant angles defined in \autoref{fig:ecliptics} that were used to relate the inertial ecliptic to the ICRF fundamental plane.}
\setlength{\tabcolsep}{2mm}
\begin{tabular}{c@{\hspace{1cm}}ccc}
   \toprule
 & $\Delta\epsilon$ & $\phi$ & $\psi $ \\
 & $''$ & \text{mas} & \text{mas} \\
\midrule
IERS(SOFA) & 0.412819 & 52.928  & 41.775    \\
JPL &   0.448000  & 0  & 0    \\
MPC &   0.448000  & 0  & 0    \\
Astorb& 0.448000  & 0 &  0 \\
Gaia &  0.411000  & 55.420  & 0    \\[8pt]
J2000&  0.406000    & -  &- \\
 \bottomrule
 &&&\\[-3pt]
 $\epsilon = 84381\arcsec + \Delta\epsilon$ &&&\\
\end{tabular}
\label{tab:icrf_ecliptic}
\end{table}

Nothing is fundamentally wrong with these multiple realisations of the ecliptic, this is just inconvenient. This all stems from the fact that the ecliptic concept is ambiguous when high accuracy is required because the plane orthogonal to the mean angular momentum of the Earth-Moon system depends on what is 
included in this mean, over which time it is averaged, whether long-period wiggles from planetary perturbations are removed or not, and so on. A less physical definition can appear as well with a rigid link to the ICRF based on conventional angles, in a similar way as the Galactic coordinates are related to the equatorial frame at J2000, with no attempt to recreate a physical definition accurately. 

In principle, an unambiguous definition together with numerical constants have been provided in 2006 by the IAU Working Group on Precession and Ecliptic \citep{2006CeMDA..94..351H} and the ensuing IAU Resolution B1 in 2006. However, a unique realisation is currently not agreed, and even within the Gaia Consortium, the ecliptic has been inherited from Hipparcos, and change in standards in a large data-processing system is always weighed against the loss of continuity and is decided only when unavoidable. It is important for the users to be aware of these differences and to be on their guard when using and propagating orbits with a target accuracy lower than 100 mas. The differences between osculating elements may be deceptive and may not reveal  real differences in the orbits. When properly transformed into state vectors in the ICRF, they may look much more similar than from their elliptic elements.

 \autoref{tab:icrf_ecliptic} clearly shows that all the major providers of orbits use the same definition, in which the ecliptic is very simply related to the ICRF equator by a single rotation about the ICRF X-axis, with an angle equal to the obliquity $\epsilon = 84381.4480 \arcsec$.
 We have adopted this convention in the following sections to compute the orbital elements and compare them to other solutions. However, to let the users select their preferred option, and above all, to avoid incorrect assumptions about the ecliptic that is used for the orbital elements, we decided that primary \gaia results will be published as state vectors in ICRF instead of osculating elements. Nevertheless, the covariance matrix between osculating elements is provided because it is independent (to $\approx 10^{-8}$) of the small differences between the ecliptic frames of \autoref{tab:icrf_ecliptic}, and its computation from the state vector could be tricky without an easily available standard transformation routine.

Incidentally, the Gaia astrometric data for asteroids and Solar System objects themselves can provide the link between the kinematically non-rotating  reference frame (ICRF) and a dynamically non-rotating reference frame (e.g. ECJ200, represented by this inertial ecliptic from the planetary solution, or an invariant plane).

\section{Analysis and orbit validation} \label{sect:orbeval}
\subsection{Overall statistics} \label{sect:main_stat}
The same set of asteroids as was selected for \gdrthree was used in this processing, but over a time span of 66 months instead of  the 34 months for \gdrthree. The number of asteroids with validated astrometry that were processed this pipeline is $156\,825,$ and we obtained converged solutions for $156\,762$  asteroids with an accepted state vector, or equivalently, a set of six orbital elements, at a particular reference epoch. The success rate may seem particularly high, and it is, but this results from the fact that the sources selected for this run were precisely those with a successfully computed orbit with the \gdrthree data, but with a shorter time range. We described the filters we applied in the course of the processing to accept a solution in \secrefalt{sect:orbfit}. 
   \begin{figure}[htbp]
   \centering
\includegraphics[width=0.95\hsize]{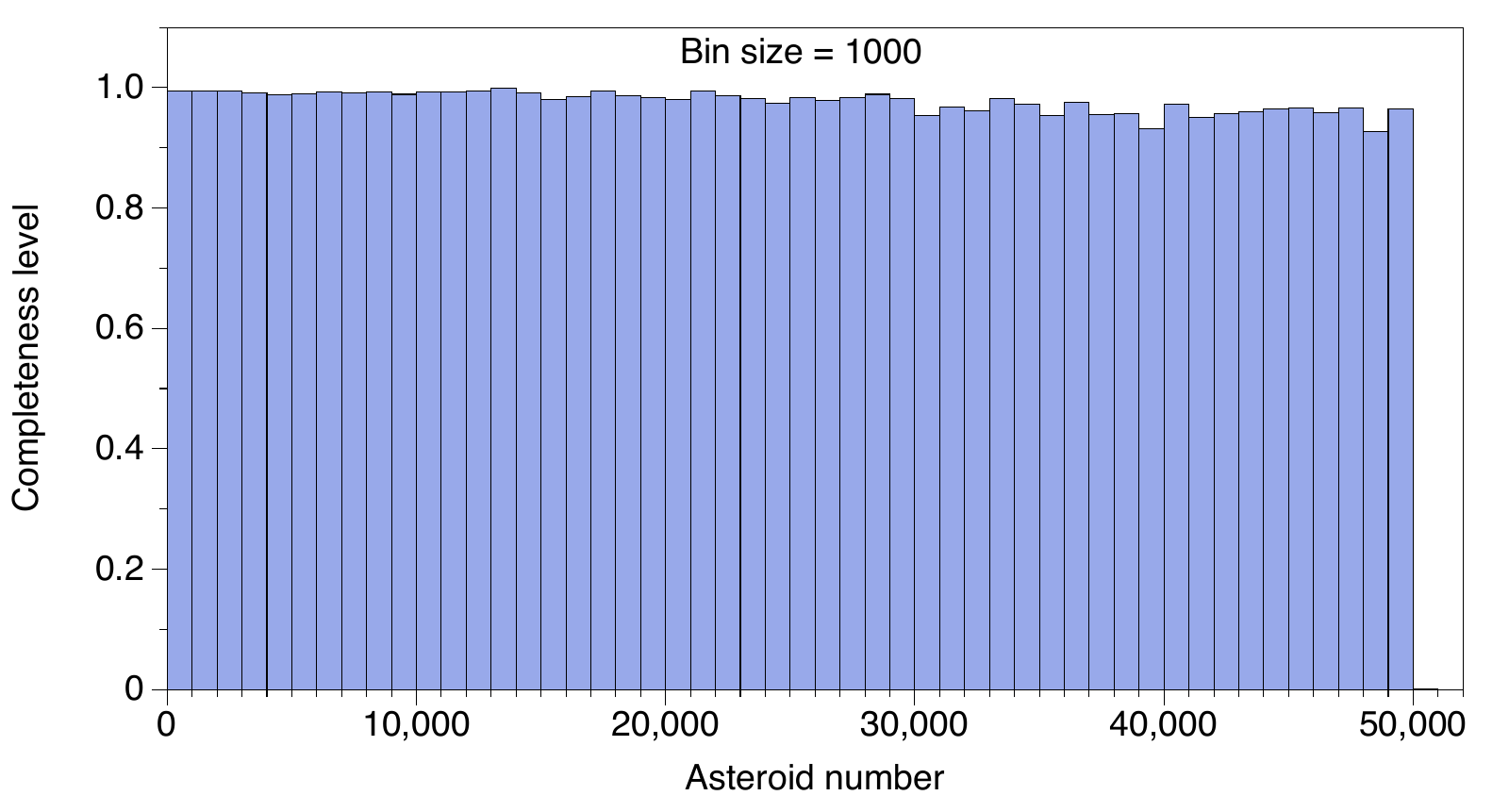} 
\includegraphics[width=0.95\hsize]{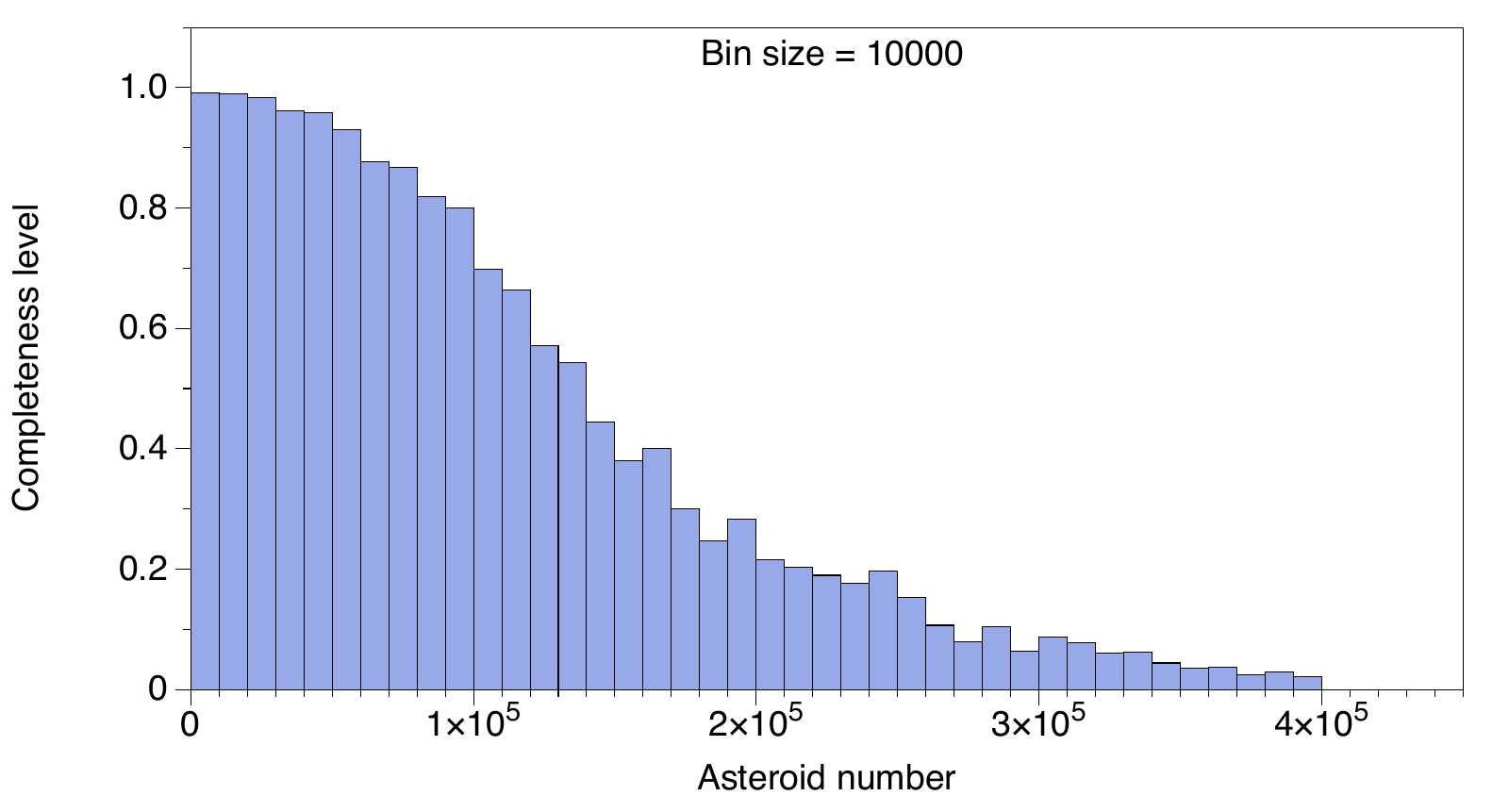} 
     \caption{Completeness level of successful orbital solutions as a function of the asteroid number per bin of 1000 (top) and 10,000 (bottom). Top: First 50\,000 numbered asteroids, with 1000 successes achievable per bin at most. Bottom: Whole set of $\approx  157\,000$ solutions per group of 10,000.} 
         \label{fig:num_sol_id}
   \end{figure}
Although in this sample, we have numbered asteroids up to  $= 400\,000$, many of the asteroids with large numbers are not observed by Gaia or are too poorly sampled for us to compute an orbit from the small number of available positions. This feature is shown in \autoref{fig:num_sol_id}, with the level of completeness as a function of the rank of the asteroid for bin of 1000 or 10,000 planets. The top plot includes up to number $=50\,000$ and is virtually 100\% complete ($99\%$ of the first $<25\,000$ planets with an orbit, and $97.12\%$ of the first 50,000). The bottom plot shows a similar histogram for the full data set, showing the sharp decrease in completeness level for smaller and fainter asteroids. It reaches below $50\%$ (5000 solutions in a bin of 10,000 planets) above  number $ 140\,000$ and reaches $10\%$ from  number $ 300\,000$. Not much more improvement is expected in the future releases (\gdrfour and \gdrfive) because the missing asteroids are just not detected and will remain so. However, the selection boundary at number $400\,000$ will be abandoned, and asteroids with larger IDs will be included in the match algorithm, but they will surely be detected with a lower completeness level.

   \begin{figure}[htbp]
   \centering
\includegraphics[width=0.95\hsize]{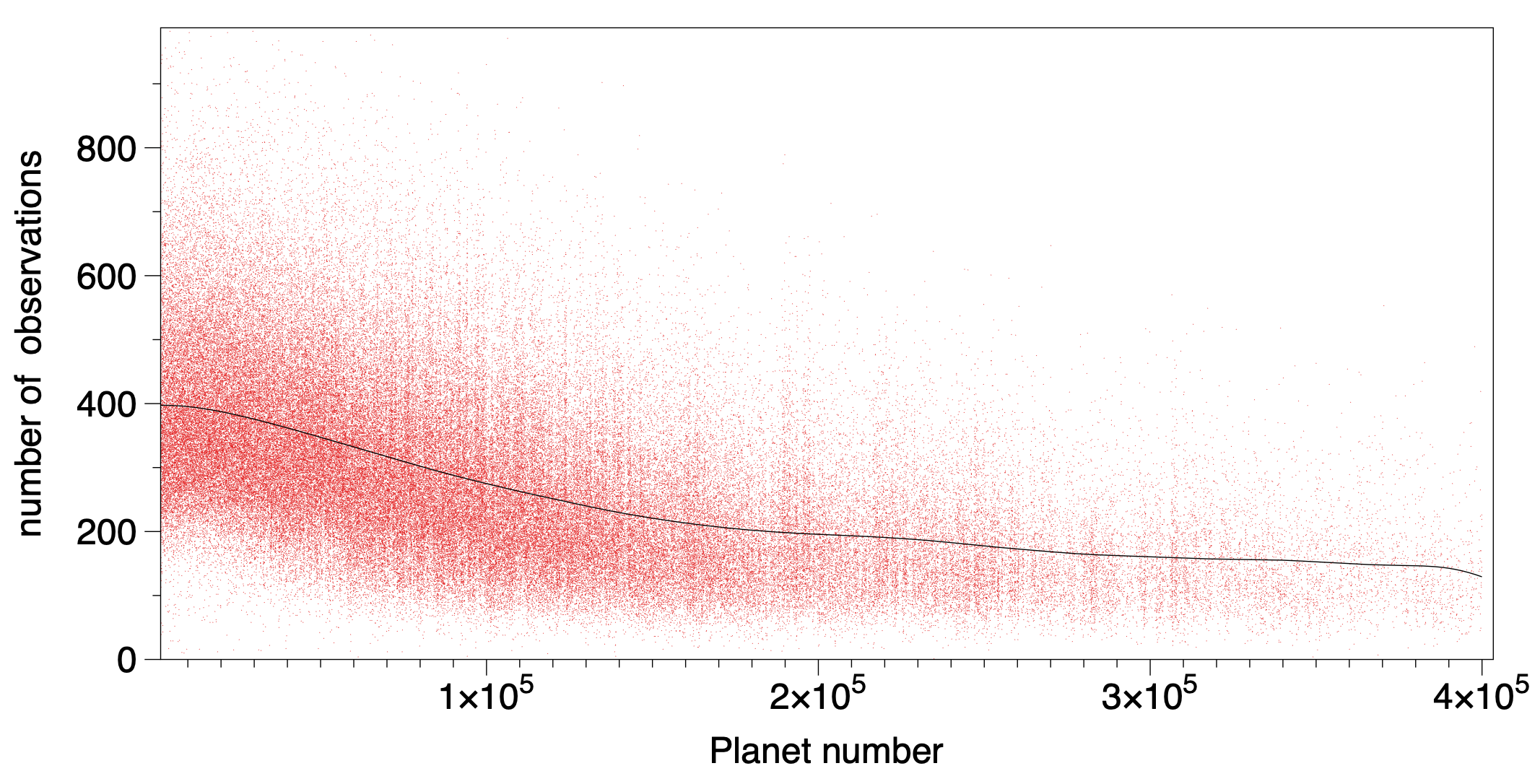} 
     \caption{Number of elementary astrometric observations as function of the number of the asteroid. The black line shows the average values. Asteroids with large numbers discovered since the year 2000  are generally smaller and fainter. A large fraction cannot be seen by the Gaia telescopes and detectors, or is seen only during some limited and more favourable orbital phases.} 
         \label{fig:num_obs_id}
   \end{figure}

   \subsection{Semi-major axis}\label{sect:semiaxe}
   Of the means contrived to quickly grasp the global quality of a new orbit solution of a large number of asteroids, the semi-major axis remains the most significant and favoured orbital parameter. It has the desirable feature over the angular parameters of being a true geometric quantity that is independent of the reference frame. This is particularly valuable when two solutions are compared for which small differences in inclination or in longitude of node may be unconnected with the orbits, but show up just as consequences of tiny rotations between reference frames. The eccentricity arguably shares this geometric nature, but the third Kepler law, linking the orbit size to the period, endows the semi-major axis with a much deeper meaning and has been taken for years as the first-rate parameter to compare solutions or to assess a solution. Above all, however, a difference in semi-major axis between two orbits has its counterpart in the mean motion and will soon show up as a secular drift in longitude, dominating any other difference from the other orbital elements and driving the accuracy of the ephemeris a few years away from the epoch. 
   
   Therefore, we follow this practice here by emphasising the semi-major axis over the angular quantities, which are less significant and are too sensitive to small differences between reference frames. The plots always take the dimensionless quantity $\sigma_a/a$ to express the relative precision, where $\sigma_a$ is the formal standard deviation of the semi-major axis computed by transforming the covariance matrix of the state vector to the matrix for the osculating elements. Similarly, differences between two solutions are shown with the relative distance $\Delta a/a$ or its unsigned value $\approx $ amplitude).

   \begin{figure}[htbp]
   \centering
\includegraphics[width=0.95\hsize]{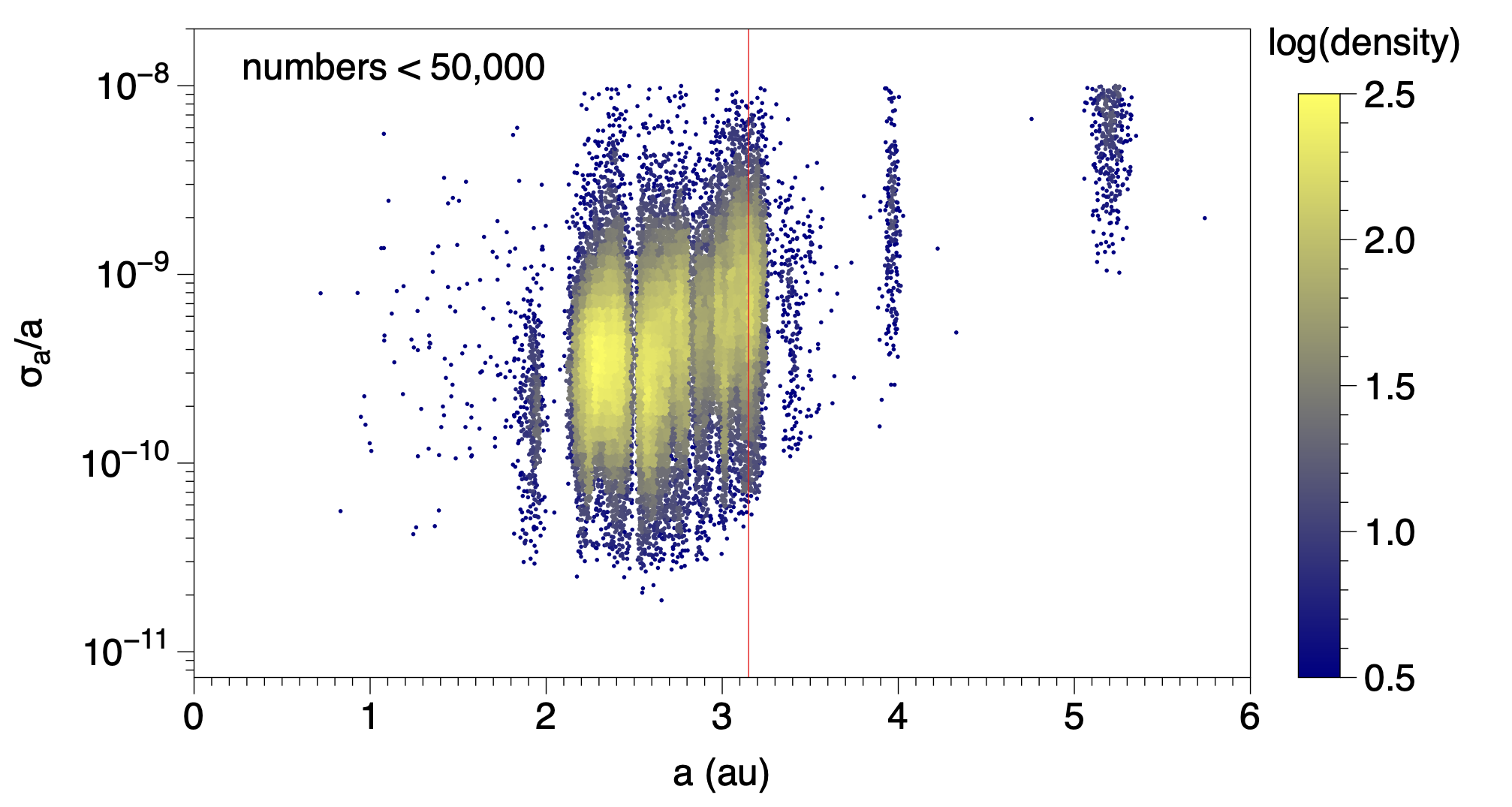} 
\includegraphics[width=0.95\hsize]{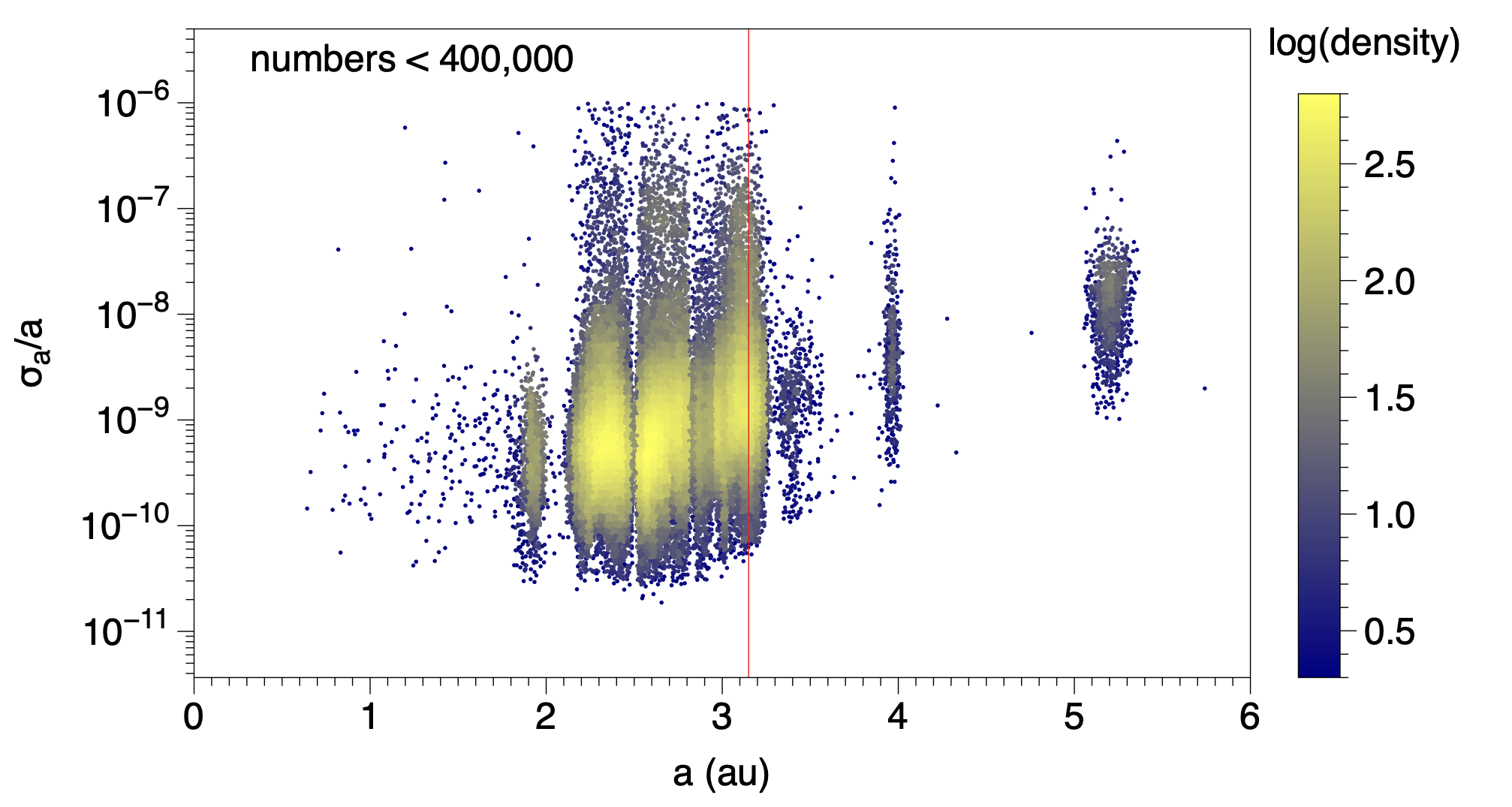} 
     \caption{Formal uncertainty of the orbits measured by $\sigma_a/a,$ where $a$ stands for the semi-major axis of the osculating orbit, and $\sigma_a$ shows its standard deviation. Top: First 50\,000 numbered asteroids. Bottom: Complete solution of 157\,000 orbits. The vertical line at $a = 3.15 \text{ au}$ corresponds to a period of 66 months, which is the longest arc in the data. The vertical scale is adapted to each set. } 
         \label{fig:siga_rel_a}
   \end{figure}

 The two plots in \autoref{fig:siga_rel_a} show the relative formal uncertainty as a function of the semi-major axis for the best solutions (top) and for the whole catalogue (bottom). The overall appearance is as expected based on the observational span of 66 months: the Trojan orbits cannot be retrieved with a quality comparable to that of the bright NEOs even after scaling by the semi-major axis. For the first group, the relative uncertainty ranges from $5\times 10^{-11}$ to $2\times 10^{-9}$ for the MBA, and the density peaks about $2\times 10^{-10}$.

\begin{figure}[htbp]
   \centering
\includegraphics[width=0.95\hsize]{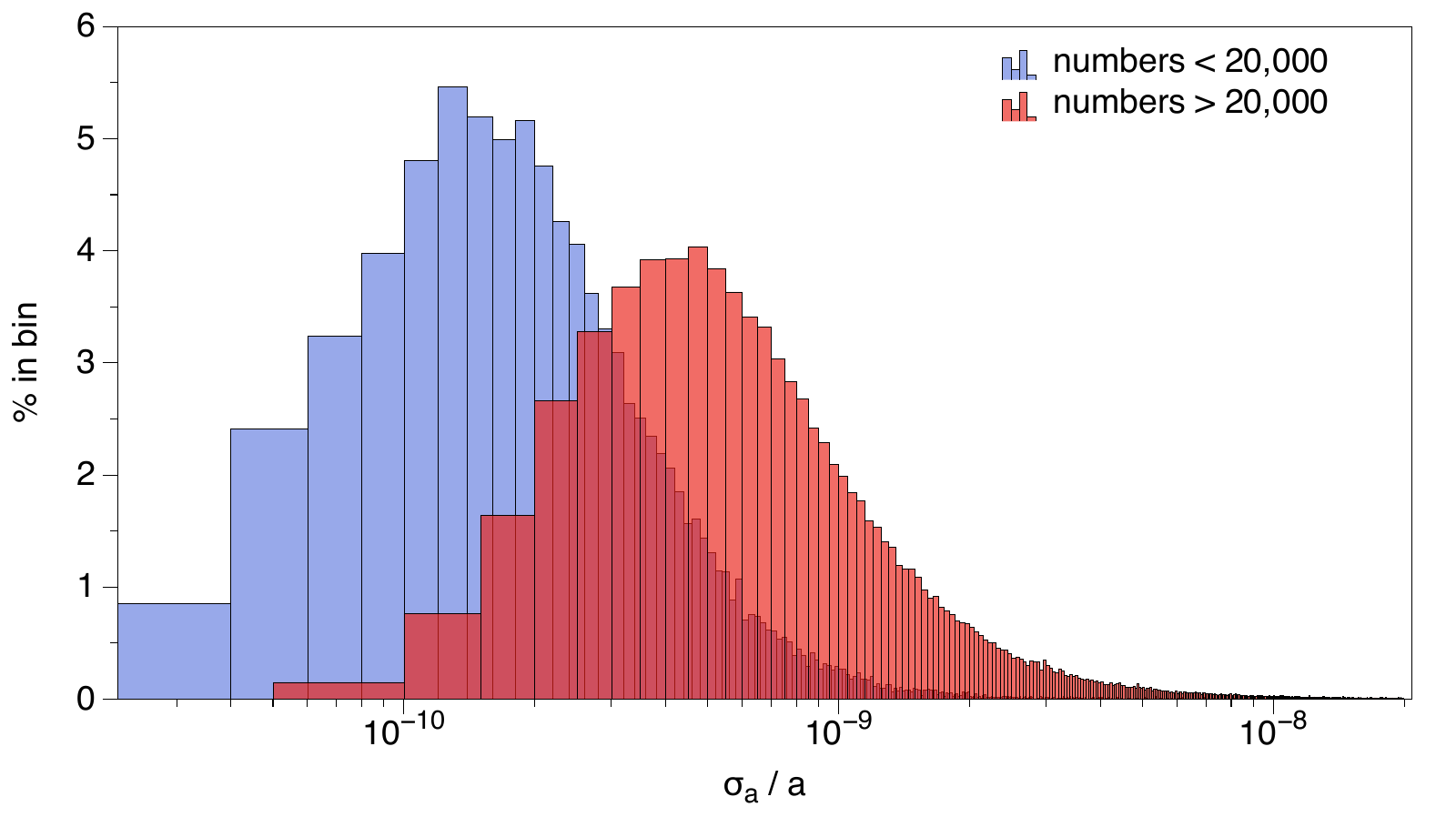} 
     \caption{Relative formal uncertainty of the Gaia orbits resulting from the fit to the observations over 66 months, divided between the bright end (numbers $ < 20\,000$) and the fainter asteroids. The bins add up to $100 \%$ for each group.}
         \label{fig: Gaia_orbits_siga}
\end{figure}

The actual distribution per bin of accuracy is shown in \autoref{fig: Gaia_orbits_siga} for the same  selections. The uncertainties are given in log-scale. The centre of the distribution at $\approx 1.5\times 10^{-10}$ for the bright population with  numbers $< 20\,000$ is better visible, as is the lower accuracy for the fainter group with a centre at $\approx 5\times 10^{-10}$ in $\sigma_a/a$. The formal uncertainty is directly derived from the fit, and the covariance matrix is scaled on the post-fit residuals.

\begin{figure}[htbp]
\centering
\includegraphics[width=0.90\hsize]{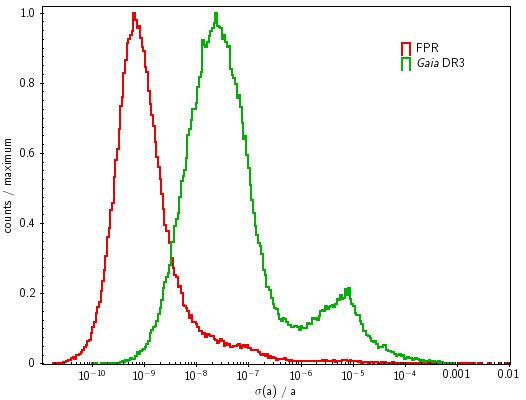}
\caption{Histogram of the relative uncertainty of the semi-major axis, normalised to maximum, for all asteroids in the middle main belt. \gdrthree\ is shown in green, and this work for the FPR is shown in red.}  
\label{fig:histoNormalisedDeltaA}
\end{figure}

\begin{figure}
   \centering
\includegraphics[width=0.95\hsize]{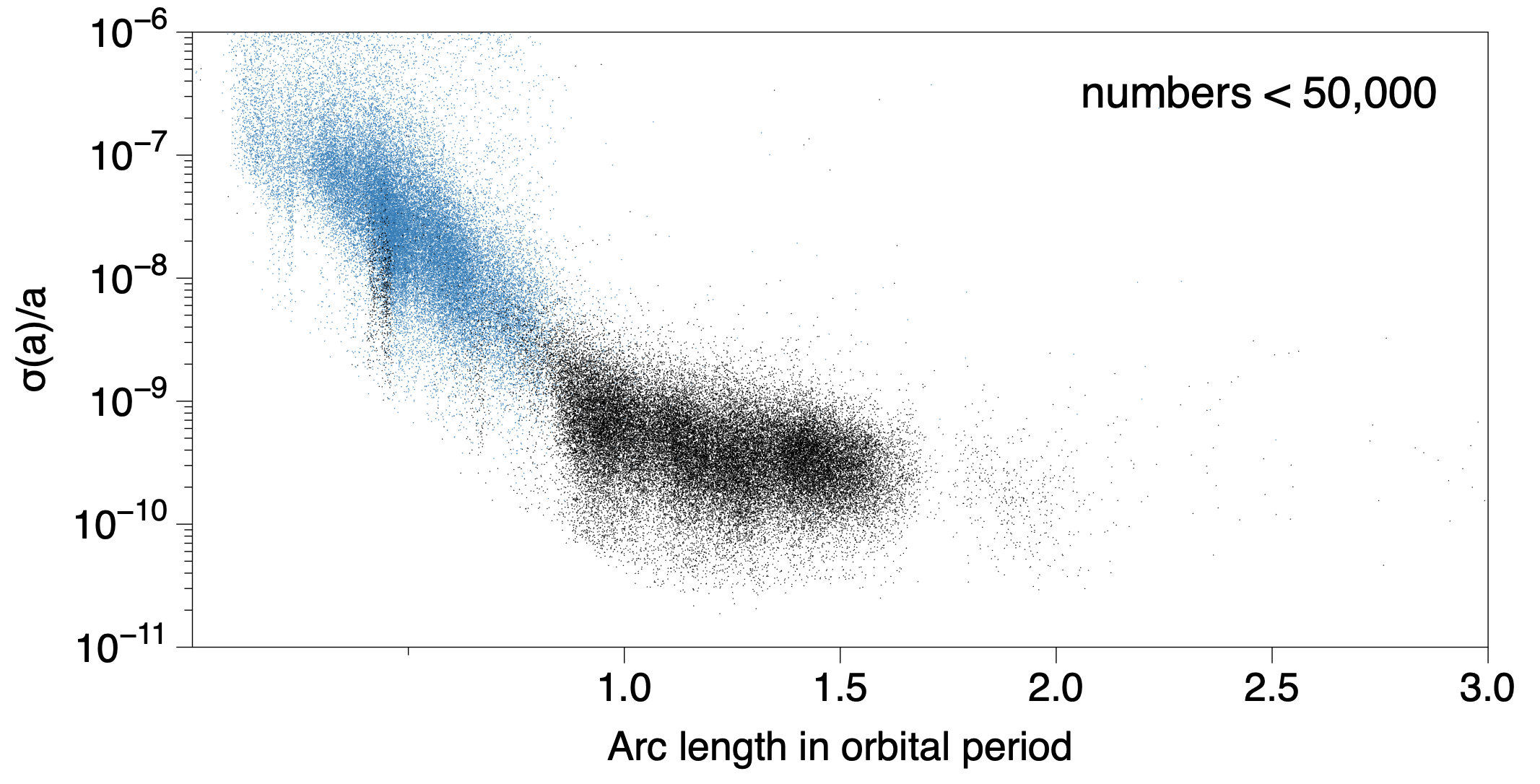} 
     \caption{Relative formal uncertainty of the Gaia orbits as a function of the arc-length coverage expressed in orbital periods. The black dots show this solution, and the light blue dots show the Gaia DR3 solution over 34 months.}
         \label{fig: Gaia_orbits_siga_arclength}
\end{figure}

Figure \ref{fig:histoNormalisedDeltaA} compares the relative uncertainty between the \gdrthree and this FPR solution. It
clearly shows the gain resulting from a longer and better-sampled arc for the orbit computation. The uncertainty in \gdrthree  was typically $\simeq 50$ times larger, and the distribution was bimodal, with a second population. The second smaller peak in the histogram stems from sources with the shortest arcs (around typically 100\,days). This second peak has now disappeared in the FPR as a direct consequence of a fuller coverage of the trajectories. More precisely, the first peak in the \gdrthree computations was at $\sim 2.6\times10^{-8}$ and the second at $\sim 6.9\times10^{-6}$ , while the distribution peaks at $\sim 8\times10^{-10}$ in this study. This means a gain between one to two orders of magnitude in the uncertainty, which is much larger than a $\simeq 1/\sqrt{2}$ improvement that would be obtained just by doubling the number of data points used in the adjustment. It essentially results from the longer observational arc, in agreement with \citet{desmars03-orbital-uncertainty}.
A final plot for this section shown in \autoref{fig: Gaia_orbits_siga_arclength} gives the formal uncertainty of the semi-major axis as a function of the arc length measured in orbital periods. The improvement as soon as the observation arc covers a full orbit is clearly visible in the steep slope on the left in the diagram, compared to the near stationary or much slower decrease when the arc is $\gtrapprox  1.2$ orbital period. In this regime, the improvement will come mainly from the combination of the larger number of observations and the extended arc length, while in the small-arc regime, the photon noise is not the main source of (in)accuracy. At mission completion, with $10.6$ years of data, even the Trojans will have nearly a full orbit coverage, and their orbit uncertainty will decrease to  the $1\times 10^{-9}$ relative uncertainty. For the purpose of comparison, the light blue dots in \autoref{fig: Gaia_orbits_siga_arclength} are plotted from the \gdrthree orbits, when only 34 months of data were used in the orbital solution, and almost no orbit had a complete orbit coverage. The improvement by a factor about 50 is outstanding in this combined plot. The main-belt orbits are not even as  good as the Trojan orbits in the 66-month solution.

 \subsection{Comparisons to other solutions} \label{sect:evaluation}
 As mentioned earlier, several other sources of orbital elements are built on  different observational material, different time spans, and different dynamical models. Although we  compared the solution from Gaia to the Astorb \citep{2022A&C....4100661M}, MPC, and JPL \citep{2021AJ....161..105P} solutions, only the comparisons to JPL orbits are discussed in detail in this paper, and MPC is mentioned only in passing.
\subsubsection{Comparison to JPL orbits}
\label{sect:JPL_compar}
The JPL service named Small-Body Data Base Query\footnote{See \url{https://ssd.jpl.nasa.gov/tools/sbdb_query.html}} allowed us to obtain orbital elements without a truncation in the output for a sample with a reasonable size. This was done for the first $50\,000$ numbered asteroids, which is enough to have a good sample covering a wide range of semi-major axes. The elements are provided at the epoch JD 2460000.5 TDB (25 February 2023) with all the digits, including a safety margin beyond the true precision. The epoch is offset by about 5 years from the Gaia mean epoch in our solution, however. A propagation was therefore needed of one of the files, or for both files midway. We chose to propagate the JPL reference data to the Gaia epoch, that is, backward 5 years. This propagation itself is the source of an additional difference between the orbits becausee it is hard to certify that an accuracy better than $\approx 1\times 10^{-10}$ is maintained over 5 years of numerical integration, and above all, that the \gaia and JPL dynamical models are close enough to reach this performance. This propagation for Gaia was made outside the main pipeline with an independent code. The dynamical model includes the complete form of the EIH equations and gravitational perturbations from 14 asteroids. 

To ascertain that no significant deviation arose during our numerical integration, we tested this propagation on a much smaller set of asteroids with JPL orbits requested at the Gaia epoch. We were able to reproduce JPL integration at a level of $8\times 10^{-10}$ in the differences $\Delta a/a$ after 5 years. This is acceptable for the comparison, although we know that many of the Gaia orbits are  better than this limit (at least in formal uncertainty). It is likely that these residual deviations mainly hold for the small differences between the two dynamical models (not for the same perturbing asteroids, the same masses, or the same Solar System ephemeris, etc.) and not from the numerical integration proper.

    \begin{figure}[htbp]
   \centering
\includegraphics[width=0.95\hsize]{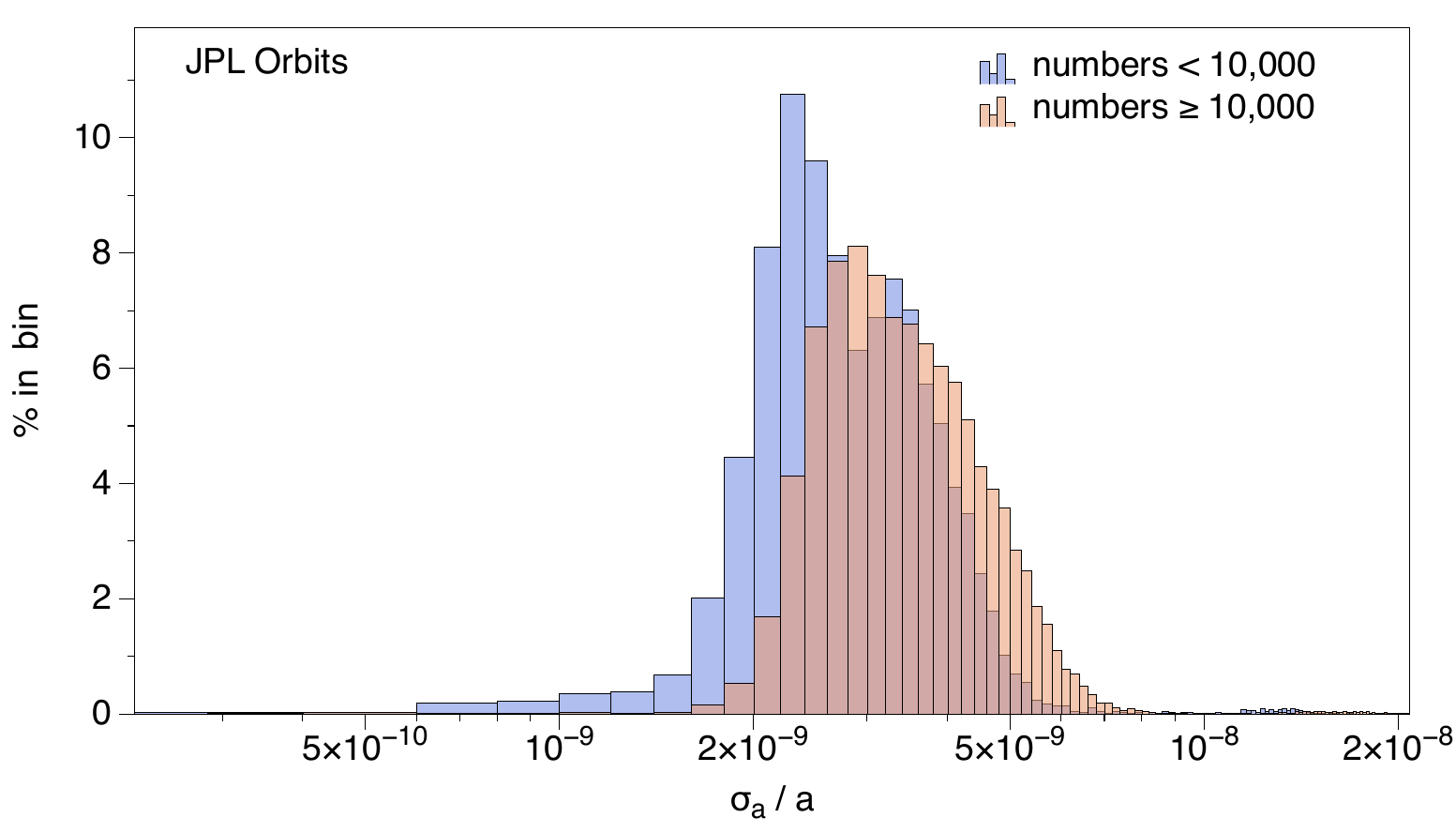} 
     \caption{Relative uncertainty of the reference orbits  computed by the JPL before propagation to the mean Gaia epoch. The bins add up to $100 \%$ for each group.} 
         \label{fig:jpl_rel_sig_a}
   \end{figure}

We now describe how \gaia used TCB for its time tagging, as explained in \secrefalt{sect:unit_timelength}. For a meaningful comparison with JPL or MPC, the same units must be used. Starting with the Gaia TCB-compatible state vector, a standard TCB-to-TDB scaling is performed as
\begin{equation}\label{eq:scaling_SV}
\begin{aligned}
    a_\text{TDB} &= (1-LB)\, a_\text{TCB}\\
    \mathbf{X}_\text{TDB} &= (1-LB)\, \mathbf{X}_\text{TCB}\\
    \mathbf{V}_\text{TDB} &=  \mathbf{V}_\text{TCB}
\end{aligned}
,\end{equation}
as discussed in \cite{klionerUnitsRelativisticTime2010} or \cite{2008A&A...478..951K}. Here, $L_B = 1.550519768  \times 10^{-8}$ is a constant defined in the system of astronomical units and is equal to the rate $\left(dTCB/dTDB-1\right)$.
The scaling  must not be applied again to the semi-major axis that was computed from the rescaled state vector. The effect is already included in the spatial components of the state vector.

\begin{figure}[htbp]
   \centering
\includegraphics[width=0.95\hsize]{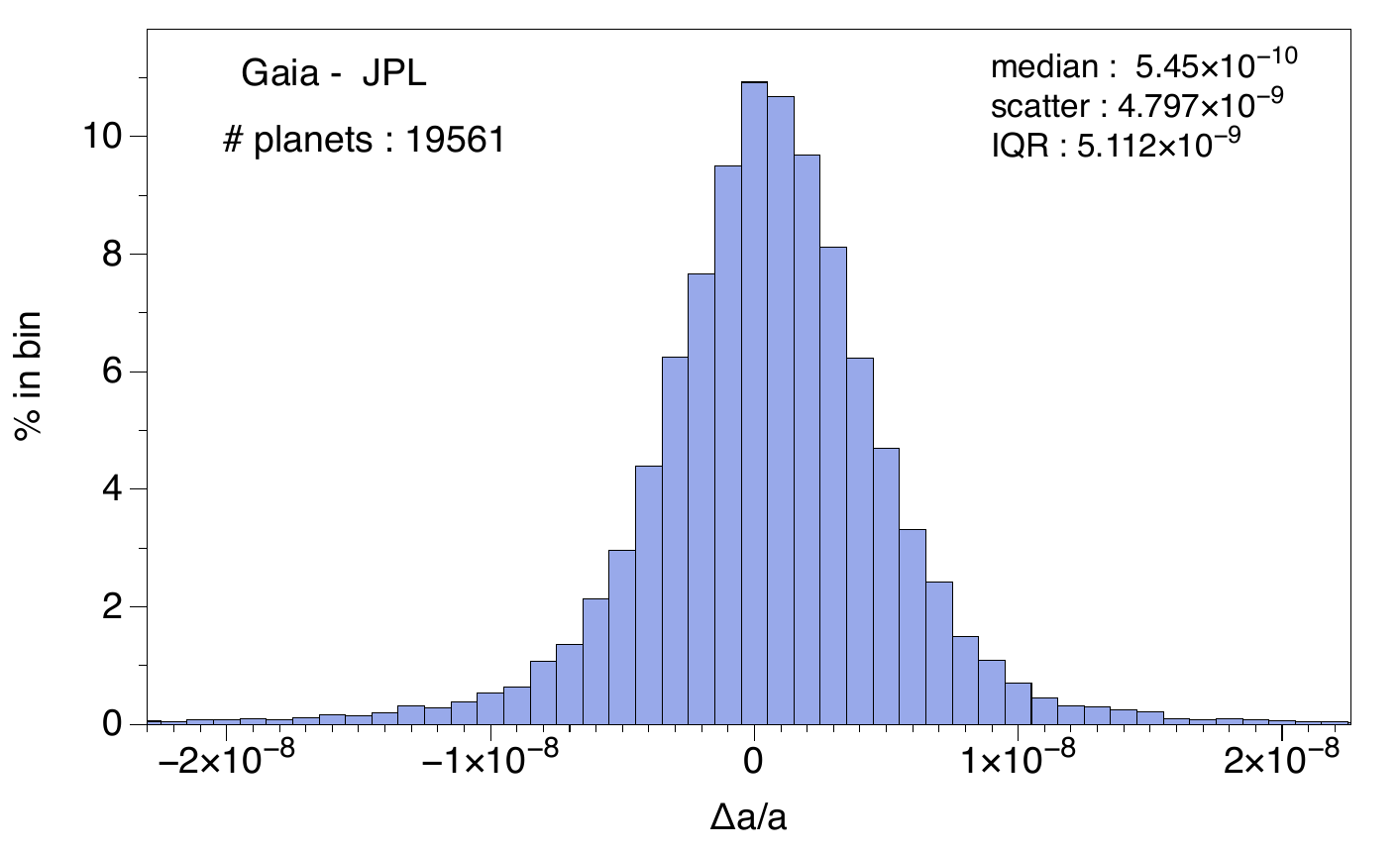} 
     \caption{Differences in semi-major axis (as $\Delta a/a$) between the \gaia orbital solution and the JPL orbits transformed to the \gaia epoch for the first $20\,000$ asteroids. There are $19\, 822$ \gaia solutions, $261$ of which are outside the plot boundaries. }
         \label{fig: Gaia_jpl_rel_del_a}     
   \end{figure}

\begin{figure}[htbp]
   \centering
\includegraphics[width=0.95\hsize]{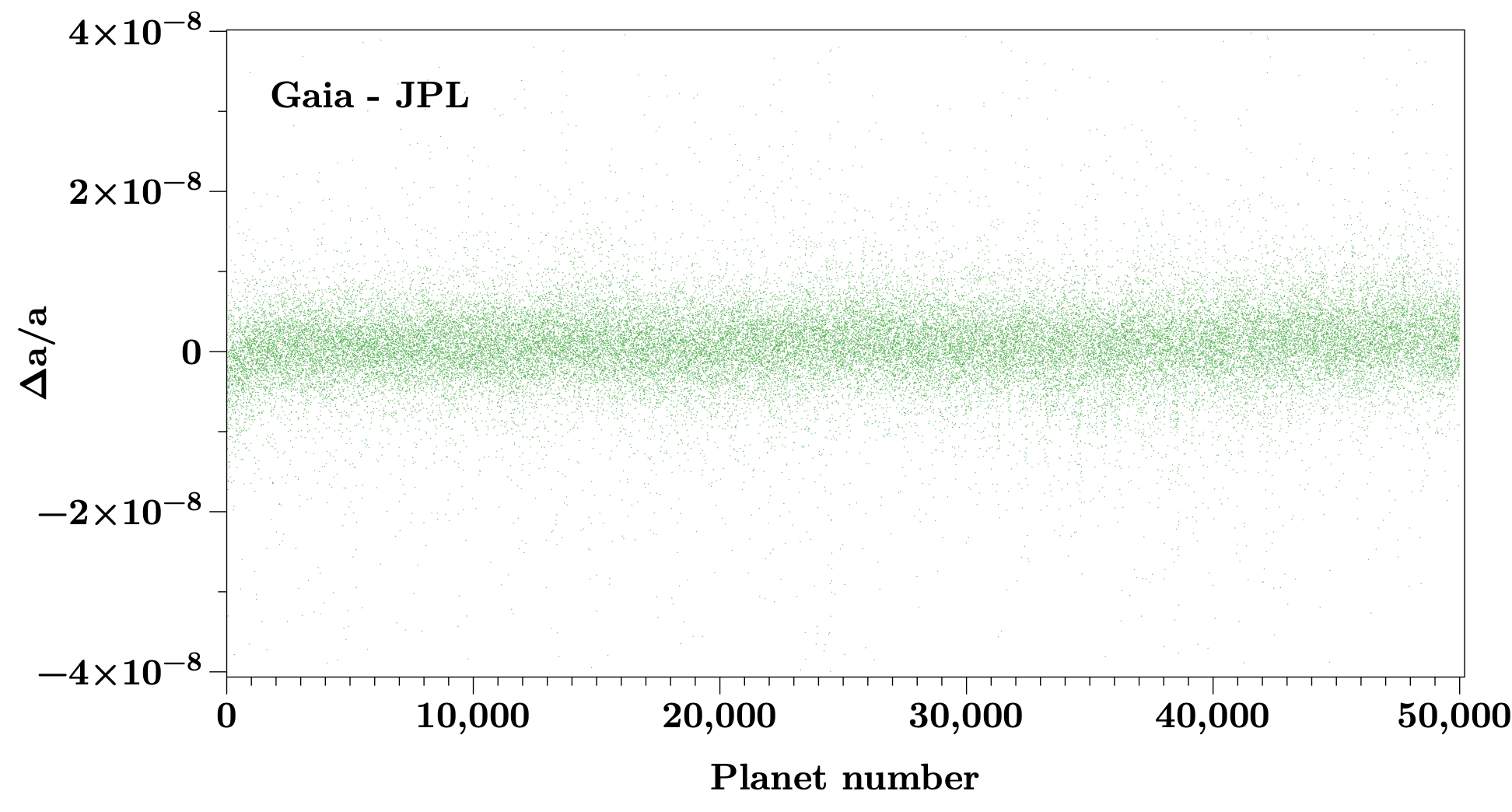} 
\caption{Differences in semi-major axis (as $\Delta a/a$) between the \gaia orbital solution and the JPL orbits for the first $50\,000$ asteroids. The bias increases slowly from $1.3\times 10^{-10}$  to  $9.5\times 10^{-10}$ from the left to the right.}
         \label{fig: Gaia_jpl_rel_del_a_scat}
   \end{figure}
   
   \begin{figure}[htbp]
   \centering   
\includegraphics[width=0.95\hsize]{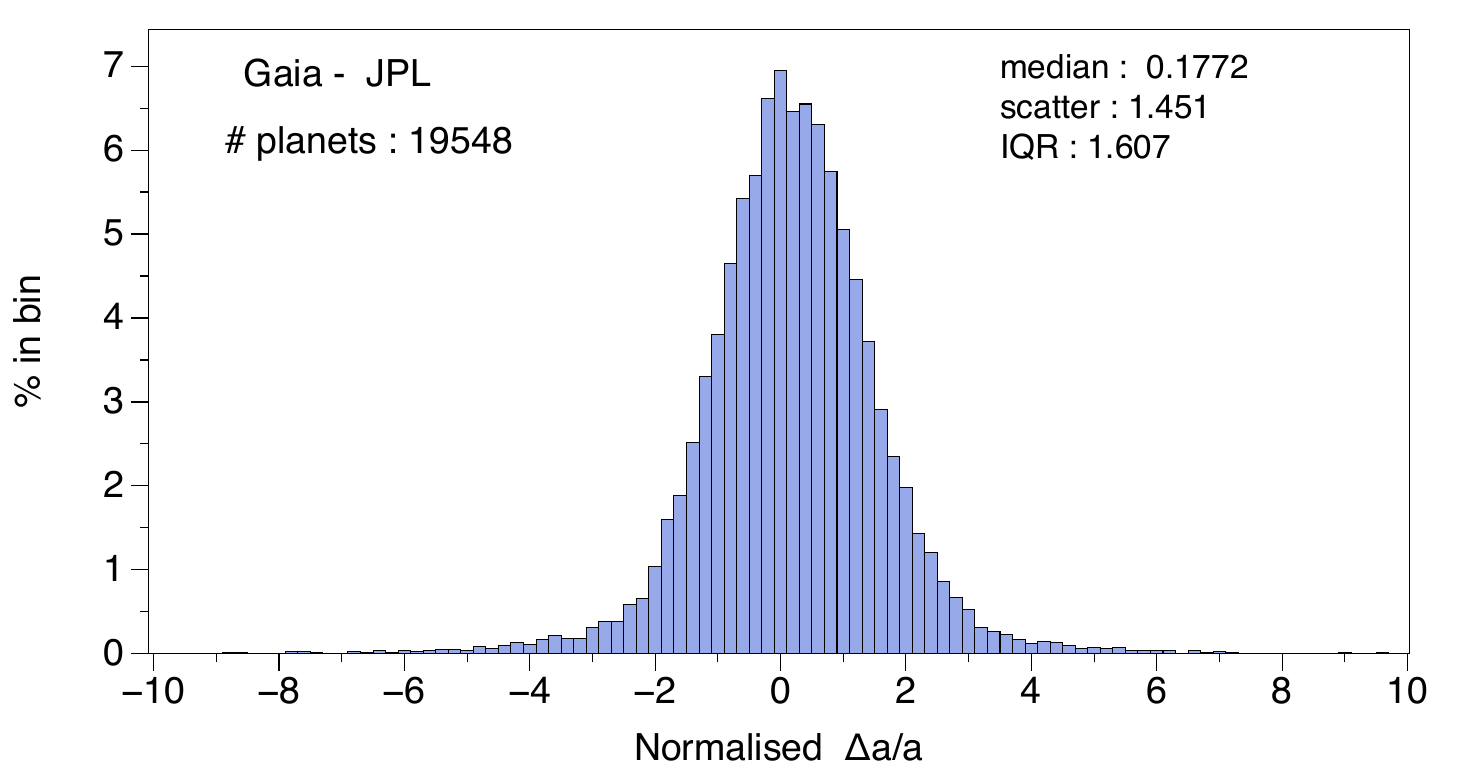} 
     \caption{Normalised differences in semi-major axis  between the Gaia orbital solution and the JPL orbits transformed to the Gaia epoch for the first $20\,000$ asteroids. Compared to \autoref{fig: Gaia_jpl_rel_del_a} , 13 bodies lie outside the$[-10,10]$ interval.}
         \label{fig:Gaia_jpl_norm_del_a}
\end{figure}  

 \begin{figure}[htbp]
   \centering
\includegraphics[width=0.95\hsize]{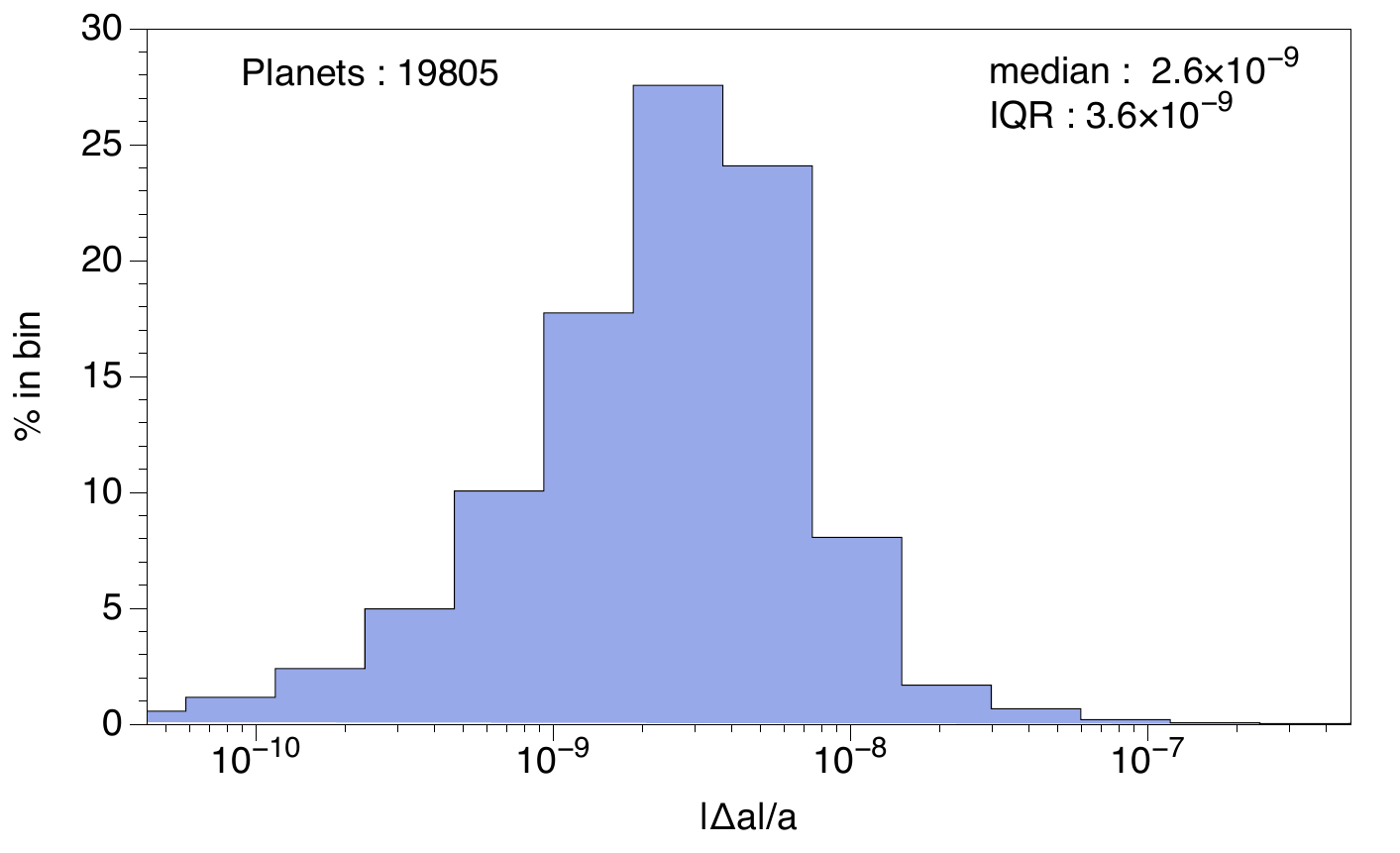} 
     \caption{Absolute values of the differences in semi-major axis  between the Gaia orbital solution and the JPL orbits transformed to the Gaia epoch for the first $20\,000$ asteroids. \autoref{fig:num_sol_id} shows $19,822$ Gaia solutions with $ID \leq 20\,000$. The difference of $19,805$ comes from the largest outliers of \autoref{fig:Gaia_jpl_outliers}.}
         \label{fig:Gaia_jpl_rel_del_a_star}
\end{figure}   
When the two sets of osculating elements are brought to a common epoch, the differences in semi-major axes as shown in \autoref{fig: Gaia_jpl_rel_del_a} are obtained in the form of the dimensionless quantity $\Delta a/a$, with a few summary statistics. They are shown in \autoref{fig: Gaia_jpl_rel_del_a_scat} with a scatter plot for the first 50,000 numbered planets. The agreement is outstanding, even though a slight bias of $5.5\times 10^{-10}$ remains between the two solutions for the first 20,000 numbered planets. Expressed in angle, when $\Delta a$ is taken as a positional uncertainty, which is close to the osculating epoch before the drift in mean anomaly takes over,   this is
equivalent to 0.1~mas. This is a very small deviation given the total independence in the construction of the two solutions. The differences in the dynamical model, in the masses, and above all, in the set of observations (none in common) drive the fit over two very distinct time spans. For a first full Gaia solution, this exceeds the most optimistic expectations. The bias is even smaller at $1.3\times 10^{-10}$ for the first $5000$ planets and grows to $8.5\times 10^{-10}$ when the full set of $50,000$ is considered. This slight increase in asteroid number is hardly noticeable in the scatter in \autoref{fig: Gaia_jpl_rel_del_a_scat}. The scatter about the mean in this plot is otherwise remarkably uniform for the range of selected planets.
Its value at $\sim 5\times 10^{-9}$ agrees with the formal uncertainty of either catalogue, as shown in Figs.~\ref{fig: Gaia_orbits_siga} -\ref{fig:jpl_rel_sig_a} . It shows that the true accuracy of the solutions is not much different from the formal accuracy. We are probably very close to the limit of what can be learnt from this comparison, which is as interesting for JPL as it is for \gaia.

The normalised differences computed as $ \Delta a/\sqrt{\sigma(a)_\text{Gaia}^2 +\sigma(a)_\text{JPL}^2 } $ are shown in \autoref{fig:Gaia_jpl_norm_del_a} for the same set of asteroids. No change was applied to JPL-reported uncertainties to account for the propagation from the JPL to the Gaia epoch. The bias, albeit normalised, has the same origin as in \autoref{fig: Gaia_jpl_rel_del_a} and is statistically significant. The scatter at $ \sim 1.45$ is not a robust estimate and takes too much weight from the tails. The inter-quartile range (IQR) for a normal distribution  (IQR $= 1.35 \sigma$) gives a more robust estimate of the central width at $1.607/1.35 \approx 1.2$  and  decreases to $1.13$ with the first $50,000$ numbered planets and even $1.08$ with the numbered planets in $[25,000, 50,000]$. This is a very satisfactory result and indicates that the JPL uncertainties are realistic and fully compatible with the scatter of $\Delta a$ between the two solutions. Because the formal errors for Gaia are smaller than those in JPL, the normalised plot is much less sensitive to an inflation factor that should be applied to \gaia uncertainties, unless this factor is as large as 10, in which case, the random uncertainties would be fully incompatible with the post-fit residuals.

\subsubsection{Comparison to MPC orbits} 
\label{sect:MPC_compar}
The Minor Planet Center at Harvard (MPC) has recently provided access to its orbits with full accuracy, including the uncertainty on each element.  A similar comparison was made for the MPC solution for $50\,000$ orbits, and it confirmed the absence of significant bias ($< 3\times 10^{-9}$ in this case) and a scatter just twice as large as with JPL solution ($\approx 1.1\times 10^{-8}$) for $\Delta a/a$. These differences between the comparison to JPL and MPC might be traced to the fact that JPL used more observations than MPC in their fit, and in particular, they used some of the \gdrtwo\ observations. This good complement demonstrates the perfect consistency between three independent fits and establishes that the \gaia solution is currently better than MPC and at least at the level of JPL. Based strictly on formal uncertainties, the \gaia solution appears even better, but we remain cautious about this claim because systematics could be larger and our arc length remains limited. The real test would be to compare the \gaia orbits to accurate observations (e.g. occultations) well outside the time range of the fit. This is a work in progress. The first results presented in \secrefalt{sect:compar_occult} show that this is a work in itself , and much more work remains to be done  for this goal to be reached.

\subsection{Outliers from this comparison}
\begin{figure}[htbp]
   \centering
\includegraphics[width=0.95\hsize]{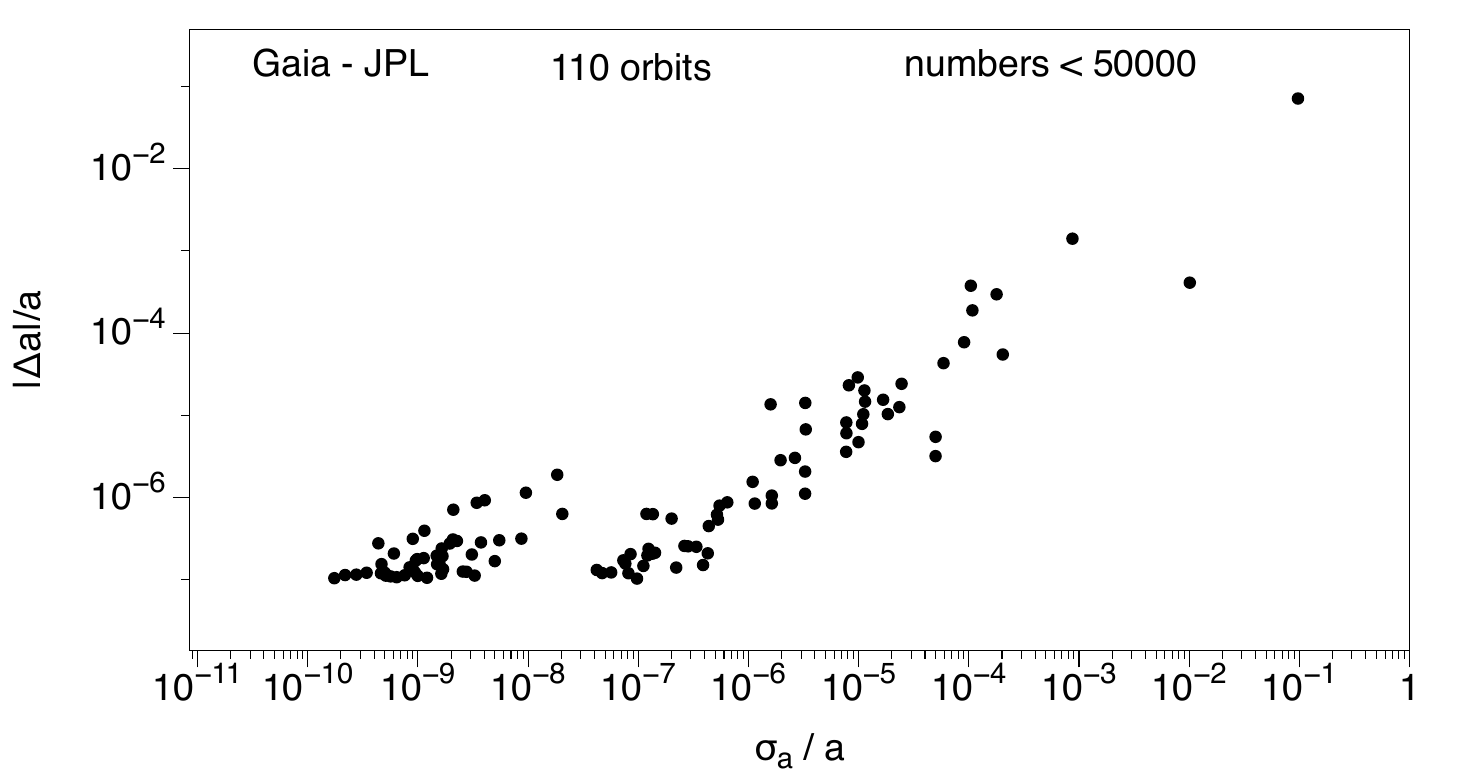} 
     \caption{Anomalous differences in semi-major axis between the Gaia orbital solution  and the JPL orbits. The plot shows the relative difference as a function of the Gaia formal uncertainty $\sigma_a/a$ for all solutions with $|\Delta a/a |> 1\times 10^{-7}$. The left group ($\sigma_a/a < 5\times 10^{-8}$) is just the small statistical right tail of the distribution of \autoref{fig:Gaia_jpl_rel_del_a_star} and is determined by the selection threshold for the plot. The  group on the right with $\sigma_a/a > 1\times 10^{-7}$ comprises solutions with a poor fit performance and is not an issue: these orbits could have been filtered out in the pipeline and were excluded from the delivery.} 
         \label{fig:Gaia_jpl_outliers}
   \end{figure}

The histogram in \autoref{fig:Gaia_jpl_rel_del_a_star} has a vertical linear scale that prevents us from seeing the extended tails when the relative differences between \gaia and JPL orbits are greater than $1\times 10^{-7}$ or $1\times 10^{-6}$. The  orbital solution of Gaia has very few filters during the iterative process or in the final output. If the solution has converged, the orbit is output, regardless of the formal uncertainty. Given the possibility of very small effective arcs for some asteroids, which are also associated with the small number of detections, incorrect orbits, if rare, may be obtained in the \gaia solution. These are more likely than in JPL, which was built upon  a much longer time range for the first $50\,000$ numbered asteroids considered in this comparison. Setting the level of anomalous differences at $1\times 10^{-7}$ for the difference  $|\Delta a |/a$, deviations that exceed this threshold occur in only 110 orbits. This is indeed an  astonishingly small number. It is close to $0.2\%$ of the set and it is well accounted for in the residual analysis, as we show below. Many of these orbits may have been filtered out from the \gaia solution based on their anomalous formal uncertainty that is shwon in \autoref{fig:Gaia_jpl_outliers}.

This plot shows two well separated groups. On the left, about 50 orbits with Gaia solutions that are acceptable and have a statistically significant difference with JPL. This can be viewed as the normal extended tail of the distribution, and no conclusion can be drawn without considering all solutions individually. The right part is more interesting and shows a strict relation between the departure between the two orbits (\gaia and JPL) and the increasingly lower quality of the \gaia solutions.  These $\approx 60$ orbits could be removed for the published set without damage, but it is very satisfactory to see this clear feature for such a small number of orbits. All the others except for one agree better with the JPL solution than $1.7\times 10^{-6}$ in relative difference in semi-major axis, and they agree usually  much better than this limit, as shown in \autoref{fig:Gaia_jpl_rel_del_a_star}. Essentially, there are no unexplained anomalous orbits in the set we used for the comparison. \gaia and JPL can both be proud of this.
   
\subsection{Other orbital elements}\label{sect:other_elements}

\begin{figure}
   \centering
\includegraphics[width=0.95\hsize]{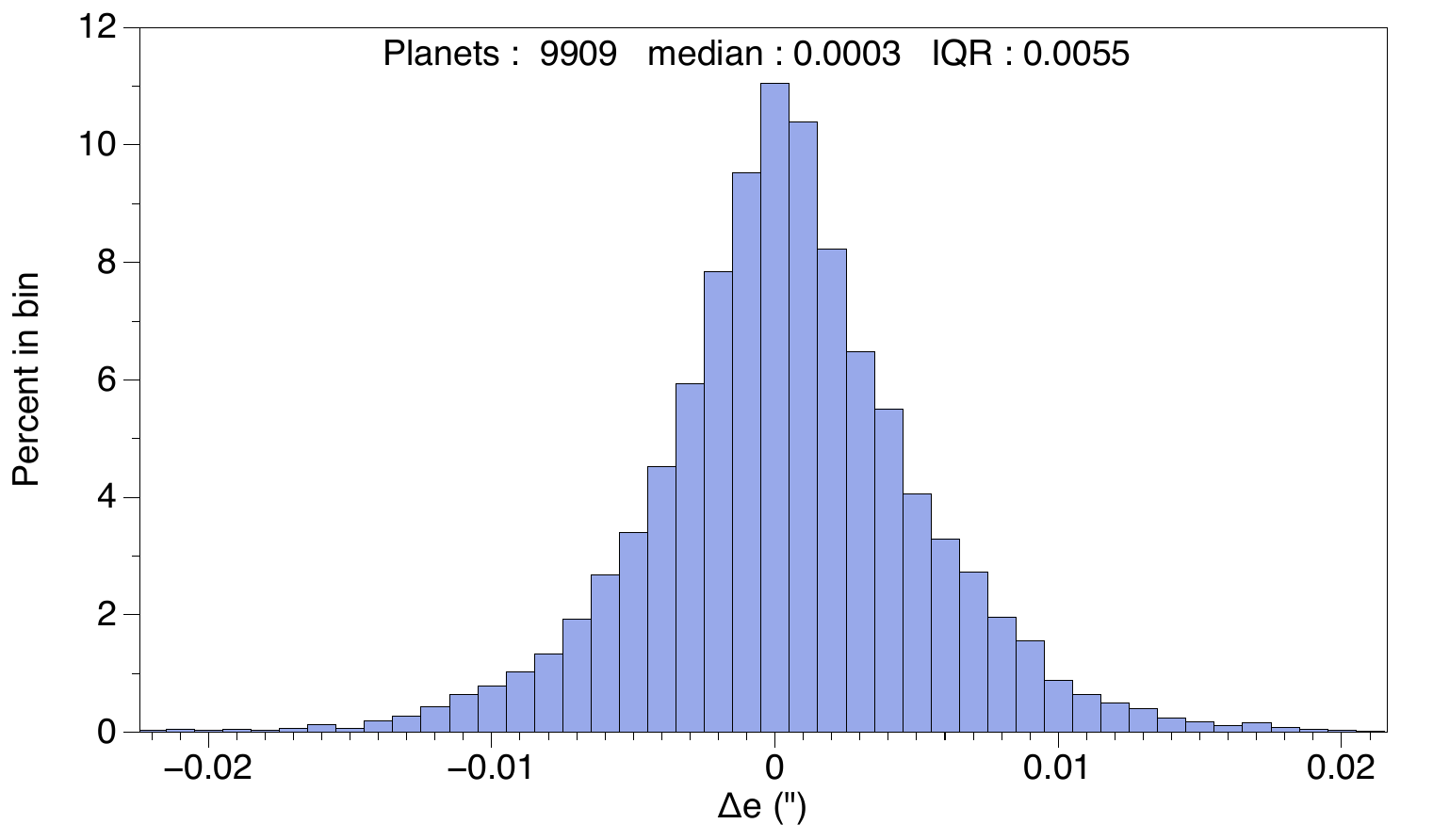} 
\includegraphics[width=0.95\hsize]{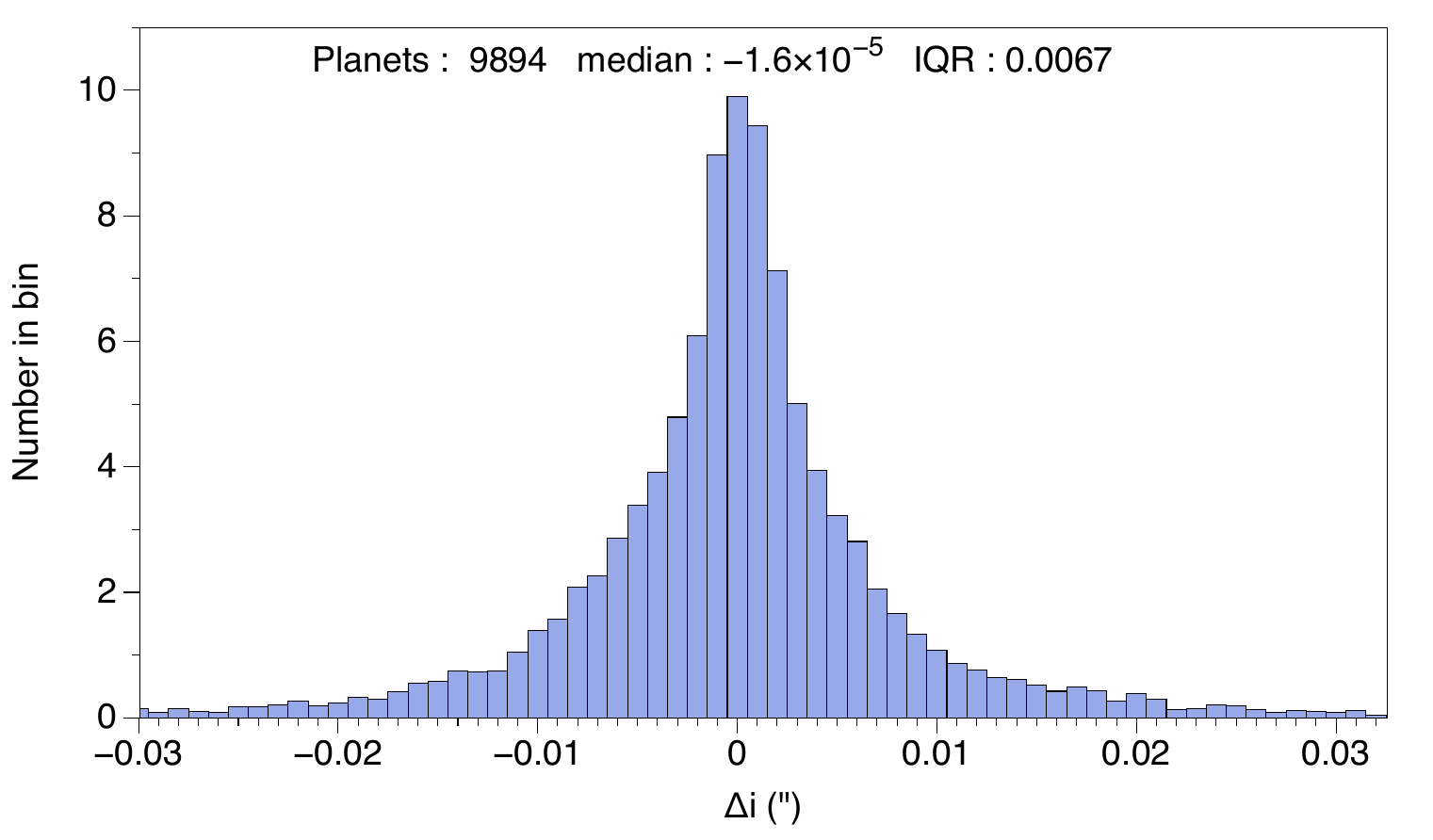} 
     \caption{Differences in eccentricity (expressed in arcsec)  and inclination  between the Gaia and JPL solutions for the first 10\,000 numbered asteroids. }
         \label{fig:Gaia_jpl_ecc_inc}
\end{figure}  
The comparison of the other orbital elements (eccentricity and the angular elements) is briefly mentioned now with additional plots. The main point is to draw attention to the details of the reference ecliptic used by either group. Any difference in the definition of the ecliptic (both the plane and the origin of longitude) would result in spurious systematic deviations that are unrelated to a real difference between the orbits.  As said earlier, osculating elements of Gaia have been compared to the same ecliptic plane as JPL, with origin of longitude at the node with the ICRF equator as defined in \autoref{fig:ecliptics}.

We show in \autoref{fig:Gaia_jpl_ecc_inc} histograms of the difference in orbital eccentricity and inclination found between the Gaia and JPL solutions for the first $10\,000$ numbered asteroids. Angular units are used, with dimensionless $\Delta e$ taken as radians and  converted into arcseconds (as if $a \Delta e$ were a position shift and  $(a \Delta e)/a = \Delta e$ were an angular shift). The two solutions are not biased, and the scatter about the mean is typically 5 to 10 mas or $< 5\times 10^{-8}$ , with a robust Gaussian scatter of 40 and 43 mas, respectively). This is larger than the equivalent angular deviation seen with the semi-major axis, but it probably gives a good upper bound for the true quality of the orbits. The prominent tails in the inclination are not Gaussian, however.

\begin{figure}
   \centering
\includegraphics[width=0.95\hsize]{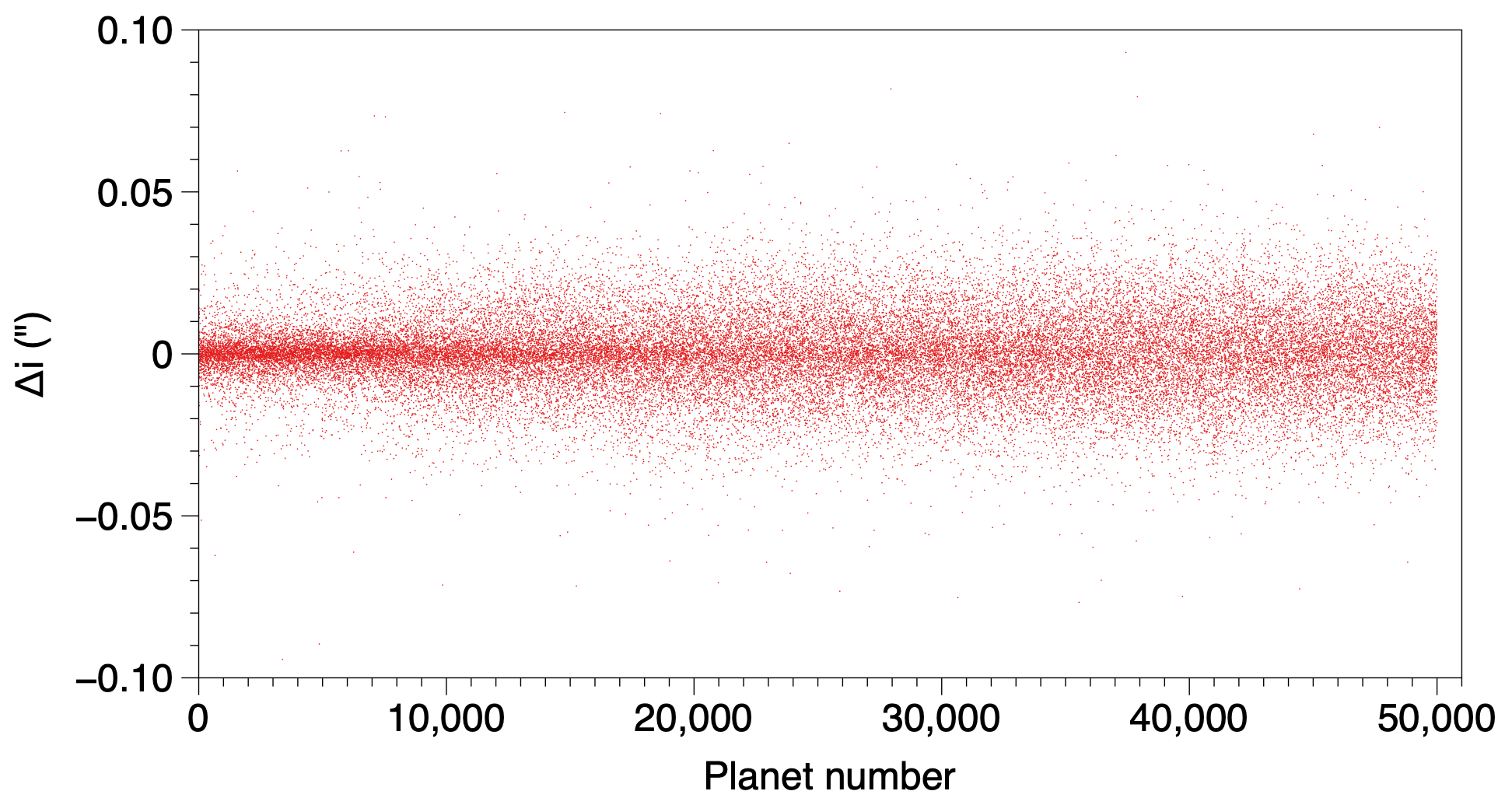} 
     \caption{Differences in inclination  between the Gaia and JPL solutions. The core is bounded with $\pm 10$\,mas, but there are extended non-Gaussian tails. }
         \label{fig:Gaia_jpl_angular_inc}
\end{figure}

\begin{figure}
   \centering
\includegraphics[width=0.95\hsize]{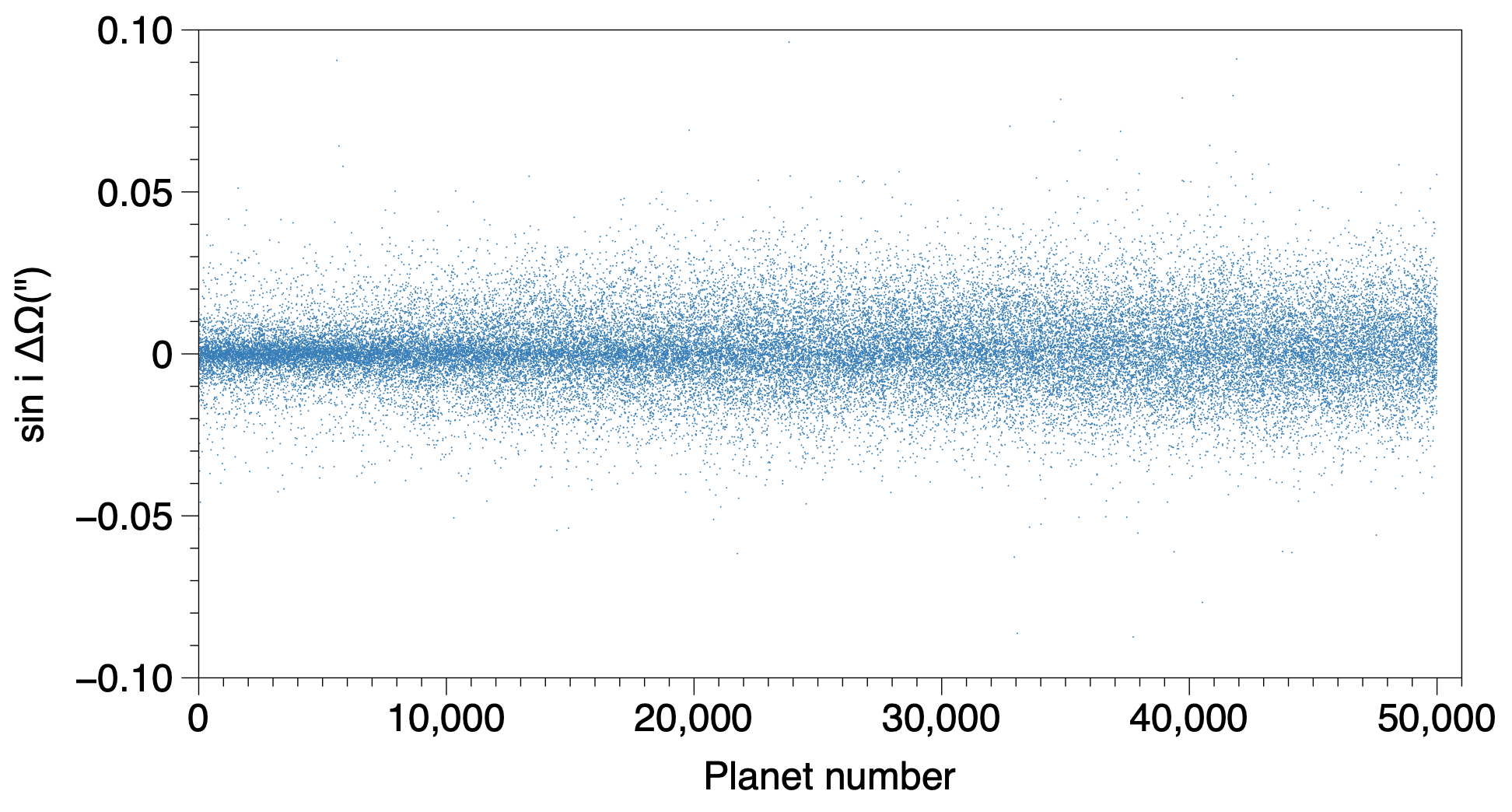} 
     \caption{Differences in longitude of node (as $\Delta \Omega \sin i$ in arcsec) between the Gaia and JPL solutions for the first 50,000 numbered minor planets.}
         \label{fig:Gaia_jpl_angular_node}
\end{figure}

\begin{figure}
   \centering
\includegraphics[width=0.95\hsize]{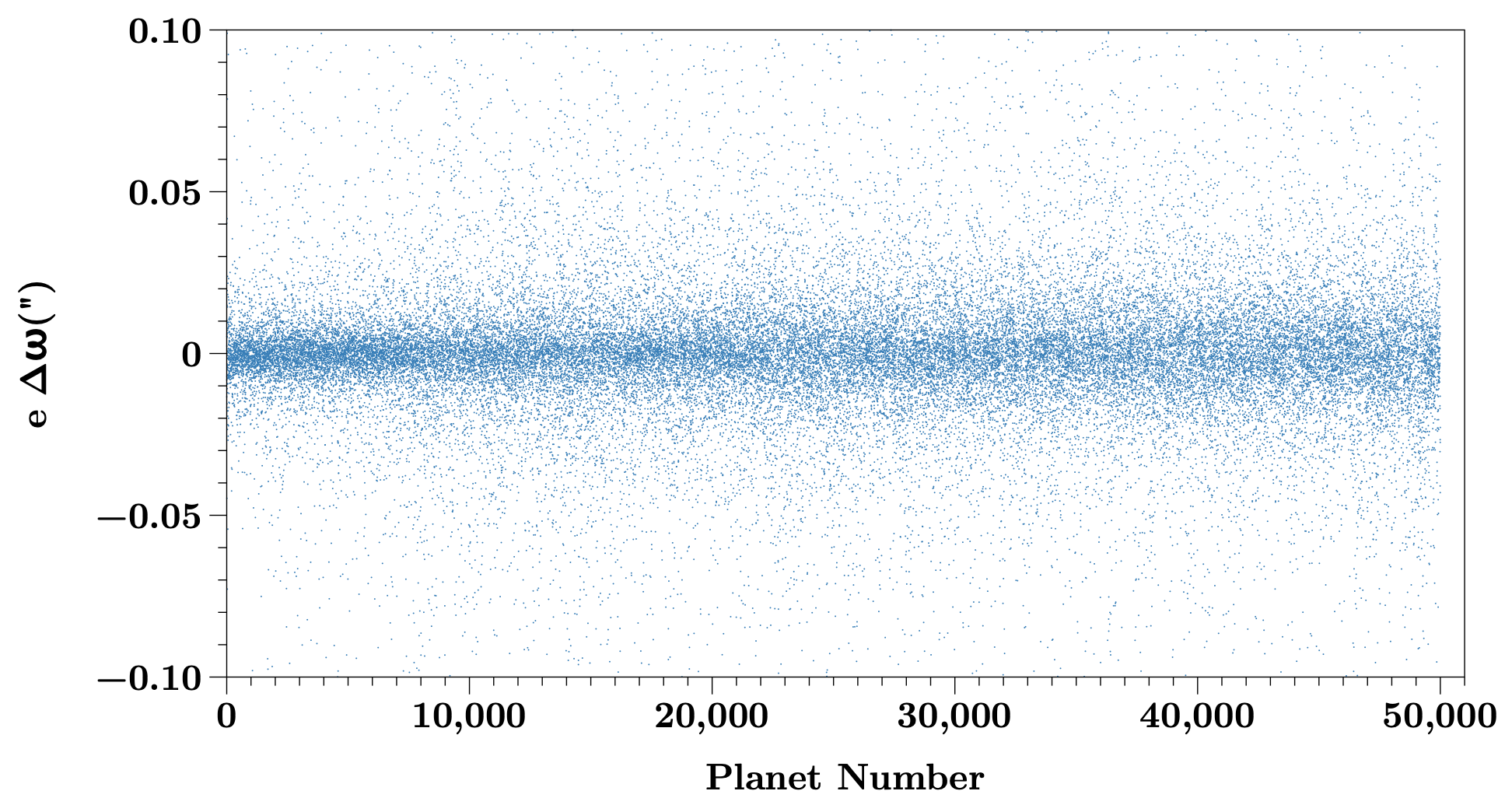} 
     \caption{Differences in argument of perihelion (as  $e \Delta \omega$ in arcsec) between the Gaia and JPL solutions for the first 50,000 numbered minor planets.}
         \label{fig:Gaia_jpl_perihelion}
\end{figure}

The following comparisons  in \autoref{fig:Gaia_jpl_angular_inc}, \autoref{fig:Gaia_jpl_angular_node}, and
\autoref{fig:Gaia_jpl_perihelion}  show scatter plots of the differences between \gaia $\text{and}$ JPL for the three angular elements inclination, longitude of node, and argument of perihelion. The last two are scaled with the inclination and the eccentricity to show the equivalent of a displacement on the sky rather than the difference in the less meaningful coordinates. The three plots are very similar. They have a core within 5 mas that is surrounded by an extended tail that increases with the rank of the asteroid.  An additional analysis shows that the variation in inclination has a systematic sine wave of $\sim 10$\,mas amplitude with the longitude of node. This is a clear sign of a reference-frame effect. However, this effect is hardly visible for asteroid numbers $< 10,000$ and becomes more visible with larger numbers.

\subsection{Comparison to stellar occultations}
\label{sect:compar_occult}

A final severe test of the quality of orbits published in the FPR is provided by stellar occultations. In principle, positions computed by using FPR orbits at the epoch of an observed occultation of a star present in the Gaia catalogue must match the astrometry derived from the event. 

\begin{figure}
   \centering
\includegraphics[width=0.95\hsize]{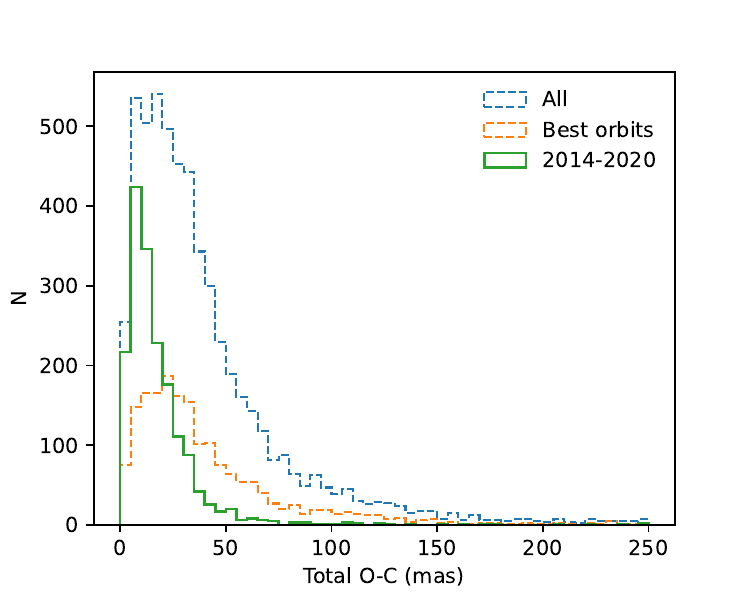} 
\includegraphics[width=0.95\hsize]{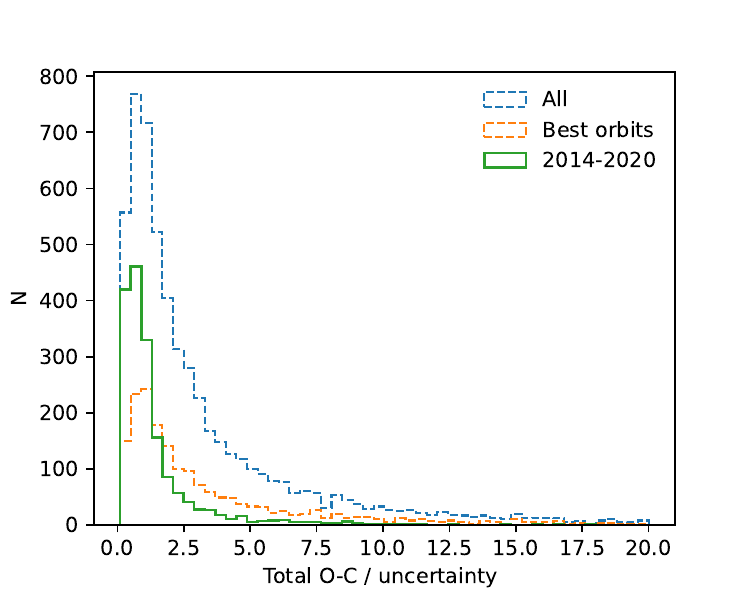} 
     \caption{Histogram of the absolute total O$-$C (distance on the sky) between the predicted and observed positions derived from stellar occultations. The bottom panel presents the O$-$C values normalised by the uncertainty of the occultation astrometry for each event. The curve labelled "2014-2020" includes only occultations around those years, while "best orbit" considers only orbits with $\sigma_a$<1.6$\times$10$^{-10}$~au.}
         \label{fig:occult_res_histogram}
\end{figure}  

We set up our test for the objects with an orbit in the FPR and  that were successfully observed by occultation. The selected data set resulted in 978 asteroids, associated with 5,774 astrometric measurements by stellar occultations. Large and bright asteroids with better orbits in the MPC or JPL data bases dominate the sample because their occultations are easier to predict.  In addition, because they are larger than the faint ones, their occultations last longer and are easier to catch.

We  propagated the FPR orbits to the epoch of each observed occultation and computed the corresponding position and its difference (O$-$C) in the directions of the equatorial frame with respect to the observed astrometry. 
We limit our study to occultations by main-belt asteroids, which are the largest majority of the sample for occultation astrometry and Gaia orbits. 
In the following, we discuss the total O$-$C, that is, the distance on the tangent sky plane between the observed and the computed position. 

The global distribution of the O$-$C values (Fig.~\ref{fig:occult_res_histogram}) reveals a large spread that reaches several hundred~milliarcseconds. Most observations lie below $\approx 50$~mas. This corresponds to the order of magnitude of the expected uncertainty for the dominating fraction of occultations with a small number of observers. In this situation, the cross-track astrometric error is dominated by the apparent size of the asteroid \citep{ferreira2022}.

However, when only occultation events contemporary to the Gaia mission (same figure) are selected, a clear subset of much smaller O$-$C appears. It peaks at about 10~mas and has a tail of high values that is substantially suppressed. 

We can then verify whether an equivalent improvement can be obtained by selecting only occultations that are related to objects with very high-quality orbits. We set a threshold on the maximum allowed $\sigma_a$  in such a way that the same number of occultation events as for the 2014-2020 period was selected (1819). The obtained curve shows that this selection is not sufficient to eliminate the high residuals, and the peak of the distribution does not move appreciably with respect to the overall distribution.

The apparent asteroid size (mentioned above as the main source of uncertainty) and other factors contributing  to errors of occultation astrometry were factored into the total error budget following the guidelines illustrated by \cite{herald2020}. A result that is more independent of the details of this error model can be obtained by dividing the O$-$C values by the uncertainty associated with each position (Fig.~\ref{fig:occult_res_histogram}, bottom panel). The obtained distribution is self-consistent and peaks around unity. The selection of the Gaia period clearly stands out as the best sample. 

\begin{figure}
   \centering
\includegraphics[width=0.95\hsize]{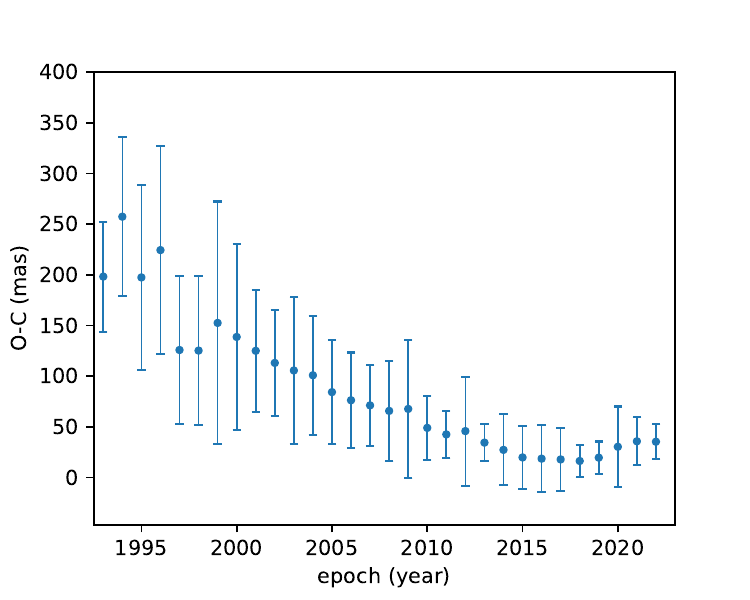} 
     \caption{Absolute values of O$-$C for the occultation data set averaged on yearly bases. The error bars correspond to the standard deviation. }
         \label{fig:res_time}
\end{figure} 

This evidence highlights a trend with time that is revealed by Fig.~\ref{fig:res_time}. Uncertainties in the occultation astrometry in the 1990s may be affected by worse statistics, but later observations are rather homogeneous \citep[Fig. 2 in ][]{ferreira2022}. The minimum around the period of nominal operations of Gaia that was used to obtain the FPR orbits is thus not due to the properties of the errors in asteroid occultations, but is clearly related to the time interval covered by the Gaia data. Interestingly, this result allows us to quantify a degradation in the ephemeris of $\sim 10$~mas/year. This clearly only holds for the asteroid sample that is observed via occultations, which is biased towards bright asteroids. Gaia astrometry is excellent but not fully optimal for these sources. 

At this stage, we also investigated this time trend for occultations involving objects with only the best orbits (same selection as in Fig.~\ref{fig:occult_res_histogram}). We found no significant difference. This probably means that the degradation in our sample of orbits is more sensitive to the limited observing arc than to the internal differences in quality.

\section{Conclusion}
We have reported a new orbital solution for nearly $160,000$ asteroids built on the Gaia astrometric measurements collected over $66$ months of operation and showed its main properties in terms of accuracy and reliability. Comparisons with the best existing orbits published by the JPL or the MPC have shown the impressive consistency between these solutions, which arises from different observation sets and time coverage and dynamical modelling. We established the absence of systematic differences larger than  $5\times 10^{-10}$ in relative difference in the semi-major axes.  

This Gaia FPR orbital solution is the first within the \gaia releases whose time coverage is at least equal to the orbital period of  main-belt asteroids. This feature is needed to ensure the accuracy of the orbit solution. These expectations were mostly shown to be correct in this paper, with an impressive overall improvement compared to \gdrthree solution by a factor almost 50 (see \autoref{fig: Gaia_orbits_siga_arclength}), whereas the number of observations was just twice as large. 

An important issue that arose in the course of this work is that institutes that publish orbital elements should agree on the reference ecliptic (same plane and same origin of longitude), while different realisations are available that differ from each other by several tens milliarcseconds. With the current quality reached at JPL, MPC, NEODyS, and Gaia, differences as small as this matter. Tiny deviations of few $1\times 10^{-8}$ in dimensionless parameters or a few  milliarcseconds in the angular elements on otherwise perfectly agreeing orbits are easily created with small differences in reference frames. With  sub-milliarcsecond astrometry and orbits approaching $1\times 10^{-10}$ in relative accuracy from JPL or \gaia, the community needs to coordinate, possibly with IAU or IERS, in order to agree about the definition and realisation of the ecliptic, physical or conventional, and about its relation to ICRF and J2000. This goes much beyond the orbits of small bodies because it must be consistent with the precession-nutation model the user will apply to obtain the positions in the equator of date.

The next step will be the future Gaia data releases, starting with \gdrfour. They will complete the current set in two ways: (i) the photometric and spectrophotometric data, and data for comets and planetary satellites not analysed here; (ii) a much larger set of asteroids, about twice as large in number, because this has been limited in the FPR to the same set as selected for \gdrthree, solely because the astrometric solution for the remainder was not yet available. 

A few years from today, the Gaia mission will finally provide data over approximately ten\,years, which will double the time interval of the FPR data we used here, and which will improve the quality, precision, and accuracy of the orbit determination of small Solar System bodies further. With this trend, Trojan orbits are expected to be as good as those of the main-belt asteroids are today. As of June 2023, about 85\% of the expected data are already safely archived on the ground for the upcoming processing. 

\begin{acknowledgements}
We are grateful to an anonymous referee of their kind remarks and constructive comments on several key-points of the paper. Their review has helped significantly improve the organisation and clarity of the paper.

This work presents results from the European Space Agency (ESA) space mission \gaia. \gaia\ data are being processed by the \gaia\ Data Processing and Analysis Consortium (DPAC). Funding for the DPAC is provided by national institutions, in particular the institutions participating in the \gaia\ MultiLateral Agreement (MLA). The \gaia\ mission website is \url{https://www.cosmos.esa.int/gaia}. The \gaia\ archive website is \url{https://archives.esac.esa.int/gaia}.
\end{acknowledgements} 

\bibliographystyle{aa} 
\bibliography{general, dpac, orbits} 

\begin{appendix}

\section{}\label{ssec:appendixA}

\subsection*{Access FPR asteroid data in the Gaia archive}

The data used in this article can be found at the Gaia mission archive of ESA\footnote{\url{https://gea.esac.esa.int/archive/}}. By navigating to the ''(Advanced) ADQL tab'', the tree of the data base can be expanded. Asteroid data are in the \texttt{focused\_product\_release} branch, in tables named \texttt{sso\_source} (orbits) and \texttt{sso\_observation} (epoch astrometry).

The meaning of the fields for the astrometry is extensively explained in the \gdrthree documentation and the related article \citep{DR3-DPACP-150}. The orbital elements have their equivalent in the \texttt{gaiadr3.sso\_orbits} table, with all the relevant details provided in this article. The residuals used to derive data statistics can be computed starting from the published orbits.

All the Solar System objects present in the FPR were also part of \gdrthree. However, as the FPR processing is completely independent of the previous processing, the inverse is not strictly true and a small number of \gdrthree sources may be missing in the FPR. These differences are very marginal, however.

We provide below an example of an ADQL query that can be used to retrieve the state vector and semi-major axis uncertainty for all the valid orbital solutions, also excluding objects for which the heliocentric orbit cannot be computed (natural planetary satellites): 

{\small\begin{verbatim}
SELECT number_mp, denomination, 
osc_epoch,
ESDC_ARRAY_ELEMENT(h_state_vector,1) AS x,
ESDC_ARRAY_ELEMENT(h_state_vector,2) AS y,
ESDC_ARRAY_ELEMENT(h_state_vector,3) AS z,
ESDC_ARRAY_ELEMENT(h_state_vector,4) AS vx,
ESDC_ARRAY_ELEMENT(h_state_vector,5) AS vy,
ESDC_ARRAY_ELEMENT(h_state_vector,6) AS vz,
SQRT(ESDC_ARRAY_ELEMENT
   (orbital_elements_var_covar_matrix, 1))
   AS sigma_a
FROM gaiafpr.sso_source WHERE ESDC_ARRAY_ELEMENT
(orbital_elements_var_covar_matrix,1)>0.
ORDER BY number_mp ASC
\end{verbatim}
}

The output is sorted by asteroid number and contains 156,762 orbital solutions. The content of the whole covariance matrix can be added by a similar method. The filter for the first element of the covariance matrix is sometimes negative when the orbit solution is poor.

All the astrometric measurements for a given object can be retrieved based on their identifiers. For instance, for (704)~Interamnia,

{\small\begin{verbatim}
SELECT * FROM gaiafpr.sso_observation 
WHERE number_mp=704.    
\end{verbatim}
}

To retrieve the astrometry for all asteroids associated with the best valid orbits with $\sigma_a$<$1\times 10^{-10}$~au, we can write

{\small\begin{verbatim}
SELECT astrom.*
FROM gaiafpr.sso_observation AS astrom
INNER JOIN
gaiafpr.sso_source AS so USING(number_mp)
WHERE ESDC_ARRAY_ELEMENT
(so.orbital_elements_var_covar_matrix,1)>0.
AND SQRT(ESDC_ARRAY_ELEMENT
   (so.orbital_elements_var_covar_matrix,1))<1.e-10 
AND astrom.is_rejected = 'false'
\end{verbatim}
}

Finally, in the following example, we compute the time span covered by the observations of each asteroid,

{\small\begin{verbatim}
SELECT number_mp, MAX(epoch)-MIN(epoch) AS t_delta_days
FROM (SELECT astrom.number_mp, astrom.epoch
FROM gaiafpr.sso_observation
AS astrom INNER JOIN gaiafpr.sso_source
AS so USING(number_mp)) AS subquery 
WHERE(number_mp>0) GROUP BY number_mp.     
\end{verbatim}
}

\section{}\label{ssec:appendixB}

\subsection*{Transformation of the state vector into ecliptic}
The state vector resulting from the orbit fitting is a 6D vector giving the initial condition at the epoch (one per body) as a combination of the position vector and the velocity vector into a single state vector. The origin is heliocentric, and the axes are aligned to the ICRF axes. The units are au and au/day, and the scaling of \autoref{eq:scaling_SV} must be applied to obtain the TDB compatible values. By reference to \autoref{fig:ecliptics} and \autoref{tab:icrf_ecliptic}, the coordinates of the same vector in the heliocentric ecliptic frame are given by
\begin{align}
    \mathbf{r}_\text{ecl} &=\mathcal{R}_z(\psi)\,\mathcal{R}_x(\epsilon) \,\mathcal{R}_z(-\phi) \,\mathbf{r}_\text{ICRF} \\
    \mathbf{v}_\text{ecl} &=\mathcal{R}_z(\psi)\,\mathcal{R}_x(\epsilon) \,\mathcal{R}_z(-\phi) \,\mathbf{v}_\text{ICRF,} 
\end{align}
where $\mathcal{R}_{x,y,z}(\alpha)$ denotes the passive rotation of angle $\alpha$ about the axes $x,$  $y,$ or $z$. In matrix form, this is\begin{align*}
\mathcal{R}_x(\alpha) &= \begin{pmatrix*}[r]
1 & 0&0\\
0 & \cos\alpha & \sin\alpha\\
0 &-\sin\alpha & \cos\alpha
\end{pmatrix*}\\
\mathcal{R}_z(\alpha) &= \begin{pmatrix*}[r]
\cos\alpha & \sin\alpha &0\\
-\sin\alpha & \cos\alpha&0\\
0 & 0& 1
\end{pmatrix*}.
\end{align*}
With the state vector given in the ecliptic frame (more precisely, in one of the possible choices of the ecliptic), the transformation into osculating elements is common to all the cases and is normally available with reliable routines in any group handling orbits to perform the transformation in both directions for elliptical or hyperbolic orbits. However, the solar mass parameter must also be included in in this transformation to ensure that it agrees with the mass parameter that is used in the force model.

The user who needs to compute a good precision (e.g. lower than 10 mas) position in the mean equator of the date must be aware that some precession routines may automatically include some of the rotations above in the form of a frame bias to relate the ICRF frame to the J2000 equatorial frame.
\section{}\label{ssec:appendixC}

The \gaia\ mission and data processing have financially been supported by (in alphabetical order by country)
\begin{itemize}
\item the Algerian Centre de Recherche en Astronomie, Astrophysique et G\'{e}ophysique of Bouzareah Observatory;
\item the Austrian Fonds zur F\"{o}rderung der wissenschaftlichen Forschung (FWF) Hertha Firnberg Programme through grants T359, P20046, and P23737;
\item the BELgian federal Science Policy Office (BELSPO) through various PROgramme de D\'{e}veloppement d'Exp\'{e}riences scientifiques (PRODEX) grants and the Polish Academy of Sciences - Fonds Wetenschappelijk Onderzoek through grant VS.091.16N, and the Fonds de la Recherche Scientifique (FNRS), and the Research Council of Katholieke Universiteit (KU) Leuven through grant C16/18/005 (Pushing AsteRoseismology to the next level with TESS, GaiA, and the Sloan DIgital Sky SurvEy -- PARADISE);  
\item the Brazil-France exchange programmes Funda\c{c}\~{a}o de Amparo \`{a} Pesquisa do Estado de S\~{a}o Paulo (FAPESP) and Coordena\c{c}\~{a}o de Aperfeicoamento de Pessoal de N\'{\i}vel Superior (CAPES) - Comit\'{e} Fran\c{c}ais d'Evaluation de la Coop\'{e}ration Universitaire et Scientifique avec le Br\'{e}sil (COFECUB);
\item the Chilean Agencia Nacional de Investigaci\'{o}n y Desarrollo (ANID) through Fondo Nacional de Desarrollo Cient\'{\i}fico y Tecnol\'{o}gico (FONDECYT) Regular Project 1210992 (L.~Chemin);
\item the National Natural Science Foundation of China (NSFC) through grants 11573054, 11703065, and 12173069, the China Scholarship Council through grant 201806040200, and the Natural Science Foundation of Shanghai through grant 21ZR1474100;  
\item the Tenure Track Pilot Programme of the Croatian Science Foundation and the \'{E}cole Polytechnique F\'{e}d\'{e}rale de Lausanne and the project TTP-2018-07-1171 `Mining the Variable Sky', with the funds of the Croatian-Swiss Research Programme;
\item the Czech-Republic Ministry of Education, Youth, and Sports through grant LG 15010 and INTER-EXCELLENCE grant LTAUSA18093, and the Czech Space Office through ESA PECS contract 98058;
\item the Danish Ministry of Science;
\item the Estonian Ministry of Education and Research through grant IUT40-1;
\item the European Commission’s Sixth Framework Programme through the European Leadership in Space Astrometry (\href{https://www.cosmos.esa.int/web/gaia/elsa-rtn-programme}{ELSA}) Marie Curie Research Training Network (MRTN-CT-2006-033481), through Marie Curie project PIOF-GA-2009-255267 (Space AsteroSeismology \& RR Lyrae stars, SAS-RRL), and through a Marie Curie Transfer-of-Knowledge (ToK) fellowship (MTKD-CT-2004-014188); the European Commission's Seventh Framework Programme through grant FP7-606740 (FP7-SPACE-2013-1) for the \gaia\ European Network for Improved data User Services (\href{https://gaia.ub.edu/twiki/do/view/GENIUS/}{GENIUS}) and through grant 264895 for the \gaia\ Research for European Astronomy Training (\href{https://www.cosmos.esa.int/web/gaia/great-programme}{GREAT-ITN}) network;
\item the European Cooperation in Science and Technology (COST) through COST Action CA18104 `Revealing the Milky Way with \gaia (MW-Gaia)';
\item the European Research Council (ERC) through grants 320360, 647208, and 834148 and through the European Union’s Horizon 2020 research and innovation and excellent science programmes through Marie Sk{\l}odowska-Curie grant 745617 (Our Galaxy at full HD -- Gal-HD) and 895174 (The build-up and fate of self-gravitating systems in the Universe) as well as grants 687378 (Small Bodies: Near and Far), 682115 (Using the Magellanic Clouds to Understand the Interaction of Galaxies), 695099 (A sub-percent distance scale from binaries and Cepheids -- CepBin), 716155 (Structured ACCREtion Disks -- SACCRED), 951549 (Sub-percent calibration of the extragalactic distance scale in the era of big surveys -- UniverScale), and 101004214 (Innovative Scientific Data Exploration and Exploitation Applications for Space Sciences -- EXPLORE);
\item the European Science Foundation (ESF), in the framework of the \gaia\ Research for European Astronomy Training Research Network Programme (\href{https://www.cosmos.esa.int/web/gaia/great-programme}{GREAT-ESF});
\item the European Space Agency (ESA) in the framework of the \gaia\ project, through the Plan for European Cooperating States (PECS) programme through contracts C98090 and 4000106398/12/NL/KML for Hungary, through contract 4000115263/15/NL/IB for Germany, and through PROgramme de D\'{e}veloppement d'Exp\'{e}riences scientifiques (PRODEX) grant 4000127986 for Slovenia;  
\item the Academy of Finland through grants 299543, 307157, 325805, 328654, 336546, and 345115 and the Magnus Ehrnrooth Foundation;
\item the French Centre National d’\'{E}tudes Spatiales (CNES), the Agence Nationale de la Recherche (ANR) through grant ANR-10-IDEX-0001-02 for the `Investissements d'avenir' programme, through grant ANR-15-CE31-0007 for project `Modelling the Milky Way in the \gaia era’ (MOD4Gaia), through grant ANR-14-CE33-0014-01 for project `The Milky Way disc formation in the \gaia era’ (ARCHEOGAL), through grant ANR-15-CE31-0012-01 for project `Unlocking the potential of Cepheids as primary distance calibrators’ (UnlockCepheids), through grant ANR-19-CE31-0017 for project `Secular evolution of galaxies' (SEGAL), and through grant ANR-18-CE31-0006 for project `Galactic Dark Matter' (GaDaMa), the Centre National de la Recherche Scientifique (CNRS) and its SNO \gaia of the Institut des Sciences de l’Univers (INSU), its Programmes Nationaux: Cosmologie et Galaxies (PNCG), Gravitation R\'{e}f\'{e}rences Astronomie M\'{e}trologie (PNGRAM), Plan\'{e}tologie (PNP), Physique et Chimie du Milieu Interstellaire (PCMI), and Physique Stellaire (PNPS), the `Action F\'{e}d\'{e}ratrice \gaia' of the Observatoire de Paris, the R\'{e}gion de Franche-Comt\'{e}, the Institut National Polytechnique (INP) and the Institut National de Physique nucl\'{e}aire et de Physique des Particules (IN2P3) co-funded by CNES;
\item the German Aerospace Agency (Deutsches Zentrum f\"{u}r Luft- und Raumfahrt e.V., DLR) through grants 50QG0501, 50QG0601, 50QG0602, 50QG0701, 50QG0901, 50QG1001, 50QG1101, 50\-QG1401, 50QG1402, 50QG1403, 50QG1404, 50QG1904, 50QG2101, 50QG2102, and 50QG2202, and the Centre for Information Services and High Performance Computing (ZIH) at the Technische Universit\"{a}t Dresden for generous allocations of computer time;
\item the Hungarian Academy of Sciences through the Lend\"{u}let Programme grants LP2014-17 and LP2018-7 and the Hungarian National Research, Development, and Innovation Office (NKFIH) through grant KKP-137523 (`SeismoLab');
\item the Science Foundation Ireland (SFI) through a Royal Society - SFI University Research Fellowship (M.~Fraser);
\item the Israel Ministry of Science and Technology through grant 3-18143 and the Tel Aviv University Center for Artificial Intelligence and Data Science (TAD) through a grant;
\item the Agenzia Spaziale Italiana (ASI) through contracts I/037/08/0, I/058/10/0, 2014-025-R.0, 2014-025-R.1.2015, and 2018-24-HH.0 to the Italian Istituto Nazionale di Astrofisica (INAF), contract 2014-049-R.0/1/2 to INAF for the Space Science Data Centre (SSDC, formerly known as the ASI Science Data Center, ASDC), contracts I/008/10/0, 2013/030/I.0, 2013-030-I.0.1-2015, and 2016-17-I.0 to the Aerospace Logistics Technology Engineering Company (ALTEC S.p.A.), INAF, and the Italian Ministry of Education, University, and Research (Ministero dell'Istruzione, dell'Universit\`{a} e della Ricerca) through the Premiale project `MIning The Cosmos Big Data and Innovative Italian Technology for Frontier Astrophysics and Cosmology' (MITiC);
\item the Netherlands Organisation for Scientific Research (NWO) through grant NWO-M-614.061.414, through a VICI grant (A.~Helmi), and through a Spinoza prize (A.~Helmi), and the Netherlands Research School for Astronomy (NOVA);
\item the Polish National Science Centre through HARMONIA grant 2018/30/M/ST9/00311 and DAINA grant 2017/27/L/ST9/03221 and the Ministry of Science and Higher Education (MNiSW) through grant DIR/WK/2018/12;
\item the Portuguese Funda\c{c}\~{a}o para a Ci\^{e}ncia e a Tecnologia (FCT) through national funds, grants SFRH/\-BD/128840/2017 and PTDC/FIS-AST/30389/2017, and work contract DL 57/2016/CP1364/CT0006, the Fundo Europeu de Desenvolvimento Regional (FEDER) through grant POCI-01-0145-FEDER-030389 and its Programa Operacional Competitividade e Internacionaliza\c{c}\~{a}o (COMPETE2020) through grants UIDB/04434/2020 and UIDP/04434/2020, and the Strategic Programme UIDB/\-00099/2020 for the Centro de Astrof\'{\i}sica e Gravita\c{c}\~{a}o (CENTRA);  
\item the Slovenian Research Agency through grant P1-0188;
\item the Spanish Ministry of Economy (MINECO/FEDER, UE), the Spanish Ministry of Science and Innovation (MICIN), the Spanish Ministry of Education, Culture, and Sports, and the Spanish Government through grants BES-2016-078499, BES-2017-083126, BES-C-2017-0085, ESP2016-80079-C2-1-R, ESP2016-80079-C2-2-R, FPU16/03827, PDC2021-121059-C22, RTI2018-095076-B-C22, and TIN2015-65316-P (`Computaci\'{o}n de Altas Prestaciones VII'), the Juan de la Cierva Incorporaci\'{o}n Programme (FJCI-2015-2671 and IJC2019-04862-I for F.~Anders), the Severo Ochoa Centre of Excellence Programme (SEV2015-0493), and MICIN/AEI/10.13039/501100011033 (and the European Union through European Regional Development Fund `A way of making Europe') through grant RTI2018-095076-B-C21, the Institute of Cosmos Sciences University of Barcelona (ICCUB, Unidad de Excelencia `Mar\'{\i}a de Maeztu’) through grant CEX2019-000918-M, the University of Barcelona's official doctoral programme for the development of an R+D+i project through an Ajuts de Personal Investigador en Formaci\'{o} (APIF) grant, the Spanish Virtual Observatory through project AyA2017-84089, the Galician Regional Government, Xunta de Galicia, through grants ED431B-2021/36, ED481A-2019/155, and ED481A-2021/296, the Centro de Investigaci\'{o}n en Tecnolog\'{\i}as de la Informaci\'{o}n y las Comunicaciones (CITIC), funded by the Xunta de Galicia and the European Union (European Regional Development Fund -- Galicia 2014-2020 Programme), through grant ED431G-2019/01, the Red Espa\~{n}ola de Supercomputaci\'{o}n (RES) computer resources at MareNostrum, the Barcelona Supercomputing Centre - Centro Nacional de Supercomputaci\'{o}n (BSC-CNS) through activities AECT-2017-2-0002, AECT-2017-3-0006, AECT-2018-1-0017, AECT-2018-2-0013, AECT-2018-3-0011, AECT-2019-1-0010, AECT-2019-2-0014, AECT-2019-3-0003, AECT-2020-1-0004, and DATA-2020-1-0010, the Departament d'Innovaci\'{o}, Universitats i Empresa de la Generalitat de Catalunya through grant 2014-SGR-1051 for project `Models de Programaci\'{o} i Entorns d'Execuci\'{o} Parallels' (MPEXPAR), and Ramon y Cajal Fellowship RYC2018-025968-I funded by MICIN/AEI/10.13039/501100011033 and the European Science Foundation (`Investing in your future');
\item the Swedish National Space Agency (SNSA/Rymdstyrelsen);
\item the Swiss State Secretariat for Education, Research, and Innovation through the Swiss Activit\'{e}s Nationales Compl\'{e}mentaires and the Swiss National Science Foundation through an Eccellenza Professorial Fellowship (award PCEFP2\_194638 for R.~Anderson);
\item the United Kingdom Particle Physics and Astronomy Research Council (PPARC), the United Kingdom Science and Technology Facilities Council (STFC), and the United Kingdom Space Agency (UKSA) through the following grants to the University of Bristol, the University of Cambridge, the University of Edinburgh, the University of Leicester, the Mullard Space Sciences Laboratory of University College London, and the United Kingdom Rutherford Appleton Laboratory (RAL): PP/D006511/1, PP/D006546/1, PP/D006570/1, ST/I000852/1, ST/J005045/1, ST/K00056X/1, ST/\-K000209/1, ST/K000756/1, ST/L006561/1, ST/N000595/1, ST/N000641/1, ST/N000978/1, ST/\-N001117/1, ST/S000089/1, ST/S000976/1, ST/S000984/1, ST/S001123/1, ST/S001948/1, ST/\-S001980/1, ST/S002103/1, ST/V000969/1, ST/W002469/1, ST/W002493/1, ST/W002671/1, ST/W002809/1, and EP/V520342/1.
\end{itemize}

The GBOT programme  uses observations collected at (i) the European Organisation for Astronomical Research in the Southern Hemisphere (ESO) with the VLT Survey Telescope (VST), under ESO programmes
092.B-0165,
093.B-0236,
094.B-0181,
095.B-0046,
096.B-0162,
097.B-0304,
098.B-0030,
099.B-0034,
0100.B-0131,
0101.B-0156,
0102.B-0174, and
0103.B-0165;
%
%
and (ii) the Liverpool Telescope, which is operated on the island of La Palma by Liverpool John Moores University in the Spanish Observatorio del Roque de los Muchachos of the Instituto de Astrof\'{\i}sica de Canarias with financial support from the United Kingdom Science and Technology Facilities Council, and (iii) telescopes of the Las Cumbres Observatory Global Telescope Network.

L. Liberato acknowledges support by the Coordena\,c\~ao de Aperfei\,coamento de Pessoal de N\'ivel Superior – Brasil (CAPES) – Finance Code 001, also by CAPES-PRINT Process 88887.570251/2020-00.

The authors want to acknowledge Valéry Lainey (IMCCE, Paris obseervatory) for providing the ephemerides of planetary satellites, and Josselin Desmars (IMCCE, Paris observatory) for providing extensive external and quality checks on the orbit computations from the NIMA software.\\
We made use of the software products \href{http://www.starlink.ac.uk/topcat/}{TOPCAT}, \citep{taylorTOPCATSTILStarlink2005}; Matplotlib \citep{Hunter:2007};
Astropy, a community-developed core Python package for Astronomy \citep{astropy:2013, astropy:2018}.

\end{appendix}
\end{document}